\documentclass[aps,twocolumn,pra,superscriptaddress,letterpaper,nofootinbib]{revtex4-1}

\usepackage{physics}
\usepackage{nicefrac}
\usepackage[pdftex]{graphicx}
\usepackage{subcaption}
\usepackage{upgreek}
\usepackage{textcomp}

\newcommand{\hg}[1]{HG#1}
\DeclareMathOperator{\arcsinc}{arcsinc}

\makeatletter
\newcommand{\vast}{\bBigg@{4}}
\newcommand{\Vast}{\bBigg@{5}}
\makeatother

\begin{document}

\title{High Dynamic Range Spatial Mode Decomposition}

\author{A. W. Jones}
\email{ajones@star.sr.bham.ac.uk}
\affiliation{School of Physics and Astronomy and Institute for Gravitational Wave Astronomy, University of Birmingham, Edgbaston, Birmingham B15 2TT, United Kingdom}
\author{M. Wang}
\affiliation{School of Physics and Astronomy, University of Birmingham, Edgbaston, Birmingham B15 2TT, United Kingdom}
\author{C. M. Mow-Lowry}
\affiliation{School of Physics and Astronomy and Institute for Gravitational Wave Astronomy, University of Birmingham, Edgbaston, Birmingham B15 2TT, United Kingdom}
\author{A. Freise}
\affiliation{School of Physics and Astronomy and Institute for Gravitational Wave Astronomy, University of Birmingham, Edgbaston, Birmingham B15 2TT, United Kingdom}


\begin{abstract}
Accurate readout of low-power optical higher-order spatial modes is of increasing importance to the precision metrology community. Mode sensors are used to prevent mode mismatches from degrading quantum and thermal noise mitigation strategies. Direct mode analysis sensors (MODAN) are a promising technology for real time monitoring of arbitrary higher-order modes. We demonstrate MODAN with photo-diode readout to mitigate the typically low dynamic range of CCDs. We look for asymmetries in the response our sensor to break degeneracies in the relative alignment of the MODAN and photo-diode and consequently improve the dynamic range of the mode sensor. We provide a tolerance analysis and show methodology that can be applied for sensors beyond first order spatial modes.
\end{abstract}
\maketitle
%%%%%%%%%%%%%%%%%%%%%%%%%%  body  %%%%%%%%%%%%%%%%%%%%%%%%%%
\section{Introduction}
Two fundamentally limiting noise sources in ground based interferometric gravitational wave (GW) detectors and optical clocks are thermal noise~\cite{Matei17,hafiz2019} and quantum (projection) noise~\cite{AdvancedLIGO15,AdvancedVirgo15,Ludlow15}. Advanced GW detectors, operating at high power, have implemented a squeezed vacuum as a quantum noise reduction technique~\cite{Grote2013,Acernese2019,Tse2019}. There are proposals to use a spatial Higher-Order Mode (HOM) as the carrier beam to mitigate thermal noise~\cite{Mours06,Chelkowski09,Ast2019}.

Squeezed vacuum is very sensitive to optical loss, thus requiring careful sensing and control of the 6 mode matching parameters (Vertical Axis Translation, Vertical Axis Tilt, Horizontal Axis Translation, Horizontal Axis Tilt, Waist Size and Waist Position) between the optical resonators. Furthermore, the high power used can lead to Parametric Instabilities~\cite{Biscans19}. Lastly, if a HOM is used as a carrier, mismatches cause extensive scattering into other modes, which places tight requirements on mirror astigmatism~\cite{Sorazu13}; mitigation strategies include \textit{in-situ} actuation~\cite{Allocca15}. 

GW detectors use interference between reflected first order modes and RF sidebands for minimization of resonator translation and tilt mismatches~\cite{anderson84} which is well developed~\cite{Babusci97,Slagmolen2005} and references therein. Direct detection of waist position and size mismatch is less well developed, but of increasing importance~\cite{Brooks15}. Such methods include: Bulls Eye photo detectors~\cite{mueller2000}, Mode Converters~\cite{fabian_qpd}, Hartmann Sensors~\cite{Brooks07} and the clipped photo-diode array discussed in~\cite{miller14} could be modified to be a direct mismatch sensor. Sensors beyond second order include scanning, lock in and Spatial Light Modulator (SLM) based phase cameras~\cite{Agatsuma19,cao2019,Ralph17}, as well as optical cavities~\cite{KSWD07,Takeno11,Bogan15}.

In contrast, direct mode analysis sensors (MODANs) (proposed~\cite{Golub82}) extract the phase and amplitude for each of higher order mode~\cite{Kaiser09} breaking degeneracy between modes of the same order. When used with an SLM, MODANs provide an independent, adjustable reference mode basis and do not need a reference beam~\cite{Flamm12}. The resulting sensor output can be readily and intuitively compared against models, offering substantial insight into the structure of the beam and easing mode matching.
%%%%
\begin{figure}
\centering
\includegraphics{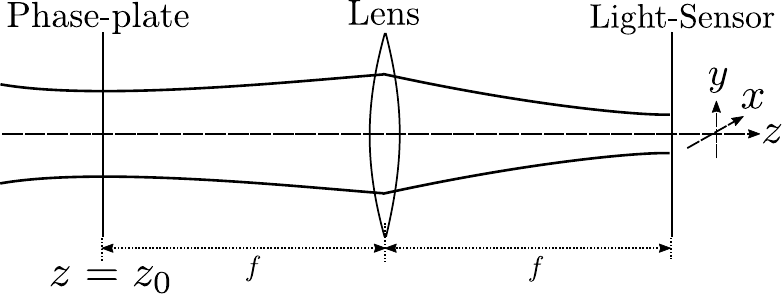}
\caption{Optical Convolution System. The light is incident on a DOE resulting in the field just after the DOE being, $U(x,y) = U_\text{in}(x,y)T(x,y)$. The light propagates a distance of $2f$ to a light-sensor (we use photodiode masked by a pinhole), with a lens of focal length $f$ placed half way between the sensor and the DOE.}
\label{fig:2f}
\end{figure}%
%%%%

Recent proposals~\cite{Kaiser09,Flamm12,dudley14} encode witness diffraction orders onto the diffractive optical element (DOE) and use a CCD as a light-sensor. This allows calibration of the relative alignment of between the CCD and DOE but limits the dynamic range. CCD blooming and streaking from light scattered by the phase-pattern limits the exposure time and dark noise is typically high.

This paper demonstrates MODAN with commercial low noise, high dynamic range, high bandwidth photo-diodes and 1064\,nm wavelength light. A pinhole of 5\,\textmu m aperture radius is used as a spatial filter to extract the signal from the scattered light. The relative alignment of the DOE and pinhole-photodiode assembly (referred to as light-sensor) is then explored by scanning the alignment of beam with respect to the phase-pattern, and positioning the light-sensor to eliminate asymmetries in the response of the system. A subsequent analytic calculation confirms the validity of this approach and is further used to develop a tolerance analysis for the pinhole aperture.

This work demonstrates the feasibility of high-dynamic-range mode-decomposition, an enabling technology for quantum and thermal noise reduction strategies. It can easily be extended to multi-branch MODANs. Furthermore, we note that our methodology is similar to mode division multiplexing with Multi-Mode Fibers~\cite{Wang18}, which is of increasing interest for increasing communications bandwidth~\cite{Richardson13}.

\section{Mode analyzers}
\label{sec:lownoise}
The methodology of mode decomposition is discussed extensively in~\cite{Forbes16}. In summary, the device consists of an optical convolution processor preceded by a DOE as shown in Fig.~\ref{fig:2f}. For given scalar input field $U_{in}(x,y,z_0)$, DOE transmission function, $T(x,y)$, and lens focal length, $f$, the field at the light-sensor is,
\begin{align}
U(x,&y,z_0+2f) \approx \frac{\exp\left(i\left(2kf + \frac{\pi}{2}\right)\right)}{f\lambda}
\int\int \mathrm{d}\xi \mathrm{d}\eta \nonumber \\
& U_{in}(\xi,\eta,z_0)T(\xi,\eta) \exp\left(\frac{-ik}{f}\left(\xi x + \eta y \right)\right),
\label{eq:2f}
\end{align}
as determined by repeated application of the Rayleigh Sommerfeld equation and all parameters defined as per~\cite{Bond2017}. We employ the modal model by setting,
\begin{align}
T(\xi,\eta) &= b_{n,m}u^*_{n,m}(\xi,\eta)\\
U_{in}(\xi,\eta,z_0) &= \sum_{n',m',}a_{n',m'}u_{n',m'}(x,y,z_0)e^{i(\omega t + kz_0)}.
\end{align}
Further, we assume that the mode basis functions, $u$, form a complete, orthonormal basis set and recognize the inner product; and neglect common phase factors, then the on-axis field at the sensor is,
\begin{align}
U(0,0,z_0+2f)
&\approx \sqrt{g_e}\frac{a_{n,m}b_{n,m}}{f\lambda}e^{i\omega t},
\end{align}
where $a_{n,m}$ is the amplitude of the mode (dimensions square-root power), and $g_e$ is the grating power efficiency (dimensionless) and $b_{n,m}$ (dimensions length) normalizes $T$. During detection the inter-modal phase information is typically lost, but, by designing the phase pattern to overlap two fields, $T^{\cos} = u_{n_0,m_0} + u_{n_1,m_1}$ and $T^{\sin} = u_{n_0,m_0} + iu_{n_1,m_1}$ the inter-modal phases can be recovered\cite{Kaiser09,Forbes16}.
\section{Experimental design}
\label{sec:ed}

\begin{figure}
\centering
\includegraphics{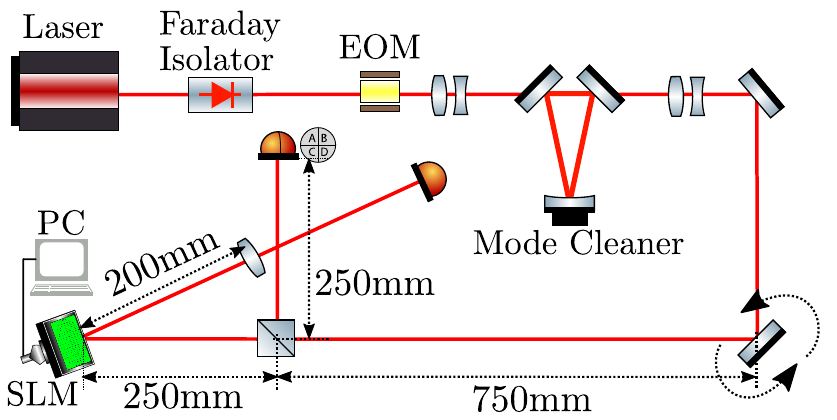}
\caption{Simplified Experimental Layout. The light is first filtered though an optical cavity to generate a high purity \hg{00} mode. A pair of steering mirrors then add controlled misalignment to the beam. The light is split between the MODAN under test and a witness QPD. The SLM is configured to display phase-pattern, $T(x,y)$ and works in reflection. Extraneous lens, waveplates and mirrors are not shown.}
\label{fig:exlayout}
\end{figure}%
The experimental layout is shown in Fig.~\ref{fig:exlayout}. % and a definition of the relative alignment is shown in Fig.~\ref{fig:geometry}. 
A laser source excites the eigenmodes of a resonator producing a high-purity spatially fundamental (HG) mode~\cite{Ast2019}. A triangular resonator, based on a Pre-Mode-Cleaner design~\cite{UeharaSPIE,WUGBKSS98}, was used due to its natural HG basis, good mode separation and $\pi$ radians Gouy phase difference between \hg{01} and \hg{10} modes. The light is then incident on a steering mirror before being split between a witness Quadrant PhotoDiode (QPD) and a MODAN. The beam radius at the SLM was $w_\text{SLM} = 1.2$\,mm.

\begin{figure}
	\centering
	\includegraphics{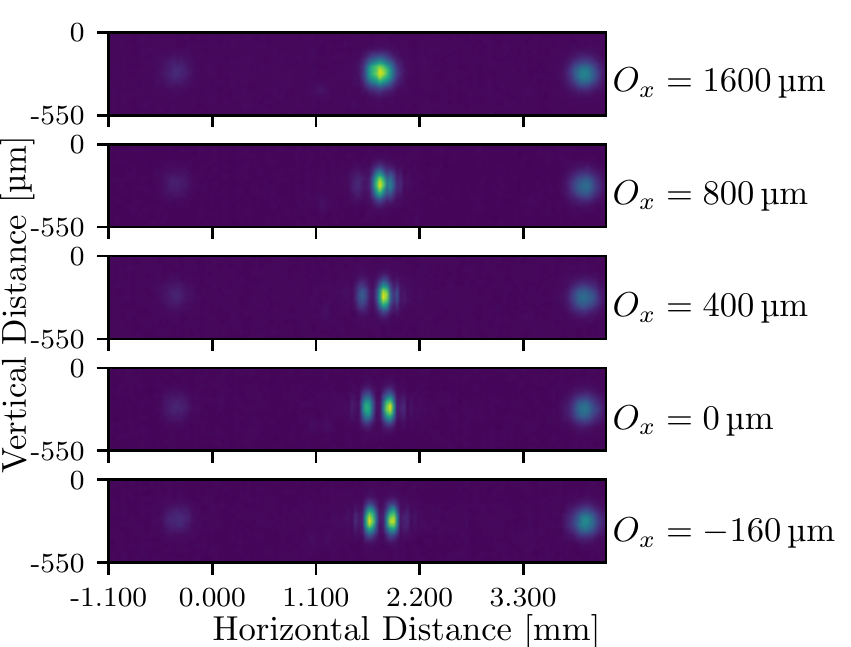}
	\caption{Camera images for several phase-pattern offsets. $O_x$ is the phase-pattern offsets with respect to the SLM. The central spot is the first diffraction order, with the specular and the second diffraction orders either side.}
	\label{fig:cmos}
\end{figure}

For small excitations of \hg{10} relevant to GW detectors, we misaligned this beam relative to the phase-pattern origin, since it can be described as an aligned beam with a small excitation of first order modes~\cite{anderson84}. This misalignment could either: be added in software, with the beam centered on the SLM (e.g. Fig.~\ref{fig:cmos}); or, using a steering mirror, with the phase-pattern origin centered on the SLM (e.g. Fig.~\ref{fig:fit}).

\begin{figure}
	\centering
	\includegraphics{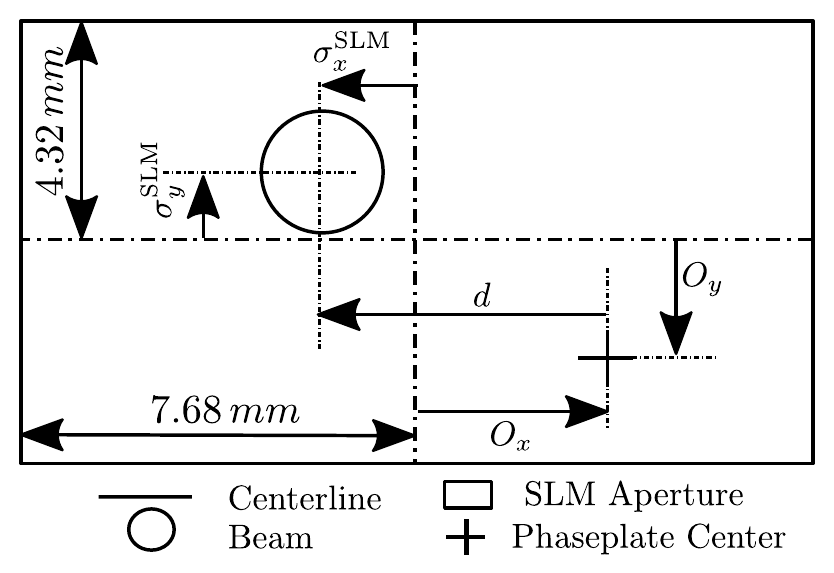}
	\caption{SLM Geometry to scale. The solid circle illustrates the point at which the power of the spatially fundamental beam falls to $1/e^2$ of peak intensity. $\sigma^{SLM}_{x,y}$, describes the position of the beam with respect to the SLM, $O_{x,y}$ describes the offset in software between the phase-pattern center and the SLM center and $d$ describes the relative $x$ offset between the phase-pattern origin and the beam.}
	\label{fig:geometry}
\end{figure}
\begin{figure}
	\centering
	\includegraphics{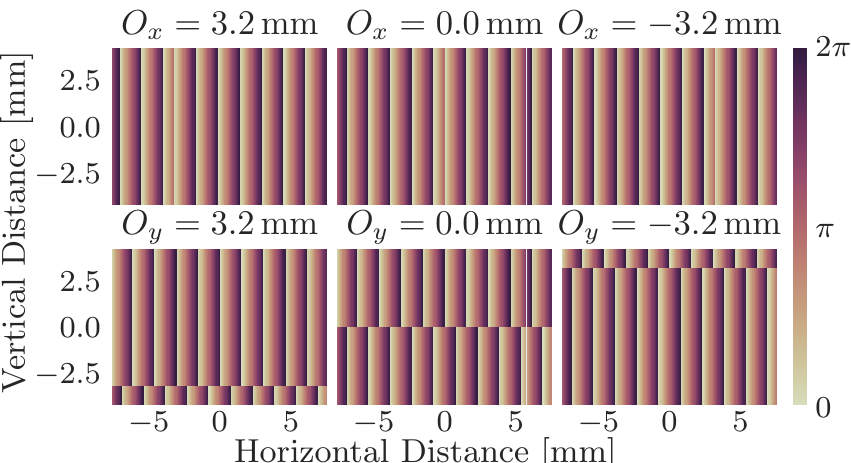}
	\caption{Phase-patterns with various software offsets. Upper patterns are $T^{PO}_{10}$ and lower pattern are $T^{PO}_{01}$. The grating period has been increased from 80\,\textmu m (10 pixels) which was used in the experiment, to 1536\,\textmu m and the number of pixels decreased by a factor 10 in both directions, to provide a legible figure.}
	\label{fig:phaseplates}
\end{figure}
A blazed grating was added to the phase-pattern and programmed onto a liquid crystal SLM (HOLOEYE PLUTO-2-NIR-015). This grating separated light which interacted with the MODAN from specular reflections.

HG phase only patterns were designed with transmission function,
\begin{align}
T^{PO}_{n,m}(x,y) = \exp \Bigg(&
	i \mod \bigg[\arg\left(u_{n,m}(x,y,z) \right)\\
		&+ \frac{2\pi (x\cos(\phi_s) + y\sin(\phi_s)}{\Lambda_s}\;,\; 2\pi\bigg]\Bigg)\nonumber,
	\label{eq:tpo}
\end{align}
where: $\phi_s$, is the grating angle; $\Lambda_s$, is the grating period; $u_{n,m}(x,y,z) \equiv u_{n}(x,z)u_{m}(y,z)$; and, 
\begin{align}
u_{n}(x,z_0) = &\left(\frac{2}{\pi}\right)^\frac{1}{4}\sqrt{\frac{\exp\left(i(2n+1)\Psi(z_0)\right)}{2^nn! w_0}}\nonumber \\
&\times \, H_n\left(\frac{\sqrt{2}x}{w_0}\right)
\exp\left( -\frac{x^2 + y^2}{w_0^2}\right),
\end{align}
is the spatial mode distribution function at the waist. All other parameters are defined as per~\cite{Bond2017}. This pattern was compared in simulation to a phase-pattern produced with phase and effective amplitude encoding~\cite{Bolduc13}. The transmission function was,
\begin{align}
T^{PA}_{n,m}(x,y) = \exp i 
	\Bigg(&
		\mathcal{M}(x,y)
		\mod \bigg[
			\mathcal{F}(x,y) \\&
			+ \frac{2\pi (x\cos(\phi_s) + y\sin(\phi_s)}{\Lambda_s}\;,\; 2\pi
		\bigg]
	\Bigg),\nonumber
	\label{eq:tpa}
\end{align}
where,
\begin{align}
\mathcal{M} &= 1 + \frac{\arcsinc \abs{u_{n,m}(x,y,z_0)}}{\pi}\\
\mathcal{F} &= \arg \left(u_{n,m}(x,y,z_0)\right) - \pi \mathcal{M}.
\end{align}
Aside from an overall reduction in grating efficiency when using $T^{PA}$, the features in our results obtained by FFT simulation~\cite{Brown2017} and experimentally were very similar (e.g. Fig~\ref{fig:fit}).

\section{Effect of a mis-positioned light-sensor}
\label{sec:effects}
\begin{figure*}[htb]
	\centering
	\includegraphics[width=\textwidth]{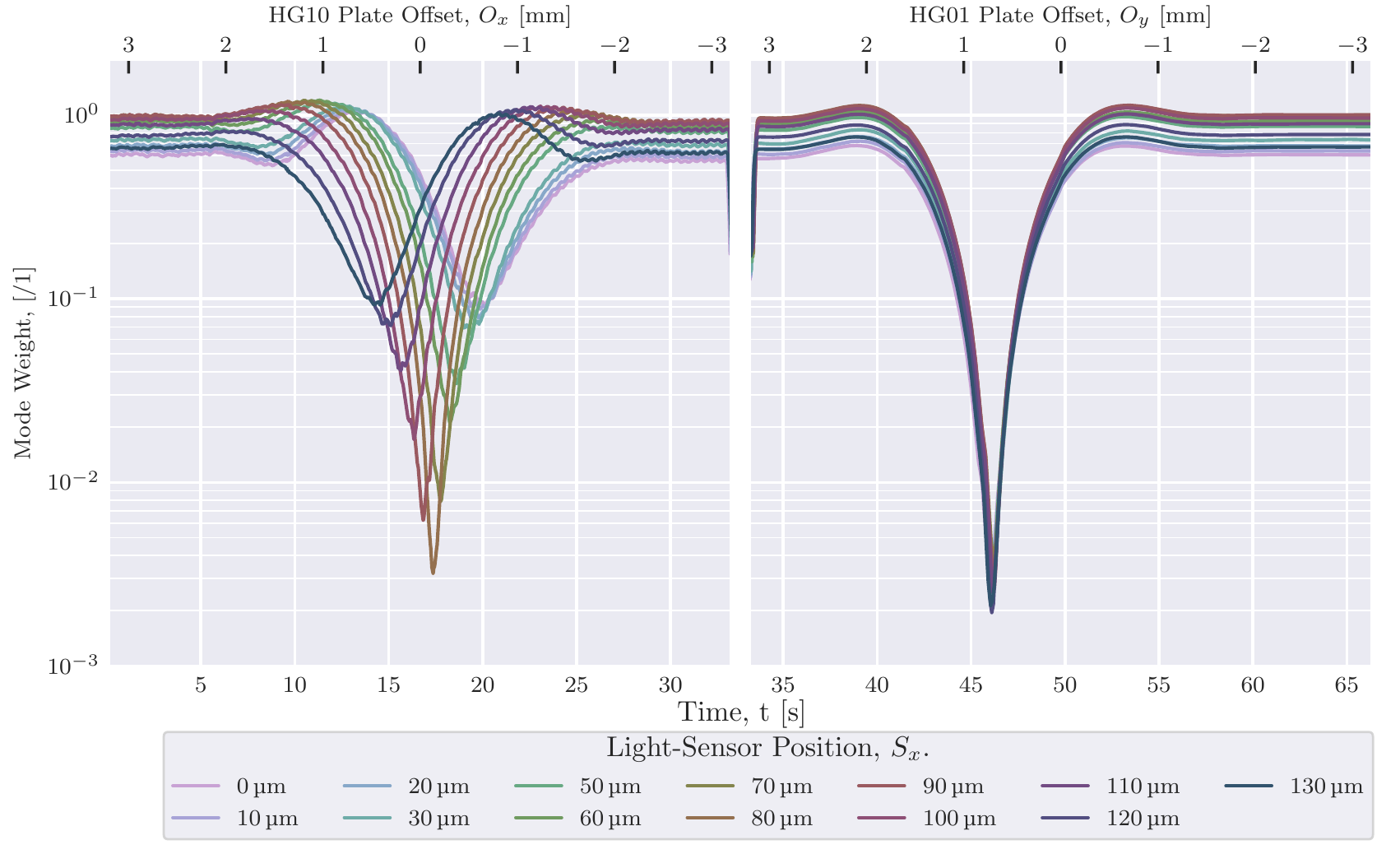}
	\caption{Light-Sensor Alignment Scan. The phase-pattern $x$ and $y$ offsets were varied in sequence while the beam remained incident on the center of the SLM (as determined with a viewing card) and the mode weights were measured. The measurement was repeated for several light-sensor $x$ positions. There is a 10\% calibration uncertainty and offset uncertainty $<3\times 10^{-5}$ for all measurements. The SLM input power was nominally $4\,mW$, which resulted in a maximum of $17\,$\textmu W on the photo-diode. The left panel shows HG10 mode weights measured with $T^{PO}_{10}(x-O_x(t),y)$, which was displayed for $33.33\,s$, followed by a blank calibration frame. The right panel shows HG01 weights measured with $T^{PO}_{01}(x,y-O_y(t))$ which was also displayed $33.33\,s$.}
	\label{fig:xscan}
\end{figure*}
The mode analyzer is a three component device, requiring careful relative alignment of each of these components for optimal performance. In this section, after preliminary alignment, we digitally scan the phase-pattern offset on the SLM while looking for asymmetries in the response of the system. By adjusting the light-sensor position (using a three axis translation stage) to eliminate the asymmetries, a high degree of alignment between the lens, phase-pattern and light-sensor is obtained, reducing TEM00 cross coupling and increasing dynamic range.

We define the possible beam and plate misalignments: $O_{x,y}$, $d$, $\sigma^\text{SLM}_{x,y}$ as per Fig.~\ref{fig:geometry}. We define, $\sigma_x^\text{QPD}$, to be the difference between the center of the SLM and the center of the QPD and $S_x$ to be the light-sensor misalignment.

The first order, HG, phase-only plates, shown in Fig.~\ref{fig:phaseplates}, do not depend on the beam parameter, and the \hg{01} and \hg{10} modes are orthogonal. Thus, by working with these plates and modes we separate horizontal alignment and vertical alignment into different measurements and mitigate beam radius mismatches, allowing a controlled study of the effect of horizontal light-sensor mis-positioning on \hg{10} readout.

For a first order phase only grating, $T^{PO}_{10}$, and misaligned TEM00 input beam, when $d > w_\text{SLM}$, little light interacts with the phase discontinuity, so the phase-pattern acts like a simple blazed grating, as shown in Fig.~\ref{fig:cmos} for $O_x = 1600$\,\textmu m. When the phase discontinuity is brought nearer the center of the beam, the device works as a mode analyzer and thus the intensity is,
\begin{align}
I \propto \abs{U(0,0,z_0+2f)}^2 
\propto \abs{a_{1,0}}^2 \propto d^2,
\end{align}
which is symmetric in $d$. 

We define the mode weight to be the ratio of mode power and input power,
\begin{align}
\rho_{n,m} = \frac{\abs{a_{n,m}}^2}{P},
\end{align}
this allows input power fluctuations to be normalized from the measurement.

Fig.~\ref{fig:xscan} shows a measurement of the mode weight, while $O_x$ is varied with a \hg{10} plate and $O_y$ with a \hg{01} plate for several light-sensor positions and constant $\sigma^\text{SLM}_x,\sigma^\text{SLM}_y$. The scan was achieved by creating a video out of several phase-patterns and displaying this on the SLM. The minima on each trace indicates the inferred beam position on the SLM.

When $S_x=80\,\upmu\text{m}$ the measured response of symmetric and shows the lowest mode weight measured $(0.34 \pm 0.03)\,\%$, implying a dynamic range $>300$. When the light-sensor is moved away from this position, the dynamic range is reduced and the response becomes asymmetric, thus incorrectly determining the HG10 mode weight.

The light-sensor $y$ position was optimized by eliminating the asymmetry in the response prior to collection of the data shown. For all light-sensor $x$ positions the response is symmetric and minima are within $(0.19 \pm 0.02)$\,\%, which is within calibration uncertainties on the beam radius and electrical gain, illustrating the orthogonality of the analysis.

The zero point is determined from the dark offset on the photodiode, measured before each trace with a statistical uncertainty $<3\times 10^{-5}$ in units of mode weight. The maximum mode power is determined by fitting the data to,
\begin{align}
\rho_{1,0} &= \left(\frac{\sqrt{2}(O_x - \sigma^\text{SLM}_x)}{w_\text{SLM}}\right)^2 + P_{\sigma,1,0},\\
\rho_{0,1} &= \left(\frac{\sqrt{2}(O_y - \sigma^\text{SLM}_y)}{w_\text{SLM}}\right)^2 + P_{\sigma, 0,1},
\end{align}
in the region $\abs{d} < 0.1w_\text{SLM}$. The result of the fit is $\sigma^\text{SLM}_x = (-0.12589 \pm 5\cross 10^{-5})$\,mm, $\sigma^\text{SLM}_y = (0.73685 \pm 7\cross 10^{-5})$\,mm. $P_{\sigma,n,m}$ are then the optical offsets shown above to limit the dynamic range, this is explained in section \ref{sec:finiteapa}. A 10\% calibration uncertainty exits on the maximum mode power due to instrumentation tolerances.

The blazing was in the $x$ plane, the motion of the blazing over the SLM causes a small periodic shifts in the optimal light-sensor position which is not present in the \hg{01} scan. Additionally the data shown was filtered with a low pass filter to reduce noise cause by the refresh of the SLM and motion of the blazing.
\section{Light-sensor position error signals}
\label{sec:errorsignals}
Given that a mispositioned light-sensor can cause systematic errors in the modal readout, it is important to develop error signals to control this degree of freedom.

The mode basis is set entirely by parameters on the phase-pattern, therefore, the light-sensor must be aligned with respect to this. In a recent demonstration of direct mode analysis, four adjustment branches were produced~\cite{Kaiser09}. These adjustment branches contained the unperturbed beam and provided a coordinate reference system on the CCD. The single branch analogue of this would be to place the light-sensor at the position of maximal intensity for a mode matched ($n=n',m=m'$) input beam and phase-pattern, however, this requires assuming that the beam and phase-pattern are already matched, which is in general not true.

In the case of a \hg{00} input beam and plate, the resulting power at the light-sensor has a stationary point at the point of maximal intensity, $\left.\nicefrac{\mathrm{d}I}{\mathrm{d}x}\right|_{x=0} = 0$. Therefore small levels of light-sensor mis-positioning are difficult to detect and directional information is missing.

In contrast, the scanning method shown in Fig.~\ref{fig:xscan}, breaks the degeneracy in light-sensor and phase-pattern position by eliminating asymmetries. Thus, by continuously scanning $O_x$ and adjusting the light-sensor position to balance the response of the MODAN, the light-sensor can be aligned with respect to the beam and phase-pattern.

\begin{figure}
	\centering
	\includegraphics{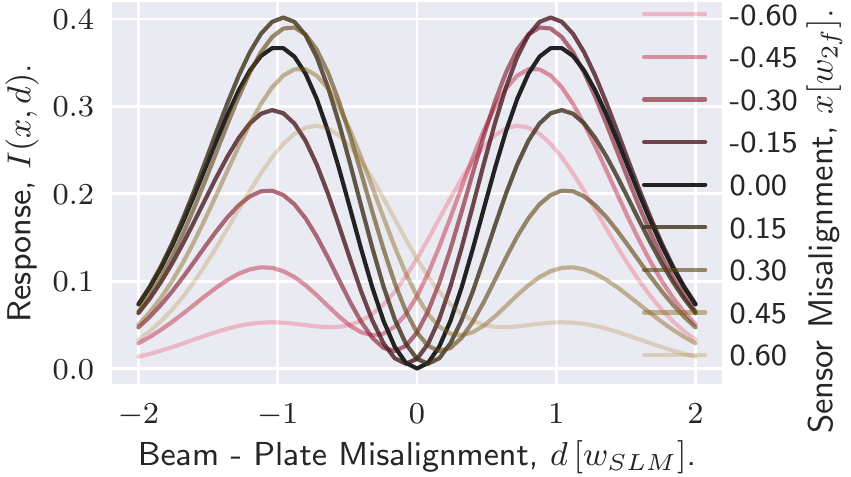}
	\caption{Ideal response of alignment MODAN to a relative misalignment between the beam and the phase-pattern, for several light-sensor positions. This is computed using Eq.~\ref{eq:pinhole_response}, with $a_0^H, a_1^H$ from \ref{eq:alpha0}, \ref{eq:alpha1} and inter-modal phase difference $\phi_0 - \phi_1 = \frac{\pi}{4}$.}
	\label{fig:theory}
\end{figure}
To analytically confirm this effect, consider Eq.~\ref{eq:2f}, use the transmission function for a phase and amplitude encoded \hg{10} plate, assume the incoming beam contains only horizontal misalignment modes, exploit the separability of the HG modes and assume the light-sensor is vertically aligned, then the field at the light-sensor is,
\begin{align}
U(x,0,z_0 + 2f)
\approx &\frac{b_{1}e^{i(2kf+\frac{\pi}{2})}}{f\lambda} \\
&\int_{-\infty}^\infty \bigg(a_0^H u_0 (\xi,z) + a_1^Hu_1(\xi,z) \bigg)\nonumber\\&\quad\quad
\bigg(u_{1}^*(\xi,z) \bigg)\exp\bigg(\frac{-ikx\xi}{f}\bigg) \mathrm{d}\xi \nonumber
\end{align}
where $b_{nm} = b_{n}^{H}b_m^V$ and similar for $a_{nm}$. We now construct the relevant ABCD matrix to describe the system as,
\begin{align}
\mathbf{M}_{2f} = 
	\begin{bmatrix}
		1 & f \\
		0 & 1
	\end{bmatrix}
	\begin{bmatrix}
		1 & 0 \\
		\frac{-1}{f} & 1
	\end{bmatrix}
	\begin{bmatrix}
		1 & f \\
		0 & 1
	\end{bmatrix}
	= \begin{bmatrix}
		0 & f \\
		\frac{-1}{f} & 0
	\end{bmatrix}.
\end{align}
We then assume the wavefront curvature at the SLM is $\infty$, $\frac{1}{R_C} = 0$, and determine that the beam radius at the DOE is,
\begin{align}
w_{2f} = \frac{\lambda f}{\pi w_{SLM}} = \frac{2f}{kw_{SLM}}. \label{eq:w2f}
\end{align}
By assuming the beam has a waist at the DOE, including the Gouy phase in the complex mode amplitudes and recognizing the $w_{2f}$ terms, we find that,
\begin{align}
U(x,0,z_0 + 2f)
\approx &\frac{b_{1}}{f\lambda} \exp \Bigg( i\big(2kf+\frac{\pi}{2}\big)- \frac{x^{2}}{2 w^2_{2f}} \Bigg)\label{eq:uf} \\ &\nonumber
\Bigg(- a_{0}^H\frac{ \sqrt{2} i x}{2w_{2f}} + a^H_{1}\left(1 - \frac{x^{2}}{w_{2f}^2}\right)\Bigg).
\end{align}
We then compute the intensity as $I = UU^*$ and find that,
\begin{align}
&I(x,0,z_0 + 2f) = \frac{\abs{b_{1}}^2}{f^2\lambda^2} e^{-x^2/w_{2f}^2}
\Bigg( \label{eq:pinhole_response} \\&
\abs{a^H_0}^2 \bigg(\frac{ \sqrt{2} x}{2w_{2f}}\bigg)^2 + \abs{a^H_1}^2 \bigg(1 - \frac{x^{2}}{w_{2f}^2}\bigg)^2 \nonumber \\& \nonumber
+ 2\abs{a^H_1} \abs{a^H_0} \bigg(1 - \frac{x^{2}}{w_{2f}^2}\bigg) \frac{ \sqrt{2} x}{2w_{2f}}\sin\left(\arg \left(a^H_0\right)-\arg \left(a^H_1\right)\right)\Bigg).
\end{align}
As we would expect, the sensitivity to misalignments is normalized by the waist size at the light-sensor, and this gives us an important insight when choosing a focal length for low noise mode analyzers.

We then note some interesting effects: $a_{0}^H$ couples into the signal, and there is a reduction in $a_{1}^H$, which are both proportional to the square of the waist normalized light-sensor mis-position. There is also a global reduction in total intensity which is exponentially sensitive to waist normalized light-sensor mis-position. Lastly and most importantly, there is interference between the zeroth and first order modes, which is proportional to the sine of the inter-modal phase difference; due to the factor $i$ acquired by the $u_{0}$ beam in Eq.~\ref{eq:uf}. This interference shifts the apparent minima by a small amount proportional to the light-sensor mis-position and causes the asymmetry which we observe in Fig.~\ref{fig:xscan}.

We can then compute the relevant mode amplitudes for an offset, $d$, between the phase-pattern origin and beam~\cite{bayer-helms}, 
\begin{align}
a_0^H &= \Big\langle u_0(\xi-O_x) \Big| u_0(\xi-\sigma_x^\text{SLM}) \Big\rangle = \exp \Bigg(\frac{-d^2}{2w^2_\text{SLM}}\Bigg)
\label{eq:alpha0}\\
a_1^H &= \Big\langle u_1(\xi-O_x) \Big| u_0(\xi-\sigma_x^\text{SLM}) \Big\rangle = -\frac{d\exp \Big(\frac{-d^2}{2w^2_\text{SLM}}\Big)}{w_\text{SLM}},\label{eq:alpha1}
\end{align}
with the inter-modal phase depending on the distance from the waist. Substituting this into Eq.~\ref{eq:pinhole_response} yields the anticipated response of the system to a beam-pattern misalignment scan at several light-sensor positions, plotted in Fig~\ref{fig:theory}. As expected, when the light-sensor is centered, the ideal response peaks when the first order mode power is maximum, $d = w_{SLM}$. Furthermore, when the pattern-beam misalignment becomes very large, $d \rightarrow \pm \infty$, or the first order mode amplitude is very small $d\rightarrow 0$ the response goes to zero. When the light-sensor becomes mis-centered, the cross talk and interference described above lead to an offset and asymmetry in the response.
\begin{figure}
	\centering
	\includegraphics{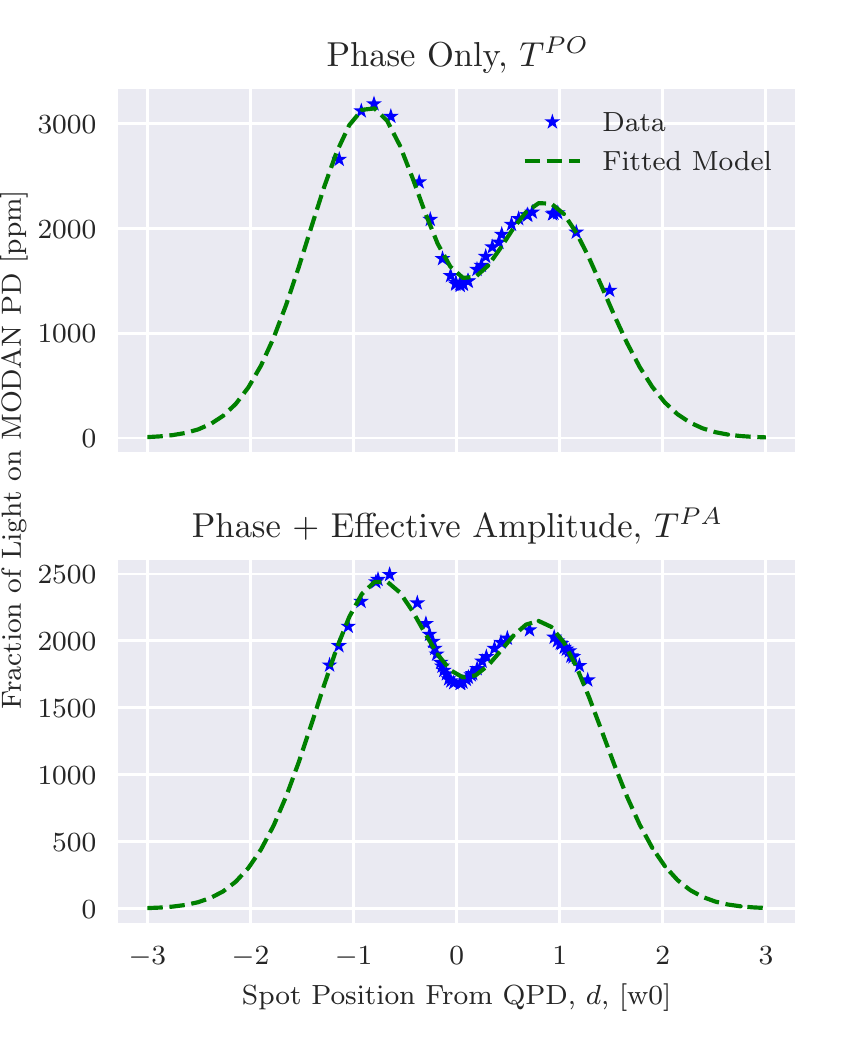}
	\caption{A steering mirror was used to scan the relative alignment between the incident light and a static phase-pattern on the SLM, a QPD was used as a witness sensor. Data could only be obtained in the region $|d| < 1$ due to the limited range of the QPD. The photo-diode offset, computed during the fit, has been added to both the data and the model. The upper and lower plots show the response for phase-pattern described by equations \ref{eq:tpo} and \ref{eq:tpa} respectively.}
	\label{fig:fit}
\end{figure}
\begin{table}
\centering
\begin{tabular}{| c | c | c |}
 \hline
 Phase-Pattern	& $T^{PO}_{10}$ & $T^{PA}_{10}$\\ \hline
 Light-Sensor Mis-position, $S_x$ [$w_{2f}$]	& $0.539 \pm 0.007$  & $0.595 \pm 0.003$ \\\hline
 Inter-modal Phase, $\phi_0 - \phi_1$ [deg] & $11 \pm 1$  & $3.8 \pm 0.4$ \\\hline
 QPD Offset, $\sigma^{QPD}_x$, [$w_{SLM}$] & $-0.027 \pm 0.015$ & $0.019 \pm 0.008$ \\\hline
\end{tabular}
\caption{Positioning offsets determined from fit.}
\label{tab:fitparams}
\end{table}

We can then fit data to Eq.~\ref{eq:pinhole_response} to determine the light-sensor offset, $S_x$, during operation. We misaligned the light-sensor position, centered the phase-pattern on the SLM ($O_x=O_y=0$) and added a small translational misalignment using a steering mirror. The light was then split between the mode analyzer and a witness QPD as shown in Fig.~\ref{fig:exlayout}.

The response of the MODAN is then plotted against the beam misalignment measured with the QPD in Fig.~\ref{fig:fit}. A Levenberg-Marquardt least squares regression~\cite{scipy,more1978levenberg} is used to extract the results shown in Table \ref{tab:fitparams}. 

Unlike Fig.~\ref{fig:xscan}, the inter-modal phase is close to zero and so the effect of the asymmetry is reduced, however, due to the large light-sensor mis-positioning, there is significant cross talk of the $a_0^H$ into the $a_1^H$ readout, leading to a reduced dynamic range. 

Thus we demonstrate, by changing the SLM to a pattern $T_{10}^{PO}$, scanning the position of the incoming beam and fitting the response, it is possible to determine the light-sensor mis-position. Here, the beam position is scanned on a stationary phase-pattern, however, it would also be possible to scan the phase-pattern position (as in Fig.~\ref{fig:xscan}) and then fit.
\section{Finite aperture effects}
\label{sec:finiteapa}
At any point other than, $x=0$, $a_0$ couples into the signal. Thus the finite size of the pixel in the CCD, or photo-diode aperture, will experience this coupling, reducing the dynamic range. We compute this effect for a centered light-sensor of radius $r_a$. The field at the light-sensor for a vertically aligned and \hg{00} incoming beam and $T^{PA}_{10}$ phase-pattern is,
\begin{align}
U(&x,y,z_0 + 2f)
\approx \frac{b_{10}e^{i(2kf+\frac{\pi}{2})}}{f\lambda} \nonumber \\&
\int_{-\infty}^\infty a_0^V u_0(\eta,z_0) u^*_0(\eta,z_0) \exp\bigg(\frac{-iky\eta}{f}\bigg)\mathrm{d}\eta,  \\\nonumber
& \int_{-\infty}^\infty \bigg(a_0^H u_0 (\xi,z) + a_1^Hu_1(\xi,z) \bigg)u_{1}^*(\xi,z)\exp\bigg(\frac{-ikx\xi}{f}\bigg) \mathrm{d}\xi. 
\end{align}
Solving, simplifying, substituting to cylindrical coordinates and integrating between $0 \leq \theta \leq 2\pi$ and $0 \leq r \leq r_a$, yields,
%\begin{widetext}
\begin{align}
P_T(r_a)= \frac{\abs{a_{0}^V}^{2} \abs{b_{10}}^{2}}{\lambda^{2} f^{2}} 
\vast[
	\frac{\pi w_{2f}^{2} \abs{a_{0}^H}^{2}}{2}
 	\vast\{\nonumber&\\
 		1
 		- \left(
 			1 + \frac{r_{a}^{2}}{w_{2f}^{2}} 
 		\right)
 		\exp\left(
 			- \frac{r_{a}^{2}}{w_{2f}^{2}}
 		\right)
 	\vast\} \label{eq:pinhole_size}\\
	+ \frac{\abs{a_{1}^H}^{2}\pi w_{2f}^{2}}{4}
	\vast\{
		\frac{
			r_{a}^{2}\exp
			\left(
				- \frac{r_{a}^{2}}{w_{2f}^{2}}
			\right)
		}{w_{2f}^{2}} 
		\left(
			1 - \frac{3 r_{a}^{2}}{2 w_{2f}^{2}}
		\right)&\nonumber\\
		+ 3 
		\left(
			1 - \exp\left(- \frac{r_{a}^{2}}{w_{2f}^{2}}\right)
		\right)
	\vast\}
\vast]. \nonumber
\end{align}
\begin{figure}
	\centering
	\includegraphics{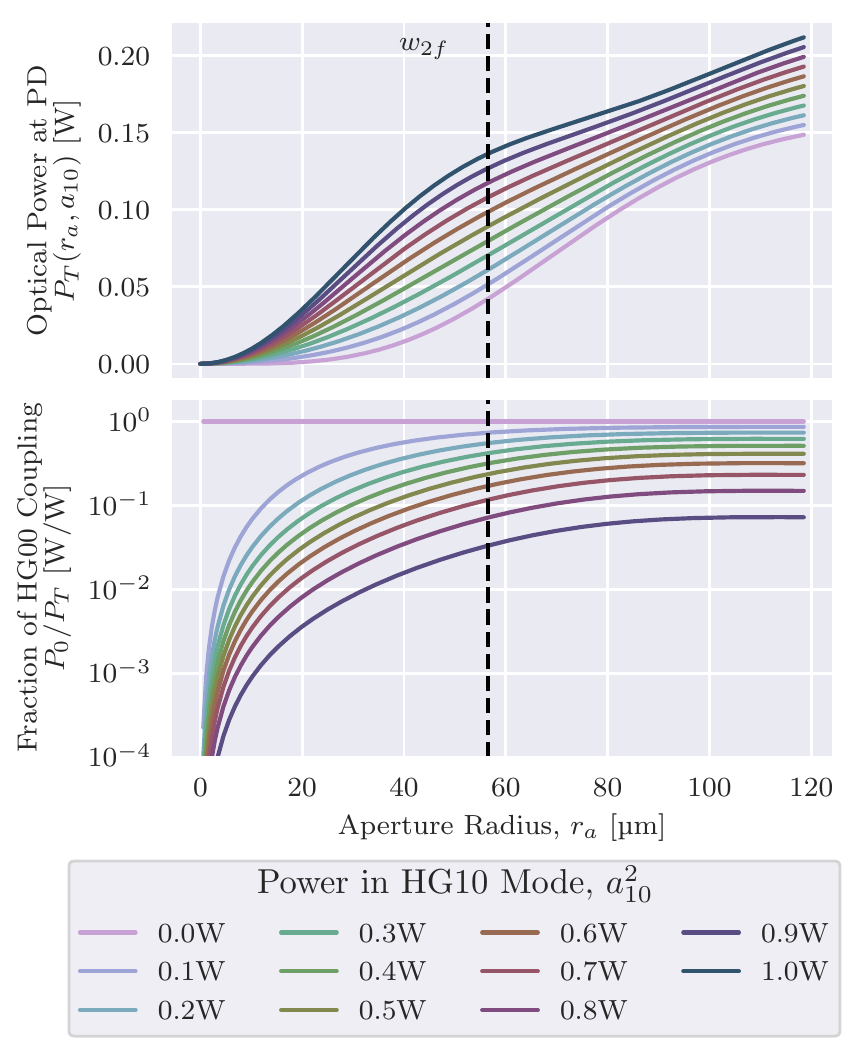}
	\caption{The upper plot shows the total optical power on the light-sensor as a function of aperture radius, for 1W total power and different amounts of \hg{10} power. The lower plot shows the fraction of this light which is crosstalk from the \hg{00} mode. The parameters used were: $\lambda =$ 1064\,nm, $f=$ 0.2\,m, $b_{10} = w_\text{SLM}=$ 1.2\,mm, $a_{00}^2=1-a_{10}^2$. $w_{2f}$ is given by Eq.~\ref{eq:w2f}.} 
	\label{fig:pinhole_size}
\end{figure}
%\end{widetext}
We note that the interference terms in Eq.~\ref{eq:pinhole_response} integrate away for a centered, finite size aperture, leaving terms that are either proportional to $a_0^H$ or $a_1^H$. Defining the crosstalk, $P_0$, to be the sum of all terms proportional to $a_0^H$ and the signal, $P_1$ to be the sum of all terms proportional to $a_1^H$.

Fig.~\ref{fig:pinhole_size} shows Eq.~\ref{eq:pinhole_response} plotted for some reasonable experimental parameters. The lightest line has all the power in the fundamental mode and the darkest line has all the power in the \hg{10} mode. When, the pinhole aperture is much smaller than the beam-size at the light-sensor, 
$r_a << w_{2f}$, the cross talk is very low $\nicefrac{P_0}{P_T}<<1$, but at the cost of reduced power. As $r_a$ increases the fraction of cross coupling rapidly increases. When $r_a = w_{2f}$, with 50:50 power split between the $a_{00}^2$ and $a_{10}^2$, 23.6\% of the light at the light-sensor is from crosstalk. 

If we then apply energy conservation by setting, $\left(a_0^H\right)^2 = 1 - \abs{a_1^H}^2$, then solve $0 = P_0 - P_1$ for $\left(a_0^H\right)^2$, we obtain the expression for the \hg{10} fraction with equal signal and crosstalk contributions,
\begin{align}
\left(a_1^H\right)^2_\text{min} = 
\left[
	\frac{
		4 w_{2f}^{2} \left(
			r_{a}^{2} - w_{2f}^{2} \exp\left(
				\frac{r_{a}^{2}}{w_{2f}^{2}}
			\right) + w_{2f}^{2}
		\right)
	}{
		3 r_{a}^{4} + 2 r_{a}^{2} w_{2f}^{2} - 10 w_{2f}^{4} \exp\left(
			\frac{
				r_{a}^{2}
			}{
				w_{2f}^{2}
			}
		\right) + 10 w_{2f}^{4}
	}
\right].
\end{align}
When evaluated for our experimental parameters $r_a = 5\upmu\mathrm{m}$, $w_{2f} = 54\,\upmu\mathrm{m}$, then $\left(a_1^H\right)^2_\text{min} = 0.002$, which, to within calibration errors, matches the minima in the HG01 response, $P_{\sigma,0,1}$, in Fig.~\ref{fig:xscan}.
\section{Considerations for higher order MODANs}
In this paper, we use a pinhole and photo-diode as a light-sensor for high dynamic range mode analysis. Our analysis is restricted to first order modes due to existence of good witness sensors and ability to generate controlled small amounts of \hg{10}, however, the methods described may be used generally for higher order sensors.

Specifically, in the case of an SLM based MODAN monitoring arbitrary higher order modes, the light-sensor should be positioned using the phase-patterns and methods shown, before collecting data on other modes. The dynamic range we demonstrate is high when compared to other results (e.g. \cite{Jollivet14} and references therein), suggesting that light-sensor alignment and aperture size are critical and can be fine tuned with the method we show.

To improve the dynamic range, the experimentalist must reduce the ratio of the photodiode or pinhole aperture and beam size at the light-sensor. Increasing beam radius is attractive, but necessarily reduces beam radius at the DOE. Stock pinholes exist down to $1\,\upmu\mathrm{m}$, but, due to power loss, photodiodes with low dark noise and high-gain are then required. Alternatively, beam radius at the light-sensor can be increased without changing the beam radius at the SLM, by increasing the focal length of the lens.

We studied the horizontal and vertical position of the light-sensor with respect to the phase-pattern, however, mode analysis requires that the longitudinal position is also tuned. The longitudinal position of the light-sensor was not tuned in this work, which introduced additional gouy phase. If the Rayleigh range is suitably large at the light-sensor, then profiling the beam may suffice. If not, then a similar approach to the one presented, scanning the beam parameter used during the phase-pattern generation and the longitudinal position of the light-sensor, may be required.

Commercial photo-diodes exist with very broad bandwidths, however, SLMs generate noise at their display refresh rates which is typically 60Hz. For a GW detector implementation, this noise can be trivially filtered because mode mismatches and parametric instability growth typically occurs at thermal timescales and parametric instabilities oscillate at kHz timescales.

\section{Conclusions}
MODAN is a promising technology for high-dynamic-range spatial-mode analysis in GW detectors. In a single branch MODAN, it is possible to increase the dynamic range by using photo-diode readout instead of a camera. Further improvements are possible by reducing the aperture of the photodiode and decreasing the beam radius at the DOE.

A relative misalignment between the photo-diode and phase-pattern causes a reduced dynamic range and introduces systematic errors. This can be characterized and eliminated by scanning the first-order Hermite Gauss mode content as shown in section \ref{sec:errorsignals}. With a suitable SLM, this scan may be done in software allowing easy calibration of the device as frequently as desired, before exploring another mode of interest.

The finite aperture of the photo-diode causes an optical offset to the measurement. Equation \ref{eq:pinhole_size} can be used to determine the optical offset and additional shot noise contributions for a range of design parameters prior to construction.

\section*{Funding}
UK Engineering and Physical Sciences Research Council (EPSRC) (EP/M013294/1, EP/T001046/1, 2161515)

\section*{Acknowledgements}
A. W. Jones thanks the Royal Astronomical Society and Institute of Physics for financial support. A. Freise has been supported by the Science and Technology Facilities Council (STFC) and by a Royal Society Wolfson Fellowship which is jointly funded by the Royal Society and the Wolfson Foundation. The authors jointly thank Maud Slangen for proofreading.

\section*{Disclosures}
The authors declare no conflicts of interest.

\section{References}
\label{sec:refs}
\bibliography{paperbib}

\begin{thebibliography}{44}%
\makeatletter
\providecommand \@ifxundefined [1]{%
 \@ifx{#1\undefined}
}%
\providecommand \@ifnum [1]{%
 \ifnum #1\expandafter \@firstoftwo
 \else \expandafter \@secondoftwo
 \fi
}%
\providecommand \@ifx [1]{%
 \ifx #1\expandafter \@firstoftwo
 \else \expandafter \@secondoftwo
 \fi
}%
\providecommand \natexlab [1]{#1}%
\providecommand \enquote  [1]{``#1''}%
\providecommand \bibnamefont  [1]{#1}%
\providecommand \bibfnamefont [1]{#1}%
\providecommand \citenamefont [1]{#1}%
\providecommand \href@noop [0]{\@secondoftwo}%
\providecommand \href [0]{\begingroup \@sanitize@url \@href}%
\providecommand \@href[1]{\@@startlink{#1}\@@href}%
\providecommand \@@href[1]{\endgroup#1\@@endlink}%
\providecommand \@sanitize@url [0]{\catcode `\\12\catcode `\$12\catcode
  `\&12\catcode `\#12\catcode `\^12\catcode `\_12\catcode `\%12\relax}%
\providecommand \@@startlink[1]{}%
\providecommand \@@endlink[0]{}%
\providecommand \url  [0]{\begingroup\@sanitize@url \@url }%
\providecommand \@url [1]{\endgroup\@href {#1}{\urlprefix }}%
\providecommand \urlprefix  [0]{URL }%
\providecommand \Eprint [0]{\href }%
\providecommand \doibase [0]{http://dx.doi.org/}%
\providecommand \selectlanguage [0]{\@gobble}%
\providecommand \bibinfo  [0]{\@secondoftwo}%
\providecommand \bibfield  [0]{\@secondoftwo}%
\providecommand \translation [1]{[#1]}%
\providecommand \BibitemOpen [0]{}%
\providecommand \bibitemStop [0]{}%
\providecommand \bibitemNoStop [0]{.\EOS\space}%
\providecommand \EOS [0]{\spacefactor3000\relax}%
\providecommand \BibitemShut  [1]{\csname bibitem#1\endcsname}%
\let\auto@bib@innerbib\@empty
%</preamble>
\bibitem [{\citenamefont {Matei}\ \emph {et~al.}(2017)\citenamefont {Matei},
  \citenamefont {Legero}, \citenamefont {H\"afner}, \citenamefont {Grebing},
  \citenamefont {Weyrich}, \citenamefont {Zhang}, \citenamefont {Sonderhouse},
  \citenamefont {Robinson}, \citenamefont {Ye}, \citenamefont {Riehle},\ and\
  \citenamefont {Sterr}}]{Matei17}%
  \BibitemOpen
  \bibfield  {author} {\bibinfo {author} {\bibfnamefont {D.~G.}\ \bibnamefont
  {Matei}}, \bibinfo {author} {\bibfnamefont {T.}~\bibnamefont {Legero}},
  \bibinfo {author} {\bibfnamefont {S.}~\bibnamefont {H\"afner}}, \bibinfo
  {author} {\bibfnamefont {C.}~\bibnamefont {Grebing}}, \bibinfo {author}
  {\bibfnamefont {R.}~\bibnamefont {Weyrich}}, \bibinfo {author} {\bibfnamefont
  {W.}~\bibnamefont {Zhang}}, \bibinfo {author} {\bibfnamefont
  {L.}~\bibnamefont {Sonderhouse}}, \bibinfo {author} {\bibfnamefont {J.~M.}\
  \bibnamefont {Robinson}}, \bibinfo {author} {\bibfnamefont {J.}~\bibnamefont
  {Ye}}, \bibinfo {author} {\bibfnamefont {F.}~\bibnamefont {Riehle}}, \ and\
  \bibinfo {author} {\bibfnamefont {U.}~\bibnamefont {Sterr}},\ }\href
  {\doibase 10.1103/PhysRevLett.118.263202} {\bibfield  {journal} {\bibinfo
  {journal} {Phys. Rev. Lett.}\ }\textbf {\bibinfo {volume} {118}},\ \bibinfo
  {pages} {263202} (\bibinfo {year} {2017})}\BibitemShut {NoStop}%
\bibitem [{\citenamefont {{Abdel-Hafiz}}\ \emph {et~al.}(2019)\citenamefont
  {{Abdel-Hafiz}}, \citenamefont {{Ablewski}}, \citenamefont {{Al-Masoudi}},
  \citenamefont {{{\'A}lvarez Mart{\'\i}nez}}, \citenamefont {{Balling}},
  \citenamefont {{Barwood}}, \citenamefont {{Benkler}}, \citenamefont
  {{Bober}}, \citenamefont {{Borkowski}}, \citenamefont {{Bowden}},
  \citenamefont {{Ciury{\l}o}}, \citenamefont {{Cybulski}}, \citenamefont
  {{Didier}}, \citenamefont {{Dole{\v{z}}al}}, \citenamefont {{D{\"o}rscher}},
  \citenamefont {{Falke}}, \citenamefont {{Godun}}, \citenamefont {{Hamid}},
  \citenamefont {{Hill}}, \citenamefont {{Hobson}}, \citenamefont
  {{Huntemann}}, \citenamefont {{Le Coq}}, \citenamefont {{Le Targat}},
  \citenamefont {{Legero}}, \citenamefont {{Lindvall}}, \citenamefont
  {{Lisdat}}, \citenamefont {{Lodewyck}}, \citenamefont {{Margolis}},
  \citenamefont {{Mehlst{\"a}ubler}}, \citenamefont {{Peik}}, \citenamefont
  {{Pelzer}}, \citenamefont {{Pizzocaro}}, \citenamefont {{Rauf}},
  \citenamefont {{Rolland }}, \citenamefont {{Scharnhorst}}, \citenamefont
  {{Schioppo}}, \citenamefont {{Schmidt}}, \citenamefont {{Schwarz}},
  \citenamefont {{{\c{S}}enel}}, \citenamefont {{Spethmann}}, \citenamefont
  {{Sterr}}, \citenamefont {{Tamm}}, \citenamefont {{Thomsen}}, \citenamefont
  {{Vianello}},\ and\ \citenamefont {{Zawada}}}]{hafiz2019}%
  \BibitemOpen
  \bibfield  {author} {\bibinfo {author} {\bibfnamefont {M.}~\bibnamefont
  {{Abdel-Hafiz}}}, \bibinfo {author} {\bibfnamefont {P.}~\bibnamefont
  {{Ablewski}}}, \bibinfo {author} {\bibfnamefont {A.}~\bibnamefont
  {{Al-Masoudi}}}, \bibinfo {author} {\bibfnamefont {H.}~\bibnamefont
  {{{\'A}lvarez Mart{\'\i}nez}}}, \bibinfo {author} {\bibfnamefont
  {P.}~\bibnamefont {{Balling}}}, \bibinfo {author} {\bibfnamefont
  {G.}~\bibnamefont {{Barwood}}}, \bibinfo {author} {\bibfnamefont
  {E.}~\bibnamefont {{Benkler}}}, \bibinfo {author} {\bibfnamefont
  {M.}~\bibnamefont {{Bober}}}, \bibinfo {author} {\bibfnamefont
  {M.}~\bibnamefont {{Borkowski}}}, \bibinfo {author} {\bibfnamefont
  {W.}~\bibnamefont {{Bowden}}}, \bibinfo {author} {\bibfnamefont
  {R.}~\bibnamefont {{Ciury{\l}o}}}, \bibinfo {author} {\bibfnamefont
  {H.}~\bibnamefont {{Cybulski}}}, \bibinfo {author} {\bibfnamefont
  {A.}~\bibnamefont {{Didier}}}, \bibinfo {author} {\bibfnamefont
  {M.}~\bibnamefont {{Dole{\v{z}}al}}}, \bibinfo {author} {\bibfnamefont
  {S.}~\bibnamefont {{D{\"o}rscher}}}, \bibinfo {author} {\bibfnamefont
  {S.}~\bibnamefont {{Falke}}}, \bibinfo {author} {\bibfnamefont {R.~M.}\
  \bibnamefont {{Godun}}}, \bibinfo {author} {\bibfnamefont {R.}~\bibnamefont
  {{Hamid}}}, \bibinfo {author} {\bibfnamefont {I.~R.}\ \bibnamefont {{Hill}}},
  \bibinfo {author} {\bibfnamefont {R.}~\bibnamefont {{Hobson}}}, \bibinfo
  {author} {\bibfnamefont {N.}~\bibnamefont {{Huntemann}}}, \bibinfo {author}
  {\bibfnamefont {Y.}~\bibnamefont {{Le Coq}}}, \bibinfo {author}
  {\bibfnamefont {R.}~\bibnamefont {{Le Targat}}}, \bibinfo {author}
  {\bibfnamefont {T.}~\bibnamefont {{Legero}}}, \bibinfo {author}
  {\bibfnamefont {T.}~\bibnamefont {{Lindvall}}}, \bibinfo {author}
  {\bibfnamefont {C.}~\bibnamefont {{Lisdat}}}, \bibinfo {author}
  {\bibfnamefont {J.}~\bibnamefont {{Lodewyck}}}, \bibinfo {author}
  {\bibfnamefont {H.~S.}\ \bibnamefont {{Margolis}}}, \bibinfo {author}
  {\bibfnamefont {T.~E.}\ \bibnamefont {{Mehlst{\"a}ubler}}}, \bibinfo {author}
  {\bibfnamefont {E.}~\bibnamefont {{Peik}}}, \bibinfo {author} {\bibfnamefont
  {L.}~\bibnamefont {{Pelzer}}}, \bibinfo {author} {\bibfnamefont
  {M.}~\bibnamefont {{Pizzocaro}}}, \bibinfo {author} {\bibfnamefont
  {B.}~\bibnamefont {{Rauf}}}, \bibinfo {author} {\bibfnamefont
  {A.}~\bibnamefont {{Rolland }}}, \bibinfo {author} {\bibfnamefont
  {N.}~\bibnamefont {{Scharnhorst}}}, \bibinfo {author} {\bibfnamefont
  {M.}~\bibnamefont {{Schioppo}}}, \bibinfo {author} {\bibfnamefont {P.~O.}\
  \bibnamefont {{Schmidt}}}, \bibinfo {author} {\bibfnamefont {R.}~\bibnamefont
  {{Schwarz}}}, \bibinfo {author} {\bibfnamefont {{\c{C}}.}~\bibnamefont
  {{{\c{S}}enel}}}, \bibinfo {author} {\bibfnamefont {N.}~\bibnamefont
  {{Spethmann}}}, \bibinfo {author} {\bibfnamefont {U.}~\bibnamefont
  {{Sterr}}}, \bibinfo {author} {\bibfnamefont {C.}~\bibnamefont {{Tamm}}},
  \bibinfo {author} {\bibfnamefont {J.~W.}\ \bibnamefont {{Thomsen}}}, \bibinfo
  {author} {\bibfnamefont {A.}~\bibnamefont {{Vianello}}}, \ and\ \bibinfo
  {author} {\bibfnamefont {M.}~\bibnamefont {{Zawada}}},\ }\href@noop {}
  {\bibfield  {journal} {\bibinfo  {journal} {arXiv e-prints}\ ,\ \bibinfo
  {eid} {arXiv:1906.11495}} (\bibinfo {year} {2019})},\ \Eprint
  {http://arxiv.org/abs/1906.11495} {arXiv:1906.11495 [physics.atom-ph]}
  \BibitemShut {NoStop}%
\bibitem [{\citenamefont {Aasi}\ \emph {et~al.}(2015)\citenamefont {Aasi},
  \citenamefont {Abbott}, \citenamefont {Abbott}, \citenamefont {Abbott},
  \citenamefont {Abernathy}, \citenamefont {Ackley}, \citenamefont {Adams},
  \citenamefont {Adams}, \citenamefont {Addesso}, \citenamefont {Adhikari},
  \citenamefont {Adya}, \citenamefont {Affeldt}, \citenamefont {Aggarwal},
  \citenamefont {Aguiar}, \citenamefont {Ain}, \citenamefont {Ajith},
  \citenamefont {Alemic}, \citenamefont {Allen}, \citenamefont {Amariutei},
  \citenamefont {Anderson}, \citenamefont {Anderson}, \citenamefont {Arai},
  \citenamefont {Araya}, \citenamefont {Arceneaux}, \citenamefont {Areeda},
  \citenamefont {Ashton}, \citenamefont {Ast}, \citenamefont {Aston},
  \citenamefont {Aufmuth}, \citenamefont {Aulbert}, \citenamefont {Aylott},
  \citenamefont {Babak}, \citenamefont {Baker}, \citenamefont {Ballmer},
  \citenamefont {Barayoga}, \citenamefont {Barbet}, \citenamefont {Barclay},
  \citenamefont {Barish}, \citenamefont {Barker}, \citenamefont {Barr},
  \citenamefont {Barsotti}, \citenamefont {Bartlett}, \citenamefont {Barton},
  \citenamefont {Bartos}, \citenamefont {Bassiri}, \citenamefont {Batch},
  \citenamefont {Baune}, \citenamefont {Behnke}, \citenamefont {Bell},
  \citenamefont {Bell}, \citenamefont {Benacquista}, \citenamefont {Bergman},
  \citenamefont {Bergmann}, \citenamefont {Berry}, \citenamefont {Betzwieser},
  \citenamefont {Bhagwat}, \citenamefont {Bhandare}, \citenamefont {Bilenko},
  \citenamefont {Billingsley}, \citenamefont {Birch}, \citenamefont {Biscans},
  \citenamefont {Biwer}, \citenamefont {Blackburn}, \citenamefont {Blackburn},
  \citenamefont {Blair}, \citenamefont {Blair}, \citenamefont {Bock},
  \citenamefont {Bodiya}, \citenamefont {Bojtos}, \citenamefont {Bond},
  \citenamefont {Bork}, \citenamefont {Born}, \citenamefont {Bose},
  \citenamefont {Brady}, \citenamefont {Braginsky}, \citenamefont {Brau},
  \citenamefont {Bridges}, \citenamefont {Brinkmann}, \citenamefont {Brooks},
  \citenamefont {Brown}, \citenamefont {Brown}, \citenamefont {Brown},
  \citenamefont {Buchman}, \citenamefont {Buikema}, \citenamefont {Buonanno},
  \citenamefont {Cadonati}, \citenamefont {Bustillo}, \citenamefont {Camp},
  \citenamefont {Cannon}, \citenamefont {Cao}, \citenamefont {Capano},
  \citenamefont {Caride}, \citenamefont {Caudill}, \citenamefont
  {Cavagli{\`{a}}}, \citenamefont {Cepeda}, \citenamefont {Chakraborty},
  \citenamefont {Chalermsongsak}, \citenamefont {Chamberlin}, \citenamefont
  {Chao}, \citenamefont {Charlton}, \citenamefont {Chen}, \citenamefont {Cho},
  \citenamefont {Cho}, \citenamefont {Chow}, \citenamefont {Christensen},
  \citenamefont {Chu}, \citenamefont {Chung}, \citenamefont {Ciani},
  \citenamefont {Clara}, \citenamefont {Clark}, \citenamefont {Collette},
  \citenamefont {Cominsky}, \citenamefont {Constancio}, \citenamefont {Cook},
  \citenamefont {Corbitt}, \citenamefont {Cornish}, \citenamefont {Corsi},
  \citenamefont {Costa}, \citenamefont {Coughlin}, \citenamefont {Countryman},
  \citenamefont {Couvares}, \citenamefont {Coward}, \citenamefont {Cowart},
  \citenamefont {Coyne}, \citenamefont {Coyne}, \citenamefont {Craig},
  \citenamefont {Creighton}, \citenamefont {Creighton}, \citenamefont {Cripe},
  \citenamefont {Crowder}, \citenamefont {Cumming}, \citenamefont {Cunningham},
  \citenamefont {Cutler}, \citenamefont {Dahl}, \citenamefont {Canton},
  \citenamefont {Damjanic}, \citenamefont {Danilishin}, \citenamefont
  {Danzmann}, \citenamefont {Dartez}, \citenamefont {Dave}, \citenamefont
  {Daveloza}, \citenamefont {Davies}, \citenamefont {Daw}, \citenamefont
  {DeBra}, \citenamefont {Pozzo}, \citenamefont {Denker}, \citenamefont {Dent},
  \citenamefont {Dergachev}, \citenamefont {DeRosa}, \citenamefont {DeSalvo},
  \citenamefont {Dhurandhar}, \citenamefont {D{\textasciiacute}{\i}az},
  \citenamefont {Palma}, \citenamefont {Dojcinoski}, \citenamefont {Dominguez},
  \citenamefont {Donovan}, \citenamefont {Dooley}, \citenamefont {Doravari},
  \citenamefont {Douglas}, \citenamefont {Downes}, \citenamefont {Driggers},
  \citenamefont {Du}, \citenamefont {Dwyer}, \citenamefont {Eberle},
  \citenamefont {Edo}, \citenamefont {Edwards}, \citenamefont {Edwards},
  \citenamefont {Effler}, \citenamefont {Eggenstein}, \citenamefont {Ehrens},
  \citenamefont {Eichholz}, \citenamefont {Eikenberry}, \citenamefont {Essick},
  \citenamefont {Etzel}, \citenamefont {Evans}, \citenamefont {Evans},
  \citenamefont {Factourovich}, \citenamefont {Fairhurst}, \citenamefont {Fan},
  \citenamefont {Fang}, \citenamefont {Farr}, \citenamefont {Farr},
  \citenamefont {Favata}, \citenamefont {Fays}, \citenamefont {Fehrmann},
  \citenamefont {Fejer}, \citenamefont {Feldbaum}, \citenamefont {Ferreira},
  \citenamefont {Fisher}, \citenamefont {Frei}, \citenamefont {Freise},
  \citenamefont {Frey}, \citenamefont {Fricke}, \citenamefont {Fritschel},
  \citenamefont {Frolov}, \citenamefont {Fuentes-Tapia}, \citenamefont {Fulda},
  \citenamefont {Fyffe}, \citenamefont {Gair}, \citenamefont {Gaonkar},
  \citenamefont {Gehrels}, \citenamefont {Gergely{\textasciiacute}},
  \citenamefont {Giaime}, \citenamefont {Giardina}, \citenamefont {Gleason},
  \citenamefont {Goetz}, \citenamefont {Goetz}, \citenamefont {Gondan},
  \citenamefont {Gonz{\'{a}}lez}, \citenamefont {Gordon}, \citenamefont
  {Gorodetsky}, \citenamefont {Gossan}, \citenamefont {Go{\ss}ler},
  \citenamefont {GrÃ€f}, \citenamefont {Graff}, \citenamefont {Grant},
  \citenamefont {Gras}, \citenamefont {Gray}, \citenamefont {Greenhalgh},
  \citenamefont {Gretarsson}, \citenamefont {Grote}, \citenamefont {Grunewald},
  \citenamefont {Guido}, \citenamefont {Guo}, \citenamefont {Gushwa},
  \citenamefont {Gustafson}, \citenamefont {Gustafson}, \citenamefont {Hacker},
  \citenamefont {Hall}, \citenamefont {Hammond}, \citenamefont {Hanke},
  \citenamefont {Hanks}, \citenamefont {Hanna}, \citenamefont {Hannam},
  \citenamefont {Hanson}, \citenamefont {Hardwick}, \citenamefont {Harry},
  \citenamefont {Harry}, \citenamefont {Hart}, \citenamefont {Hartman},
  \citenamefont {Haster}, \citenamefont {Haughian}, \citenamefont {Hee},
  \citenamefont {Heintze}, \citenamefont {Heinzel}, \citenamefont {Hendry},
  \citenamefont {Heng}, \citenamefont {Heptonstall}, \citenamefont {Heurs},
  \citenamefont {Hewitson}, \citenamefont {Hild}, \citenamefont {Hoak},
  \citenamefont {Hodge}, \citenamefont {Hollitt}, \citenamefont {Holt},
  \citenamefont {Hopkins}, \citenamefont {Hosken}, \citenamefont {Hough},
  \citenamefont {Houston}, \citenamefont {Howell}, \citenamefont {Hu},
  \citenamefont {Huerta}, \citenamefont {Hughey}, \citenamefont {Husa},
  \citenamefont {Huttner}, \citenamefont {Huynh}, \citenamefont {Huynh-Dinh},
  \citenamefont {Idrisy}, \citenamefont {Indik}, \citenamefont {Ingram},
  \citenamefont {Inta}, \citenamefont {Islas}, \citenamefont {Isler},
  \citenamefont {Isogai}, \citenamefont {Iyer}, \citenamefont {Izumi},
  \citenamefont {Jacobson}, \citenamefont {Jang}, \citenamefont {Jawahar},
  \citenamefont {Ji}, \citenamefont {Jim{\'{e}}nez-Forteza}, \citenamefont
  {Johnson}, \citenamefont {Jones}, \citenamefont {Jones}, \citenamefont {Ju},
  \citenamefont {Haris}, \citenamefont {Kalogera}, \citenamefont {Kandhasamy},
  \citenamefont {Kang}, \citenamefont {Kanner}, \citenamefont {Katsavounidis},
  \citenamefont {Katzman}, \citenamefont {Kaufer}, \citenamefont {Kaufer},
  \citenamefont {Kaur}, \citenamefont {Kawabe}, \citenamefont {Kawazoe},
  \citenamefont {Keiser}, \citenamefont {Keitel}, \citenamefont {Kelley},
  \citenamefont {Kells}, \citenamefont {Keppel}, \citenamefont {Key},
  \citenamefont {Khalaidovski}, \citenamefont {Khalili}, \citenamefont
  {Khazanov}, \citenamefont {Kim}, \citenamefont {Kim}, \citenamefont {Kim},
  \citenamefont {Kim}, \citenamefont {Kim}, \citenamefont {King}, \citenamefont
  {King}, \citenamefont {Kinzel}, \citenamefont {Kissel}, \citenamefont
  {Klimenko}, \citenamefont {Kline}, \citenamefont {Koehlenbeck}, \citenamefont
  {Kokeyama}, \citenamefont {Kondrashov}, \citenamefont {Korobko},
  \citenamefont {Korth}, \citenamefont {Kozak}, \citenamefont {Kringel},
  \citenamefont {Krishnan}, \citenamefont {Krueger}, \citenamefont {Kuehn},
  \citenamefont {Kumar}, \citenamefont {Kumar}, \citenamefont {Kuo},
  \citenamefont {Landry}, \citenamefont {Lantz}, \citenamefont {Larson},
  \citenamefont {Lasky}, \citenamefont {Lazzarini}, \citenamefont {Lazzaro},
  \citenamefont {Le}, \citenamefont {Leaci}, \citenamefont {Leavey},
  \citenamefont {Lebigot}, \citenamefont {Lee}, \citenamefont {Lee},
  \citenamefont {Lee}, \citenamefont {Leong}, \citenamefont {Levin},
  \citenamefont {Levine}, \citenamefont {Lewis}, \citenamefont {Li},
  \citenamefont {Libbrecht}, \citenamefont {Libson}, \citenamefont {Lin},
  \citenamefont {Littenberg}, \citenamefont {Lockerbie}, \citenamefont
  {Lockett}, \citenamefont {Logue}, \citenamefont {Lombardi}, \citenamefont
  {Lormand}, \citenamefont {Lough}, \citenamefont {Lubinski}, \citenamefont
  {LÃŒck}, \citenamefont {Lundgren}, \citenamefont {Lynch}, \citenamefont
  {Ma}, \citenamefont {Macarthur}, \citenamefont {MacDonald}, \citenamefont
  {Machenschalk}, \citenamefont {MacInnis}, \citenamefont {Macleod},
  \citenamefont {Maga{\~{n}}a-Sandoval}, \citenamefont {Magee}, \citenamefont
  {Mageswaran}, \citenamefont {Maglione}, \citenamefont {Mailand},
  \citenamefont {Mandel}, \citenamefont {Mandic}, \citenamefont {Mangano},
  \citenamefont {Mansell}, \citenamefont {M{\'{a}}rka}, \citenamefont
  {M{\'{a}}rka}, \citenamefont {Markosyan}, \citenamefont {Maros},
  \citenamefont {Martin}, \citenamefont {Martin}, \citenamefont {Martynov},
  \citenamefont {Marx}, \citenamefont {Mason}, \citenamefont {Massinger},
  \citenamefont {Matichard}, \citenamefont {Matone}, \citenamefont {Mavalvala},
  \citenamefont {Mazumder}, \citenamefont {Mazzolo}, \citenamefont {McCarthy},
  \citenamefont {McClelland}, \citenamefont {McCormick}, \citenamefont
  {McGuire}, \citenamefont {McIntyre}, \citenamefont {McIver}, \citenamefont
  {McLin}, \citenamefont {McWilliams}, \citenamefont {Meadors}, \citenamefont
  {Meinders}, \citenamefont {Melatos}, \citenamefont {Mendell}, \citenamefont
  {Mercer}, \citenamefont {Meshkov}, \citenamefont {Messenger}, \citenamefont
  {Meyers}, \citenamefont {Miao}, \citenamefont {Middleton}, \citenamefont
  {Mikhailov}, \citenamefont {Miller}, \citenamefont {Miller}, \citenamefont
  {Millhouse}, \citenamefont {Ming}, \citenamefont {Mirshekari}, \citenamefont
  {Mishra}, \citenamefont {Mitra}, \citenamefont {Mitrofanov}, \citenamefont
  {Mitselmakher}, \citenamefont {Mittleman}, \citenamefont {Moe}, \citenamefont
  {Mohanty}, \citenamefont {Mohapatra}, \citenamefont {Moore}, \citenamefont
  {Moraru}, \citenamefont {Moreno}, \citenamefont {Morriss}, \citenamefont
  {Mossavi}, \citenamefont {Mow-Lowry}, \citenamefont {Mueller}, \citenamefont
  {Mueller}, \citenamefont {Mukherjee}, \citenamefont {Mullavey}, \citenamefont
  {Munch}, \citenamefont {Murphy}, \citenamefont {Murray}, \citenamefont
  {Mytidis}, \citenamefont {Nash}, \citenamefont {Nayak}, \citenamefont
  {Necula}, \citenamefont {Nedkova}, \citenamefont {Newton}, \citenamefont
  {Nguyen}, \citenamefont {Nielsen}, \citenamefont {Nissanke}, \citenamefont
  {Nitz}, \citenamefont {Nolting}, \citenamefont {Normandin}, \citenamefont
  {Nuttall}, \citenamefont {Ochsner}, \citenamefont {O'Dell}, \citenamefont
  {Oelker}, \citenamefont {Ogin}, \citenamefont {Oh}, \citenamefont {Oh},
  \citenamefont {Ohme}, \citenamefont {Oppermann}, \citenamefont {Oram},
  \citenamefont {O'Reilly}, \citenamefont {Ortega}, \citenamefont
  {O'Shaughnessy}, \citenamefont {Osthelder}, \citenamefont {Ott},
  \citenamefont {Ottaway}, \citenamefont {Ottens}, \citenamefont {Overmier},
  \citenamefont {Owen}, \citenamefont {Padilla}, \citenamefont {Pai},
  \citenamefont {Pai}, \citenamefont {Palashov}, \citenamefont {Pal-Singh},
  \citenamefont {Pan}, \citenamefont {Pankow}, \citenamefont {Pannarale},
  \citenamefont {Pant}, \citenamefont {Papa}, \citenamefont {Paris},
  \citenamefont {Patrick}, \citenamefont {Pedraza}, \citenamefont {Pekowsky},
  \citenamefont {Pele}, \citenamefont {Penn}, \citenamefont {Perreca},
  \citenamefont {Phelps}, \citenamefont {Pierro}, \citenamefont {Pinto},
  \citenamefont {Pitkin}, \citenamefont {Poeld}, \citenamefont {Post},
  \citenamefont {Poteomkin}, \citenamefont {Powell}, \citenamefont {Prasad},
  \citenamefont {Predoi}, \citenamefont {Premachandra}, \citenamefont
  {Prestegard}, \citenamefont {Price}, \citenamefont {Principe}, \citenamefont
  {Privitera}, \citenamefont {Prix}, \citenamefont {Prokhorov}, \citenamefont
  {Puncken}, \citenamefont {PÃŒrrer}, \citenamefont {Qin}, \citenamefont
  {Quetschke}, \citenamefont {Quintero}, \citenamefont {Quiroga}, \citenamefont
  {Quitzow-James}, \citenamefont {Raab}, \citenamefont {Rabeling},
  \citenamefont {Radkins}, \citenamefont {Raffai}, \citenamefont {Raja},
  \citenamefont {Rajalakshmi}, \citenamefont {Rakhmanov}, \citenamefont
  {Ramirez}, \citenamefont {Raymond}, \citenamefont {Reed}, \citenamefont
  {Reid}, \citenamefont {Reitze}, \citenamefont {Reula}, \citenamefont {Riles},
  \citenamefont {Robertson}, \citenamefont {Robie}, \citenamefont {Rollins},
  \citenamefont {Roma}, \citenamefont {Romano}, \citenamefont {Romanov},
  \citenamefont {Romie}, \citenamefont {Rowan}, \citenamefont {RÃŒdiger},
  \citenamefont {Ryan}, \citenamefont {Sachdev}, \citenamefont {Sadecki},
  \citenamefont {Sadeghian}, \citenamefont {Saleem}, \citenamefont {Salemi},
  \citenamefont {Sammut}, \citenamefont {Sandberg}, \citenamefont {Sanders},
  \citenamefont {Sannibale}, \citenamefont {Santiago-Prieto}, \citenamefont
  {Sathyaprakash}, \citenamefont {Saulson}, \citenamefont {Savage},
  \citenamefont {Sawadsky}, \citenamefont {Scheuer}, \citenamefont {Schilling},
  \citenamefont {Schmidt}, \citenamefont {Schnabel}, \citenamefont {Schofield},
  \citenamefont {Schreiber}, \citenamefont {Schuette}, \citenamefont {Schutz},
  \citenamefont {Scott}, \citenamefont {Scott}, \citenamefont {Sellers},
  \citenamefont {Sengupta}, \citenamefont {Sergeev}, \citenamefont {Serna},
  \citenamefont {Sevigny}, \citenamefont {Shaddock}, \citenamefont {Shahriar},
  \citenamefont {Shaltev}, \citenamefont {Shao}, \citenamefont {Shapiro},
  \citenamefont {Shawhan}, \citenamefont {Shoemaker}, \citenamefont {Sidery},
  \citenamefont {Siemens}, \citenamefont {Sigg}, \citenamefont {Silva},
  \citenamefont {Simakov}, \citenamefont {Singer}, \citenamefont {Singer},
  \citenamefont {Singh}, \citenamefont {Sintes}, \citenamefont {Slagmolen},
  \citenamefont {Smith}, \citenamefont {Smith}, \citenamefont {Smith},
  \citenamefont {Smith-Lefebvre}, \citenamefont {Son}, \citenamefont {Sorazu},
  \citenamefont {Souradeep}, \citenamefont {Staley}, \citenamefont {Stebbins},
  \citenamefont {Steinke}, \citenamefont {Steinlechner}, \citenamefont
  {Steinlechner}, \citenamefont {Steinmeyer}, \citenamefont {Stephens},
  \citenamefont {Steplewski}, \citenamefont {Stevenson}, \citenamefont {Stone},
  \citenamefont {Strain}, \citenamefont {Strigin}, \citenamefont {Sturani},
  \citenamefont {Stuver}, \citenamefont {Summerscales}, \citenamefont {Sutton},
  \citenamefont {Szczepanczyk}, \citenamefont {Szeifert}, \citenamefont
  {Talukder}, \citenamefont {Tanner}, \citenamefont {T{\'{a}}pai},
  \citenamefont {Tarabrin}, \citenamefont {Taracchini}, \citenamefont {Taylor},
  \citenamefont {Tellez}, \citenamefont {Theeg}, \citenamefont
  {Thirugnanasambandam}, \citenamefont {Thomas}, \citenamefont {Thomas},
  \citenamefont {Thorne}, \citenamefont {Thorne}, \citenamefont {Thrane},
  \citenamefont {Tiwari}, \citenamefont {Tomlinson}, \citenamefont {Torres},
  \citenamefont {Torrie}, \citenamefont {Traylor}, \citenamefont {Tse},
  \citenamefont {Tshilumba}, \citenamefont {Ugolini}, \citenamefont
  {Unnikrishnan}, \citenamefont {Urban}, \citenamefont {Usman}, \citenamefont
  {Vahlbruch}, \citenamefont {Vajente}, \citenamefont {Valdes}, \citenamefont
  {Vallisneri}, \citenamefont {van Veggel}, \citenamefont {Vass}, \citenamefont
  {Vaulin}, \citenamefont {Vecchio}, \citenamefont {Veitch}, \citenamefont
  {Veitch}, \citenamefont {Venkateswara}, \citenamefont {Vincent-Finley},
  \citenamefont {Vitale}, \citenamefont {Vo}, \citenamefont {Vorvick},
  \citenamefont {Vousden}, \citenamefont {Vyatchanin}, \citenamefont {Wade},
  \citenamefont {Wade}, \citenamefont {Wade}, \citenamefont {Walker},
  \citenamefont {Wallace}, \citenamefont {Walsh}, \citenamefont {Wang},
  \citenamefont {Wang}, \citenamefont {Wang}, \citenamefont {Ward},
  \citenamefont {Warner}, \citenamefont {Was}, \citenamefont {Weaver},
  \citenamefont {Weinert}, \citenamefont {Weinstein}, \citenamefont {Weiss},
  \citenamefont {Welborn}, \citenamefont {Wen}, \citenamefont {Wessels},
  \citenamefont {Westphal}, \citenamefont {Wette}, \citenamefont {Whelan},
  \citenamefont {Whitcomb}, \citenamefont {White}, \citenamefont {Whiting},
  \citenamefont {Wilkinson}, \citenamefont {Williams}, \citenamefont
  {Williams}, \citenamefont {Williamson}, \citenamefont {Willis}, \citenamefont
  {Willke}, \citenamefont {Wimmer}, \citenamefont {Winkler}, \citenamefont
  {Wipf}, \citenamefont {Wittel}, \citenamefont {Woan}, \citenamefont {Worden},
  \citenamefont {Xie}, \citenamefont {Yablon}, \citenamefont {Yakushin},
  \citenamefont {Yam}, \citenamefont {Yamamoto}, \citenamefont {Yancey},
  \citenamefont {Yang}, \citenamefont {Zanolin}, \citenamefont {Zhang},
  \citenamefont {Zhang}, \citenamefont {Zhang}, \citenamefont {Zhang},
  \citenamefont {Zhao}, \citenamefont {Zhou}, \citenamefont {Zhu},
  \citenamefont {Zucker}, \citenamefont {Zuraw},\ and\ \citenamefont
  {Zweizig}}]{AdvancedLIGO15}%
  \BibitemOpen
fo\bibfield  {author} {\bibinfo {author} {\bibfnamefont {J.}~\bibnamefont
  {Aasi}}, \bibinfo {author} {\bibfnamefont {B.~P.}\ \bibnamefont {Abbott}},
  \bibinfo {author} {\bibfnamefont {R.}~\bibnamefont {Abbott}}, \bibinfo
  {author} {\bibfnamefont {T.}~\bibnamefont {Abbott}}, \bibinfo {author}
  {\bibfnamefont {M.~R.}\ \bibnamefont {Abernathy}}, \bibinfo {author}
  {\bibfnamefont {K.}~\bibnamefont {Ackley}}, \bibinfo {author} {\bibfnamefont
  {C.}~\bibnamefont {Adams}}, \bibinfo {author} {\bibfnamefont
  {T.}~\bibnamefont {Adams}}, \bibinfo {author} {\bibfnamefont
  {P.}~\bibnamefont {Addesso}}, \bibinfo {author} {\bibfnamefont {R.~X.}\
  \bibnamefont {Adhikari}}, \bibinfo {author} {\bibfnamefont {V.}~\bibnamefont
  {Adya}}, \bibinfo {author} {\bibfnamefont {C.}~\bibnamefont {Affeldt}},
  \bibinfo {author} {\bibfnamefont {N.}~\bibnamefont {Aggarwal}}, \bibinfo
  {author} {\bibfnamefont {O.~D.}\ \bibnamefont {Aguiar}}, \bibinfo {author}
  {\bibfnamefont {A.}~\bibnamefont {Ain}}, \bibinfo {author} {\bibfnamefont
  {P.}~\bibnamefont {Ajith}}, \bibinfo {author} {\bibfnamefont
  {A.}~\bibnamefont {Alemic}}, \bibinfo {author} {\bibfnamefont
  {B.}~\bibnamefont {Allen}}, \bibinfo {author} {\bibfnamefont
  {D.}~\bibnamefont {Amariutei}}, \bibinfo {author} {\bibfnamefont {S.~B.}\
  \bibnamefont {Anderson}}, \bibinfo {author} {\bibfnamefont {W.~G.}\
  \bibnamefont {Anderson}}, \bibinfo {author} {\bibfnamefont {K.}~\bibnamefont
  {Arai}}, \bibinfo {author} {\bibfnamefont {M.~C.}\ \bibnamefont {Araya}},
  \bibinfo {author} {\bibfnamefont {C.}~\bibnamefont {Arceneaux}}, \bibinfo
  {author} {\bibfnamefont {J.~S.}\ \bibnamefont {Areeda}}, \bibinfo {author}
  {\bibfnamefont {G.}~\bibnamefont {Ashton}}, \bibinfo {author} {\bibfnamefont
  {S.}~\bibnamefont {Ast}}, \bibinfo {author} {\bibfnamefont {S.~M.}\
  \bibnamefont {Aston}}, \bibinfo {author} {\bibfnamefont {P.}~\bibnamefont
  {Aufmuth}}, \bibinfo {author} {\bibfnamefont {C.}~\bibnamefont {Aulbert}},
  \bibinfo {author} {\bibfnamefont {B.~E.}\ \bibnamefont {Aylott}}, \bibinfo
  {author} {\bibfnamefont {S.}~\bibnamefont {Babak}}, \bibinfo {author}
  {\bibfnamefont {P.~T.}\ \bibnamefont {Baker}}, \bibinfo {author}
  {\bibfnamefont {S.~W.}\ \bibnamefont {Ballmer}}, \bibinfo {author}
  {\bibfnamefont {J.~C.}\ \bibnamefont {Barayoga}}, \bibinfo {author}
  {\bibfnamefont {M.}~\bibnamefont {Barbet}}, \bibinfo {author} {\bibfnamefont
  {S.}~\bibnamefont {Barclay}}, \bibinfo {author} {\bibfnamefont {B.~C.}\
  \bibnamefont {Barish}}, \bibinfo {author} {\bibfnamefont {D.}~\bibnamefont
  {Barker}}, \bibinfo {author} {\bibfnamefont {B.}~\bibnamefont {Barr}},
  \bibinfo {author} {\bibfnamefont {L.}~\bibnamefont {Barsotti}}, \bibinfo
  {author} {\bibfnamefont {J.}~\bibnamefont {Bartlett}}, \bibinfo {author}
  {\bibfnamefont {M.~A.}\ \bibnamefont {Barton}}, \bibinfo {author}
  {\bibfnamefont {I.}~\bibnamefont {Bartos}}, \bibinfo {author} {\bibfnamefont
  {R.}~\bibnamefont {Bassiri}}, \bibinfo {author} {\bibfnamefont {J.~C.}\
  \bibnamefont {Batch}}, \bibinfo {author} {\bibfnamefont {C.}~\bibnamefont
  {Baune}}, \bibinfo {author} {\bibfnamefont {B.}~\bibnamefont {Behnke}},
  \bibinfo {author} {\bibfnamefont {A.~S.}\ \bibnamefont {Bell}}, \bibinfo
  {author} {\bibfnamefont {C.}~\bibnamefont {Bell}}, \bibinfo {author}
  {\bibfnamefont {M.}~\bibnamefont {Benacquista}}, \bibinfo {author}
  {\bibfnamefont {J.}~\bibnamefont {Bergman}}, \bibinfo {author} {\bibfnamefont
  {G.}~\bibnamefont {Bergmann}}, \bibinfo {author} {\bibfnamefont {C.~P.~L.}\
  \bibnamefont {Berry}}, \bibinfo {author} {\bibfnamefont {J.}~\bibnamefont
  {Betzwieser}}, \bibinfo {author} {\bibfnamefont {S.}~\bibnamefont {Bhagwat}},
  \bibinfo {author} {\bibfnamefont {R.}~\bibnamefont {Bhandare}}, \bibinfo
  {author} {\bibfnamefont {I.~A.}\ \bibnamefont {Bilenko}}, \bibinfo {author}
  {\bibfnamefont {G.}~\bibnamefont {Billingsley}}, \bibinfo {author}
  {\bibfnamefont {J.}~\bibnamefont {Birch}}, \bibinfo {author} {\bibfnamefont
  {S.}~\bibnamefont {Biscans}}, \bibinfo {author} {\bibfnamefont
  {C.}~\bibnamefont {Biwer}}, \bibinfo {author} {\bibfnamefont {J.~K.}\
  \bibnamefont {Blackburn}}, \bibinfo {author} {\bibfnamefont {L.}~\bibnamefont
  {Blackburn}}, \bibinfo {author} {\bibfnamefont {C.~D.}\ \bibnamefont
  {Blair}}, \bibinfo {author} {\bibfnamefont {D.}~\bibnamefont {Blair}},
  \bibinfo {author} {\bibfnamefont {O.}~\bibnamefont {Bock}}, \bibinfo {author}
  {\bibfnamefont {T.~P.}\ \bibnamefont {Bodiya}}, \bibinfo {author}
  {\bibfnamefont {P.}~\bibnamefont {Bojtos}}, \bibinfo {author} {\bibfnamefont
  {C.}~\bibnamefont {Bond}}, \bibinfo {author} {\bibfnamefont {R.}~\bibnamefont
  {Bork}}, \bibinfo {author} {\bibfnamefont {M.}~\bibnamefont {Born}}, \bibinfo
  {author} {\bibfnamefont {S.}~\bibnamefont {Bose}}, \bibinfo {author}
  {\bibfnamefont {P.~R.}\ \bibnamefont {Brady}}, \bibinfo {author}
  {\bibfnamefont {V.~B.}\ \bibnamefont {Braginsky}}, \bibinfo {author}
  {\bibfnamefont {J.~E.}\ \bibnamefont {Brau}}, \bibinfo {author}
  {\bibfnamefont {D.~O.}\ \bibnamefont {Bridges}}, \bibinfo {author}
  {\bibfnamefont {M.}~\bibnamefont {Brinkmann}}, \bibinfo {author}
  {\bibfnamefont {A.~F.}\ \bibnamefont {Brooks}}, \bibinfo {author}
  {\bibfnamefont {D.~A.}\ \bibnamefont {Brown}}, \bibinfo {author}
  {\bibfnamefont {D.~D.}\ \bibnamefont {Brown}}, \bibinfo {author}
  {\bibfnamefont {N.~M.}\ \bibnamefont {Brown}}, \bibinfo {author}
  {\bibfnamefont {S.}~\bibnamefont {Buchman}}, \bibinfo {author} {\bibfnamefont
  {A.}~\bibnamefont {Buikema}}, \bibinfo {author} {\bibfnamefont
  {A.}~\bibnamefont {Buonanno}}, \bibinfo {author} {\bibfnamefont
  {L.}~\bibnamefont {Cadonati}}, \bibinfo {author} {\bibfnamefont {J.~C.}\
  \bibnamefont {Bustillo}}, \bibinfo {author} {\bibfnamefont {J.~B.}\
  \bibnamefont {Camp}}, \bibinfo {author} {\bibfnamefont {K.~C.}\ \bibnamefont
  {Cannon}}, \bibinfo {author} {\bibfnamefont {J.}~\bibnamefont {Cao}},
  \bibinfo {author} {\bibfnamefont {C.~D.}\ \bibnamefont {Capano}}, \bibinfo
  {author} {\bibfnamefont {S.}~\bibnamefont {Caride}}, \bibinfo {author}
  {\bibfnamefont {S.}~\bibnamefont {Caudill}}, \bibinfo {author} {\bibfnamefont
  {M.}~\bibnamefont {Cavagli{\`{a}}}}, \bibinfo {author} {\bibfnamefont
  {C.}~\bibnamefont {Cepeda}}, \bibinfo {author} {\bibfnamefont
  {R.}~\bibnamefont {Chakraborty}}, \bibinfo {author} {\bibfnamefont
  {T.}~\bibnamefont {Chalermsongsak}}, \bibinfo {author} {\bibfnamefont
  {S.~J.}\ \bibnamefont {Chamberlin}}, \bibinfo {author} {\bibfnamefont
  {S.}~\bibnamefont {Chao}}, \bibinfo {author} {\bibfnamefont {P.}~\bibnamefont
  {Charlton}}, \bibinfo {author} {\bibfnamefont {Y.}~\bibnamefont {Chen}},
  \bibinfo {author} {\bibfnamefont {H.~S.}\ \bibnamefont {Cho}}, \bibinfo
  {author} {\bibfnamefont {M.}~\bibnamefont {Cho}}, \bibinfo {author}
  {\bibfnamefont {J.~H.}\ \bibnamefont {Chow}}, \bibinfo {author}
  {\bibfnamefont {N.}~\bibnamefont {Christensen}}, \bibinfo {author}
  {\bibfnamefont {Q.}~\bibnamefont {Chu}}, \bibinfo {author} {\bibfnamefont
  {S.}~\bibnamefont {Chung}}, \bibinfo {author} {\bibfnamefont
  {G.}~\bibnamefont {Ciani}}, \bibinfo {author} {\bibfnamefont
  {F.}~\bibnamefont {Clara}}, \bibinfo {author} {\bibfnamefont {J.~A.}\
  \bibnamefont {Clark}}, \bibinfo {author} {\bibfnamefont {C.}~\bibnamefont
  {Collette}}, \bibinfo {author} {\bibfnamefont {L.}~\bibnamefont {Cominsky}},
  \bibinfo {author} {\bibfnamefont {M.}~\bibnamefont {Constancio}}, \bibinfo
  {author} {\bibfnamefont {D.}~\bibnamefont {Cook}}, \bibinfo {author}
  {\bibfnamefont {T.~R.}\ \bibnamefont {Corbitt}}, \bibinfo {author}
  {\bibfnamefont {N.}~\bibnamefont {Cornish}}, \bibinfo {author} {\bibfnamefont
  {A.}~\bibnamefont {Corsi}}, \bibinfo {author} {\bibfnamefont {C.~A.}\
  \bibnamefont {Costa}}, \bibinfo {author} {\bibfnamefont {M.~W.}\ \bibnamefont
  {Coughlin}}, \bibinfo {author} {\bibfnamefont {S.}~\bibnamefont
  {Countryman}}, \bibinfo {author} {\bibfnamefont {P.}~\bibnamefont
  {Couvares}}, \bibinfo {author} {\bibfnamefont {D.~M.}\ \bibnamefont
  {Coward}}, \bibinfo {author} {\bibfnamefont {M.~J.}\ \bibnamefont {Cowart}},
  \bibinfo {author} {\bibfnamefont {D.~C.}\ \bibnamefont {Coyne}}, \bibinfo
  {author} {\bibfnamefont {R.}~\bibnamefont {Coyne}}, \bibinfo {author}
  {\bibfnamefont {K.}~\bibnamefont {Craig}}, \bibinfo {author} {\bibfnamefont
  {J.~D.~E.}\ \bibnamefont {Creighton}}, \bibinfo {author} {\bibfnamefont
  {T.~D.}\ \bibnamefont {Creighton}}, \bibinfo {author} {\bibfnamefont
  {J.}~\bibnamefont {Cripe}}, \bibinfo {author} {\bibfnamefont {S.~G.}\
  \bibnamefont {Crowder}}, \bibinfo {author} {\bibfnamefont {A.}~\bibnamefont
  {Cumming}}, \bibinfo {author} {\bibfnamefont {L.}~\bibnamefont {Cunningham}},
  \bibinfo {author} {\bibfnamefont {C.}~\bibnamefont {Cutler}}, \bibinfo
  {author} {\bibfnamefont {K.}~\bibnamefont {Dahl}}, \bibinfo {author}
  {\bibfnamefont {T.~D.}\ \bibnamefont {Canton}}, \bibinfo {author}
  {\bibfnamefont {M.}~\bibnamefont {Damjanic}}, \bibinfo {author}
  {\bibfnamefont {S.~L.}\ \bibnamefont {Danilishin}}, \bibinfo {author}
  {\bibfnamefont {K.}~\bibnamefont {Danzmann}}, \bibinfo {author}
  {\bibfnamefont {L.}~\bibnamefont {Dartez}}, \bibinfo {author} {\bibfnamefont
  {I.}~\bibnamefont {Dave}}, \bibinfo {author} {\bibfnamefont {H.}~\bibnamefont
  {Daveloza}}, \bibinfo {author} {\bibfnamefont {G.~S.}\ \bibnamefont
  {Davies}}, \bibinfo {author} {\bibfnamefont {E.~J.}\ \bibnamefont {Daw}},
  \bibinfo {author} {\bibfnamefont {D.}~\bibnamefont {DeBra}}, \bibinfo
  {author} {\bibfnamefont {W.~D.}\ \bibnamefont {Pozzo}}, \bibinfo {author}
  {\bibfnamefont {T.}~\bibnamefont {Denker}}, \bibinfo {author} {\bibfnamefont
  {T.}~\bibnamefont {Dent}}, \bibinfo {author} {\bibfnamefont {V.}~\bibnamefont
  {Dergachev}}, \bibinfo {author} {\bibfnamefont {R.~T.}\ \bibnamefont
  {DeRosa}}, \bibinfo {author} {\bibfnamefont {R.}~\bibnamefont {DeSalvo}},
  \bibinfo {author} {\bibfnamefont {S.}~\bibnamefont {Dhurandhar}}, \bibinfo
  {author} {\bibfnamefont {M.}~\bibnamefont {D{\textasciiacute}{\i}az}},
  \bibinfo {author} {\bibfnamefont {I.~D.}\ \bibnamefont {Palma}}, \bibinfo
  {author} {\bibfnamefont {G.}~\bibnamefont {Dojcinoski}}, \bibinfo {author}
  {\bibfnamefont {E.}~\bibnamefont {Dominguez}}, \bibinfo {author}
  {\bibfnamefont {F.}~\bibnamefont {Donovan}}, \bibinfo {author} {\bibfnamefont
  {K.~L.}\ \bibnamefont {Dooley}}, \bibinfo {author} {\bibfnamefont
  {S.}~\bibnamefont {Doravari}}, \bibinfo {author} {\bibfnamefont
  {R.}~\bibnamefont {Douglas}}, \bibinfo {author} {\bibfnamefont {T.~P.}\
  \bibnamefont {Downes}}, \bibinfo {author} {\bibfnamefont {J.~C.}\
  \bibnamefont {Driggers}}, \bibinfo {author} {\bibfnamefont {Z.}~\bibnamefont
  {Du}}, \bibinfo {author} {\bibfnamefont {S.}~\bibnamefont {Dwyer}}, \bibinfo
  {author} {\bibfnamefont {T.}~\bibnamefont {Eberle}}, \bibinfo {author}
  {\bibfnamefont {T.}~\bibnamefont {Edo}}, \bibinfo {author} {\bibfnamefont
  {M.}~\bibnamefont {Edwards}}, \bibinfo {author} {\bibfnamefont
  {M.}~\bibnamefont {Edwards}}, \bibinfo {author} {\bibfnamefont
  {A.}~\bibnamefont {Effler}}, \bibinfo {author} {\bibfnamefont {H.-B.}\
  \bibnamefont {Eggenstein}}, \bibinfo {author} {\bibfnamefont
  {P.}~\bibnamefont {Ehrens}}, \bibinfo {author} {\bibfnamefont
  {J.}~\bibnamefont {Eichholz}}, \bibinfo {author} {\bibfnamefont {S.~S.}\
  \bibnamefont {Eikenberry}}, \bibinfo {author} {\bibfnamefont
  {R.}~\bibnamefont {Essick}}, \bibinfo {author} {\bibfnamefont
  {T.}~\bibnamefont {Etzel}}, \bibinfo {author} {\bibfnamefont
  {M.}~\bibnamefont {Evans}}, \bibinfo {author} {\bibfnamefont
  {T.}~\bibnamefont {Evans}}, \bibinfo {author} {\bibfnamefont
  {M.}~\bibnamefont {Factourovich}}, \bibinfo {author} {\bibfnamefont
  {S.}~\bibnamefont {Fairhurst}}, \bibinfo {author} {\bibfnamefont
  {X.}~\bibnamefont {Fan}}, \bibinfo {author} {\bibfnamefont {Q.}~\bibnamefont
  {Fang}}, \bibinfo {author} {\bibfnamefont {B.}~\bibnamefont {Farr}}, \bibinfo
  {author} {\bibfnamefont {W.~M.}\ \bibnamefont {Farr}}, \bibinfo {author}
  {\bibfnamefont {M.}~\bibnamefont {Favata}}, \bibinfo {author} {\bibfnamefont
  {M.}~\bibnamefont {Fays}}, \bibinfo {author} {\bibfnamefont {H.}~\bibnamefont
  {Fehrmann}}, \bibinfo {author} {\bibfnamefont {M.~M.}\ \bibnamefont {Fejer}},
  \bibinfo {author} {\bibfnamefont {D.}~\bibnamefont {Feldbaum}}, \bibinfo
  {author} {\bibfnamefont {E.~C.}\ \bibnamefont {Ferreira}}, \bibinfo {author}
  {\bibfnamefont {R.~P.}\ \bibnamefont {Fisher}}, \bibinfo {author}
  {\bibfnamefont {Z.}~\bibnamefont {Frei}}, \bibinfo {author} {\bibfnamefont
  {A.}~\bibnamefont {Freise}}, \bibinfo {author} {\bibfnamefont
  {R.}~\bibnamefont {Frey}}, \bibinfo {author} {\bibfnamefont {T.~T.}\
  \bibnamefont {Fricke}}, \bibinfo {author} {\bibfnamefont {P.}~\bibnamefont
  {Fritschel}}, \bibinfo {author} {\bibfnamefont {V.~V.}\ \bibnamefont
  {Frolov}}, \bibinfo {author} {\bibfnamefont {S.}~\bibnamefont
  {Fuentes-Tapia}}, \bibinfo {author} {\bibfnamefont {P.}~\bibnamefont
  {Fulda}}, \bibinfo {author} {\bibfnamefont {M.}~\bibnamefont {Fyffe}},
  \bibinfo {author} {\bibfnamefont {J.~R.}\ \bibnamefont {Gair}}, \bibinfo
  {author} {\bibfnamefont {S.}~\bibnamefont {Gaonkar}}, \bibinfo {author}
  {\bibfnamefont {N.}~\bibnamefont {Gehrels}}, \bibinfo {author} {\bibfnamefont
  {L.~{\'{A}}.}\ \bibnamefont {Gergely{\textasciiacute}}}, \bibinfo {author}
  {\bibfnamefont {J.~A.}\ \bibnamefont {Giaime}}, \bibinfo {author}
  {\bibfnamefont {K.~D.}\ \bibnamefont {Giardina}}, \bibinfo {author}
  {\bibfnamefont {J.}~\bibnamefont {Gleason}}, \bibinfo {author} {\bibfnamefont
  {E.}~\bibnamefont {Goetz}}, \bibinfo {author} {\bibfnamefont
  {R.}~\bibnamefont {Goetz}}, \bibinfo {author} {\bibfnamefont
  {L.}~\bibnamefont {Gondan}}, \bibinfo {author} {\bibfnamefont
  {G.}~\bibnamefont {Gonz{\'{a}}lez}}, \bibinfo {author} {\bibfnamefont
  {N.}~\bibnamefont {Gordon}}, \bibinfo {author} {\bibfnamefont {M.~L.}\
  \bibnamefont {Gorodetsky}}, \bibinfo {author} {\bibfnamefont
  {S.}~\bibnamefont {Gossan}}, \bibinfo {author} {\bibfnamefont
  {S.}~\bibnamefont {Go{\ss}ler}}, \bibinfo {author} {\bibfnamefont
  {C.}~\bibnamefont {GrÃ€f}}, \bibinfo {author} {\bibfnamefont {P.~B.}\
  \bibnamefont {Graff}}, \bibinfo {author} {\bibfnamefont {A.}~\bibnamefont
  {Grant}}, \bibinfo {author} {\bibfnamefont {S.}~\bibnamefont {Gras}},
  \bibinfo {author} {\bibfnamefont {C.}~\bibnamefont {Gray}}, \bibinfo {author}
  {\bibfnamefont {R.~J.~S.}\ \bibnamefont {Greenhalgh}}, \bibinfo {author}
  {\bibfnamefont {A.~M.}\ \bibnamefont {Gretarsson}}, \bibinfo {author}
  {\bibfnamefont {H.}~\bibnamefont {Grote}}, \bibinfo {author} {\bibfnamefont
  {S.}~\bibnamefont {Grunewald}}, \bibinfo {author} {\bibfnamefont {C.~J.}\
  \bibnamefont {Guido}}, \bibinfo {author} {\bibfnamefont {X.}~\bibnamefont
  {Guo}}, \bibinfo {author} {\bibfnamefont {K.}~\bibnamefont {Gushwa}},
  \bibinfo {author} {\bibfnamefont {E.~K.}\ \bibnamefont {Gustafson}}, \bibinfo
  {author} {\bibfnamefont {R.}~\bibnamefont {Gustafson}}, \bibinfo {author}
  {\bibfnamefont {J.}~\bibnamefont {Hacker}}, \bibinfo {author} {\bibfnamefont
  {E.~D.}\ \bibnamefont {Hall}}, \bibinfo {author} {\bibfnamefont
  {G.}~\bibnamefont {Hammond}}, \bibinfo {author} {\bibfnamefont
  {M.}~\bibnamefont {Hanke}}, \bibinfo {author} {\bibfnamefont
  {J.}~\bibnamefont {Hanks}}, \bibinfo {author} {\bibfnamefont
  {C.}~\bibnamefont {Hanna}}, \bibinfo {author} {\bibfnamefont {M.~D.}\
  \bibnamefont {Hannam}}, \bibinfo {author} {\bibfnamefont {J.}~\bibnamefont
  {Hanson}}, \bibinfo {author} {\bibfnamefont {T.}~\bibnamefont {Hardwick}},
  \bibinfo {author} {\bibfnamefont {G.~M.}\ \bibnamefont {Harry}}, \bibinfo
  {author} {\bibfnamefont {I.~W.}\ \bibnamefont {Harry}}, \bibinfo {author}
  {\bibfnamefont {M.}~\bibnamefont {Hart}}, \bibinfo {author} {\bibfnamefont
  {M.~T.}\ \bibnamefont {Hartman}}, \bibinfo {author} {\bibfnamefont {C.-J.}\
  \bibnamefont {Haster}}, \bibinfo {author} {\bibfnamefont {K.}~\bibnamefont
  {Haughian}}, \bibinfo {author} {\bibfnamefont {S.}~\bibnamefont {Hee}},
  \bibinfo {author} {\bibfnamefont {M.}~\bibnamefont {Heintze}}, \bibinfo
  {author} {\bibfnamefont {G.}~\bibnamefont {Heinzel}}, \bibinfo {author}
  {\bibfnamefont {M.}~\bibnamefont {Hendry}}, \bibinfo {author} {\bibfnamefont
  {I.~S.}\ \bibnamefont {Heng}}, \bibinfo {author} {\bibfnamefont {A.~W.}\
  \bibnamefont {Heptonstall}}, \bibinfo {author} {\bibfnamefont
  {M.}~\bibnamefont {Heurs}}, \bibinfo {author} {\bibfnamefont
  {M.}~\bibnamefont {Hewitson}}, \bibinfo {author} {\bibfnamefont
  {S.}~\bibnamefont {Hild}}, \bibinfo {author} {\bibfnamefont {D.}~\bibnamefont
  {Hoak}}, \bibinfo {author} {\bibfnamefont {K.~A.}\ \bibnamefont {Hodge}},
  \bibinfo {author} {\bibfnamefont {S.~E.}\ \bibnamefont {Hollitt}}, \bibinfo
  {author} {\bibfnamefont {K.}~\bibnamefont {Holt}}, \bibinfo {author}
  {\bibfnamefont {P.}~\bibnamefont {Hopkins}}, \bibinfo {author} {\bibfnamefont
  {D.~J.}\ \bibnamefont {Hosken}}, \bibinfo {author} {\bibfnamefont
  {J.}~\bibnamefont {Hough}}, \bibinfo {author} {\bibfnamefont
  {E.}~\bibnamefont {Houston}}, \bibinfo {author} {\bibfnamefont {E.~J.}\
  \bibnamefont {Howell}}, \bibinfo {author} {\bibfnamefont {Y.~M.}\
  \bibnamefont {Hu}}, \bibinfo {author} {\bibfnamefont {E.}~\bibnamefont
  {Huerta}}, \bibinfo {author} {\bibfnamefont {B.}~\bibnamefont {Hughey}},
  \bibinfo {author} {\bibfnamefont {S.}~\bibnamefont {Husa}}, \bibinfo {author}
  {\bibfnamefont {S.~H.}\ \bibnamefont {Huttner}}, \bibinfo {author}
  {\bibfnamefont {M.}~\bibnamefont {Huynh}}, \bibinfo {author} {\bibfnamefont
  {T.}~\bibnamefont {Huynh-Dinh}}, \bibinfo {author} {\bibfnamefont
  {A.}~\bibnamefont {Idrisy}}, \bibinfo {author} {\bibfnamefont
  {N.}~\bibnamefont {Indik}}, \bibinfo {author} {\bibfnamefont {D.~R.}\
  \bibnamefont {Ingram}}, \bibinfo {author} {\bibfnamefont {R.}~\bibnamefont
  {Inta}}, \bibinfo {author} {\bibfnamefont {G.}~\bibnamefont {Islas}},
  \bibinfo {author} {\bibfnamefont {J.~C.}\ \bibnamefont {Isler}}, \bibinfo
  {author} {\bibfnamefont {T.}~\bibnamefont {Isogai}}, \bibinfo {author}
  {\bibfnamefont {B.~R.}\ \bibnamefont {Iyer}}, \bibinfo {author}
  {\bibfnamefont {K.}~\bibnamefont {Izumi}}, \bibinfo {author} {\bibfnamefont
  {M.}~\bibnamefont {Jacobson}}, \bibinfo {author} {\bibfnamefont
  {H.}~\bibnamefont {Jang}}, \bibinfo {author} {\bibfnamefont {S.}~\bibnamefont
  {Jawahar}}, \bibinfo {author} {\bibfnamefont {Y.}~\bibnamefont {Ji}},
  \bibinfo {author} {\bibfnamefont {F.}~\bibnamefont {Jim{\'{e}}nez-Forteza}},
  \bibinfo {author} {\bibfnamefont {W.~W.}\ \bibnamefont {Johnson}}, \bibinfo
  {author} {\bibfnamefont {D.~I.}\ \bibnamefont {Jones}}, \bibinfo {author}
  {\bibfnamefont {R.}~\bibnamefont {Jones}}, \bibinfo {author} {\bibfnamefont
  {L.}~\bibnamefont {Ju}}, \bibinfo {author} {\bibfnamefont {K.}~\bibnamefont
  {Haris}}, \bibinfo {author} {\bibfnamefont {V.}~\bibnamefont {Kalogera}},
  \bibinfo {author} {\bibfnamefont {S.}~\bibnamefont {Kandhasamy}}, \bibinfo
  {author} {\bibfnamefont {G.}~\bibnamefont {Kang}}, \bibinfo {author}
  {\bibfnamefont {J.~B.}\ \bibnamefont {Kanner}}, \bibinfo {author}
  {\bibfnamefont {E.}~\bibnamefont {Katsavounidis}}, \bibinfo {author}
  {\bibfnamefont {W.}~\bibnamefont {Katzman}}, \bibinfo {author} {\bibfnamefont
  {H.}~\bibnamefont {Kaufer}}, \bibinfo {author} {\bibfnamefont
  {S.}~\bibnamefont {Kaufer}}, \bibinfo {author} {\bibfnamefont
  {T.}~\bibnamefont {Kaur}}, \bibinfo {author} {\bibfnamefont {K.}~\bibnamefont
  {Kawabe}}, \bibinfo {author} {\bibfnamefont {F.}~\bibnamefont {Kawazoe}},
  \bibinfo {author} {\bibfnamefont {G.~M.}\ \bibnamefont {Keiser}}, \bibinfo
  {author} {\bibfnamefont {D.}~\bibnamefont {Keitel}}, \bibinfo {author}
  {\bibfnamefont {D.~B.}\ \bibnamefont {Kelley}}, \bibinfo {author}
  {\bibfnamefont {W.}~\bibnamefont {Kells}}, \bibinfo {author} {\bibfnamefont
  {D.~G.}\ \bibnamefont {Keppel}}, \bibinfo {author} {\bibfnamefont {J.~S.}\
  \bibnamefont {Key}}, \bibinfo {author} {\bibfnamefont {A.}~\bibnamefont
  {Khalaidovski}}, \bibinfo {author} {\bibfnamefont {F.~Y.}\ \bibnamefont
  {Khalili}}, \bibinfo {author} {\bibfnamefont {E.~A.}\ \bibnamefont
  {Khazanov}}, \bibinfo {author} {\bibfnamefont {C.}~\bibnamefont {Kim}},
  \bibinfo {author} {\bibfnamefont {K.}~\bibnamefont {Kim}}, \bibinfo {author}
  {\bibfnamefont {N.~G.}\ \bibnamefont {Kim}}, \bibinfo {author} {\bibfnamefont
  {N.}~\bibnamefont {Kim}}, \bibinfo {author} {\bibfnamefont {Y.-M.}\
  \bibnamefont {Kim}}, \bibinfo {author} {\bibfnamefont {E.~J.}\ \bibnamefont
  {King}}, \bibinfo {author} {\bibfnamefont {P.~J.}\ \bibnamefont {King}},
  \bibinfo {author} {\bibfnamefont {D.~L.}\ \bibnamefont {Kinzel}}, \bibinfo
  {author} {\bibfnamefont {J.~S.}\ \bibnamefont {Kissel}}, \bibinfo {author}
  {\bibfnamefont {S.}~\bibnamefont {Klimenko}}, \bibinfo {author}
  {\bibfnamefont {J.}~\bibnamefont {Kline}}, \bibinfo {author} {\bibfnamefont
  {S.}~\bibnamefont {Koehlenbeck}}, \bibinfo {author} {\bibfnamefont
  {K.}~\bibnamefont {Kokeyama}}, \bibinfo {author} {\bibfnamefont
  {V.}~\bibnamefont {Kondrashov}}, \bibinfo {author} {\bibfnamefont
  {M.}~\bibnamefont {Korobko}}, \bibinfo {author} {\bibfnamefont {W.~Z.}\
  \bibnamefont {Korth}}, \bibinfo {author} {\bibfnamefont {D.~B.}\ \bibnamefont
  {Kozak}}, \bibinfo {author} {\bibfnamefont {V.}~\bibnamefont {Kringel}},
  \bibinfo {author} {\bibfnamefont {B.}~\bibnamefont {Krishnan}}, \bibinfo
  {author} {\bibfnamefont {C.}~\bibnamefont {Krueger}}, \bibinfo {author}
  {\bibfnamefont {G.}~\bibnamefont {Kuehn}}, \bibinfo {author} {\bibfnamefont
  {A.}~\bibnamefont {Kumar}}, \bibinfo {author} {\bibfnamefont
  {P.}~\bibnamefont {Kumar}}, \bibinfo {author} {\bibfnamefont
  {L.}~\bibnamefont {Kuo}}, \bibinfo {author} {\bibfnamefont {M.}~\bibnamefont
  {Landry}}, \bibinfo {author} {\bibfnamefont {B.}~\bibnamefont {Lantz}},
  \bibinfo {author} {\bibfnamefont {S.}~\bibnamefont {Larson}}, \bibinfo
  {author} {\bibfnamefont {P.~D.}\ \bibnamefont {Lasky}}, \bibinfo {author}
  {\bibfnamefont {A.}~\bibnamefont {Lazzarini}}, \bibinfo {author}
  {\bibfnamefont {C.}~\bibnamefont {Lazzaro}}, \bibinfo {author} {\bibfnamefont
  {J.}~\bibnamefont {Le}}, \bibinfo {author} {\bibfnamefont {P.}~\bibnamefont
  {Leaci}}, \bibinfo {author} {\bibfnamefont {S.}~\bibnamefont {Leavey}},
  \bibinfo {author} {\bibfnamefont {E.~O.}\ \bibnamefont {Lebigot}}, \bibinfo
  {author} {\bibfnamefont {C.~H.}\ \bibnamefont {Lee}}, \bibinfo {author}
  {\bibfnamefont {H.~K.}\ \bibnamefont {Lee}}, \bibinfo {author} {\bibfnamefont
  {H.~M.}\ \bibnamefont {Lee}}, \bibinfo {author} {\bibfnamefont {J.~R.}\
  \bibnamefont {Leong}}, \bibinfo {author} {\bibfnamefont {Y.}~\bibnamefont
  {Levin}}, \bibinfo {author} {\bibfnamefont {B.}~\bibnamefont {Levine}},
  \bibinfo {author} {\bibfnamefont {J.}~\bibnamefont {Lewis}}, \bibinfo
  {author} {\bibfnamefont {T.~G.~F.}\ \bibnamefont {Li}}, \bibinfo {author}
  {\bibfnamefont {K.}~\bibnamefont {Libbrecht}}, \bibinfo {author}
  {\bibfnamefont {A.}~\bibnamefont {Libson}}, \bibinfo {author} {\bibfnamefont
  {A.~C.}\ \bibnamefont {Lin}}, \bibinfo {author} {\bibfnamefont {T.~B.}\
  \bibnamefont {Littenberg}}, \bibinfo {author} {\bibfnamefont {N.~A.}\
  \bibnamefont {Lockerbie}}, \bibinfo {author} {\bibfnamefont {V.}~\bibnamefont
  {Lockett}}, \bibinfo {author} {\bibfnamefont {J.}~\bibnamefont {Logue}},
  \bibinfo {author} {\bibfnamefont {A.~L.}\ \bibnamefont {Lombardi}}, \bibinfo
  {author} {\bibfnamefont {M.}~\bibnamefont {Lormand}}, \bibinfo {author}
  {\bibfnamefont {J.}~\bibnamefont {Lough}}, \bibinfo {author} {\bibfnamefont
  {M.~J.}\ \bibnamefont {Lubinski}}, \bibinfo {author} {\bibfnamefont
  {H.}~\bibnamefont {LÃŒck}}, \bibinfo {author} {\bibfnamefont {A.~P.}\
  \bibnamefont {Lundgren}}, \bibinfo {author} {\bibfnamefont {R.}~\bibnamefont
  {Lynch}}, \bibinfo {author} {\bibfnamefont {Y.}~\bibnamefont {Ma}}, \bibinfo
  {author} {\bibfnamefont {J.}~\bibnamefont {Macarthur}}, \bibinfo {author}
  {\bibfnamefont {T.}~\bibnamefont {MacDonald}}, \bibinfo {author}
  {\bibfnamefont {B.}~\bibnamefont {Machenschalk}}, \bibinfo {author}
  {\bibfnamefont {M.}~\bibnamefont {MacInnis}}, \bibinfo {author}
  {\bibfnamefont {D.~M.}\ \bibnamefont {Macleod}}, \bibinfo {author}
  {\bibfnamefont {F.}~\bibnamefont {Maga{\~{n}}a-Sandoval}}, \bibinfo {author}
  {\bibfnamefont {R.}~\bibnamefont {Magee}}, \bibinfo {author} {\bibfnamefont
  {M.}~\bibnamefont {Mageswaran}}, \bibinfo {author} {\bibfnamefont
  {C.}~\bibnamefont {Maglione}}, \bibinfo {author} {\bibfnamefont
  {K.}~\bibnamefont {Mailand}}, \bibinfo {author} {\bibfnamefont
  {I.}~\bibnamefont {Mandel}}, \bibinfo {author} {\bibfnamefont
  {V.}~\bibnamefont {Mandic}}, \bibinfo {author} {\bibfnamefont
  {V.}~\bibnamefont {Mangano}}, \bibinfo {author} {\bibfnamefont {G.~L.}\
  \bibnamefont {Mansell}}, \bibinfo {author} {\bibfnamefont {S.}~\bibnamefont
  {M{\'{a}}rka}}, \bibinfo {author} {\bibfnamefont {Z.}~\bibnamefont
  {M{\'{a}}rka}}, \bibinfo {author} {\bibfnamefont {A.}~\bibnamefont
  {Markosyan}}, \bibinfo {author} {\bibfnamefont {E.}~\bibnamefont {Maros}},
  \bibinfo {author} {\bibfnamefont {I.~W.}\ \bibnamefont {Martin}}, \bibinfo
  {author} {\bibfnamefont {R.~M.}\ \bibnamefont {Martin}}, \bibinfo {author}
  {\bibfnamefont {D.}~\bibnamefont {Martynov}}, \bibinfo {author}
  {\bibfnamefont {J.~N.}\ \bibnamefont {Marx}}, \bibinfo {author}
  {\bibfnamefont {K.}~\bibnamefont {Mason}}, \bibinfo {author} {\bibfnamefont
  {T.~J.}\ \bibnamefont {Massinger}}, \bibinfo {author} {\bibfnamefont
  {F.}~\bibnamefont {Matichard}}, \bibinfo {author} {\bibfnamefont
  {L.}~\bibnamefont {Matone}}, \bibinfo {author} {\bibfnamefont
  {N.}~\bibnamefont {Mavalvala}}, \bibinfo {author} {\bibfnamefont
  {N.}~\bibnamefont {Mazumder}}, \bibinfo {author} {\bibfnamefont
  {G.}~\bibnamefont {Mazzolo}}, \bibinfo {author} {\bibfnamefont
  {R.}~\bibnamefont {McCarthy}}, \bibinfo {author} {\bibfnamefont {D.~E.}\
  \bibnamefont {McClelland}}, \bibinfo {author} {\bibfnamefont
  {S.}~\bibnamefont {McCormick}}, \bibinfo {author} {\bibfnamefont {S.~C.}\
  \bibnamefont {McGuire}}, \bibinfo {author} {\bibfnamefont {G.}~\bibnamefont
  {McIntyre}}, \bibinfo {author} {\bibfnamefont {J.}~\bibnamefont {McIver}},
  \bibinfo {author} {\bibfnamefont {K.}~\bibnamefont {McLin}}, \bibinfo
  {author} {\bibfnamefont {S.}~\bibnamefont {McWilliams}}, \bibinfo {author}
  {\bibfnamefont {G.~D.}\ \bibnamefont {Meadors}}, \bibinfo {author}
  {\bibfnamefont {M.}~\bibnamefont {Meinders}}, \bibinfo {author}
  {\bibfnamefont {A.}~\bibnamefont {Melatos}}, \bibinfo {author} {\bibfnamefont
  {G.}~\bibnamefont {Mendell}}, \bibinfo {author} {\bibfnamefont {R.~A.}\
  \bibnamefont {Mercer}}, \bibinfo {author} {\bibfnamefont {S.}~\bibnamefont
  {Meshkov}}, \bibinfo {author} {\bibfnamefont {C.}~\bibnamefont {Messenger}},
  \bibinfo {author} {\bibfnamefont {P.~M.}\ \bibnamefont {Meyers}}, \bibinfo
  {author} {\bibfnamefont {H.}~\bibnamefont {Miao}}, \bibinfo {author}
  {\bibfnamefont {H.}~\bibnamefont {Middleton}}, \bibinfo {author}
  {\bibfnamefont {E.~E.}\ \bibnamefont {Mikhailov}}, \bibinfo {author}
  {\bibfnamefont {A.}~\bibnamefont {Miller}}, \bibinfo {author} {\bibfnamefont
  {J.}~\bibnamefont {Miller}}, \bibinfo {author} {\bibfnamefont
  {M.}~\bibnamefont {Millhouse}}, \bibinfo {author} {\bibfnamefont
  {J.}~\bibnamefont {Ming}}, \bibinfo {author} {\bibfnamefont {S.}~\bibnamefont
  {Mirshekari}}, \bibinfo {author} {\bibfnamefont {C.}~\bibnamefont {Mishra}},
  \bibinfo {author} {\bibfnamefont {S.}~\bibnamefont {Mitra}}, \bibinfo
  {author} {\bibfnamefont {V.~P.}\ \bibnamefont {Mitrofanov}}, \bibinfo
  {author} {\bibfnamefont {G.}~\bibnamefont {Mitselmakher}}, \bibinfo {author}
  {\bibfnamefont {R.}~\bibnamefont {Mittleman}}, \bibinfo {author}
  {\bibfnamefont {B.}~\bibnamefont {Moe}}, \bibinfo {author} {\bibfnamefont
  {S.~D.}\ \bibnamefont {Mohanty}}, \bibinfo {author} {\bibfnamefont
  {S.~R.~P.}\ \bibnamefont {Mohapatra}}, \bibinfo {author} {\bibfnamefont
  {B.}~\bibnamefont {Moore}}, \bibinfo {author} {\bibfnamefont
  {D.}~\bibnamefont {Moraru}}, \bibinfo {author} {\bibfnamefont
  {G.}~\bibnamefont {Moreno}}, \bibinfo {author} {\bibfnamefont {S.~R.}\
  \bibnamefont {Morriss}}, \bibinfo {author} {\bibfnamefont {K.}~\bibnamefont
  {Mossavi}}, \bibinfo {author} {\bibfnamefont {C.~M.}\ \bibnamefont
  {Mow-Lowry}}, \bibinfo {author} {\bibfnamefont {C.~L.}\ \bibnamefont
  {Mueller}}, \bibinfo {author} {\bibfnamefont {G.}~\bibnamefont {Mueller}},
  \bibinfo {author} {\bibfnamefont {S.}~\bibnamefont {Mukherjee}}, \bibinfo
  {author} {\bibfnamefont {A.}~\bibnamefont {Mullavey}}, \bibinfo {author}
  {\bibfnamefont {J.}~\bibnamefont {Munch}}, \bibinfo {author} {\bibfnamefont
  {D.}~\bibnamefont {Murphy}}, \bibinfo {author} {\bibfnamefont {P.~G.}\
  \bibnamefont {Murray}}, \bibinfo {author} {\bibfnamefont {A.}~\bibnamefont
  {Mytidis}}, \bibinfo {author} {\bibfnamefont {T.}~\bibnamefont {Nash}},
  \bibinfo {author} {\bibfnamefont {R.~K.}\ \bibnamefont {Nayak}}, \bibinfo
  {author} {\bibfnamefont {V.}~\bibnamefont {Necula}}, \bibinfo {author}
  {\bibfnamefont {K.}~\bibnamefont {Nedkova}}, \bibinfo {author} {\bibfnamefont
  {G.}~\bibnamefont {Newton}}, \bibinfo {author} {\bibfnamefont
  {T.}~\bibnamefont {Nguyen}}, \bibinfo {author} {\bibfnamefont {A.~B.}\
  \bibnamefont {Nielsen}}, \bibinfo {author} {\bibfnamefont {S.}~\bibnamefont
  {Nissanke}}, \bibinfo {author} {\bibfnamefont {A.~H.}\ \bibnamefont {Nitz}},
  \bibinfo {author} {\bibfnamefont {D.}~\bibnamefont {Nolting}}, \bibinfo
  {author} {\bibfnamefont {M.~E.~N.}\ \bibnamefont {Normandin}}, \bibinfo
  {author} {\bibfnamefont {L.~K.}\ \bibnamefont {Nuttall}}, \bibinfo {author}
  {\bibfnamefont {E.}~\bibnamefont {Ochsner}}, \bibinfo {author} {\bibfnamefont
  {J.}~\bibnamefont {O'Dell}}, \bibinfo {author} {\bibfnamefont
  {E.}~\bibnamefont {Oelker}}, \bibinfo {author} {\bibfnamefont {G.~H.}\
  \bibnamefont {Ogin}}, \bibinfo {author} {\bibfnamefont {J.~J.}\ \bibnamefont
  {Oh}}, \bibinfo {author} {\bibfnamefont {S.~H.}\ \bibnamefont {Oh}}, \bibinfo
  {author} {\bibfnamefont {F.}~\bibnamefont {Ohme}}, \bibinfo {author}
  {\bibfnamefont {P.}~\bibnamefont {Oppermann}}, \bibinfo {author}
  {\bibfnamefont {R.}~\bibnamefont {Oram}}, \bibinfo {author} {\bibfnamefont
  {B.}~\bibnamefont {O'Reilly}}, \bibinfo {author} {\bibfnamefont
  {W.}~\bibnamefont {Ortega}}, \bibinfo {author} {\bibfnamefont
  {R.}~\bibnamefont {O'Shaughnessy}}, \bibinfo {author} {\bibfnamefont
  {C.}~\bibnamefont {Osthelder}}, \bibinfo {author} {\bibfnamefont {C.~D.}\
  \bibnamefont {Ott}}, \bibinfo {author} {\bibfnamefont {D.~J.}\ \bibnamefont
  {Ottaway}}, \bibinfo {author} {\bibfnamefont {R.~S.}\ \bibnamefont {Ottens}},
  \bibinfo {author} {\bibfnamefont {H.}~\bibnamefont {Overmier}}, \bibinfo
  {author} {\bibfnamefont {B.~J.}\ \bibnamefont {Owen}}, \bibinfo {author}
  {\bibfnamefont {C.}~\bibnamefont {Padilla}}, \bibinfo {author} {\bibfnamefont
  {A.}~\bibnamefont {Pai}}, \bibinfo {author} {\bibfnamefont {S.}~\bibnamefont
  {Pai}}, \bibinfo {author} {\bibfnamefont {O.}~\bibnamefont {Palashov}},
  \bibinfo {author} {\bibfnamefont {A.}~\bibnamefont {Pal-Singh}}, \bibinfo
  {author} {\bibfnamefont {H.}~\bibnamefont {Pan}}, \bibinfo {author}
  {\bibfnamefont {C.}~\bibnamefont {Pankow}}, \bibinfo {author} {\bibfnamefont
  {F.}~\bibnamefont {Pannarale}}, \bibinfo {author} {\bibfnamefont {B.~C.}\
  \bibnamefont {Pant}}, \bibinfo {author} {\bibfnamefont {M.~A.}\ \bibnamefont
  {Papa}}, \bibinfo {author} {\bibfnamefont {H.}~\bibnamefont {Paris}},
  \bibinfo {author} {\bibfnamefont {Z.}~\bibnamefont {Patrick}}, \bibinfo
  {author} {\bibfnamefont {M.}~\bibnamefont {Pedraza}}, \bibinfo {author}
  {\bibfnamefont {L.}~\bibnamefont {Pekowsky}}, \bibinfo {author}
  {\bibfnamefont {A.}~\bibnamefont {Pele}}, \bibinfo {author} {\bibfnamefont
  {S.}~\bibnamefont {Penn}}, \bibinfo {author} {\bibfnamefont {A.}~\bibnamefont
  {Perreca}}, \bibinfo {author} {\bibfnamefont {M.}~\bibnamefont {Phelps}},
  \bibinfo {author} {\bibfnamefont {V.}~\bibnamefont {Pierro}}, \bibinfo
  {author} {\bibfnamefont {I.~M.}\ \bibnamefont {Pinto}}, \bibinfo {author}
  {\bibfnamefont {M.}~\bibnamefont {Pitkin}}, \bibinfo {author} {\bibfnamefont
  {J.}~\bibnamefont {Poeld}}, \bibinfo {author} {\bibfnamefont
  {A.}~\bibnamefont {Post}}, \bibinfo {author} {\bibfnamefont {A.}~\bibnamefont
  {Poteomkin}}, \bibinfo {author} {\bibfnamefont {J.}~\bibnamefont {Powell}},
  \bibinfo {author} {\bibfnamefont {J.}~\bibnamefont {Prasad}}, \bibinfo
  {author} {\bibfnamefont {V.}~\bibnamefont {Predoi}}, \bibinfo {author}
  {\bibfnamefont {S.}~\bibnamefont {Premachandra}}, \bibinfo {author}
  {\bibfnamefont {T.}~\bibnamefont {Prestegard}}, \bibinfo {author}
  {\bibfnamefont {L.~R.}\ \bibnamefont {Price}}, \bibinfo {author}
  {\bibfnamefont {M.}~\bibnamefont {Principe}}, \bibinfo {author}
  {\bibfnamefont {S.}~\bibnamefont {Privitera}}, \bibinfo {author}
  {\bibfnamefont {R.}~\bibnamefont {Prix}}, \bibinfo {author} {\bibfnamefont
  {L.}~\bibnamefont {Prokhorov}}, \bibinfo {author} {\bibfnamefont
  {O.}~\bibnamefont {Puncken}}, \bibinfo {author} {\bibfnamefont
  {M.}~\bibnamefont {PÃŒrrer}}, \bibinfo {author} {\bibfnamefont
  {J.}~\bibnamefont {Qin}}, \bibinfo {author} {\bibfnamefont {V.}~\bibnamefont
  {Quetschke}}, \bibinfo {author} {\bibfnamefont {E.}~\bibnamefont {Quintero}},
  \bibinfo {author} {\bibfnamefont {G.}~\bibnamefont {Quiroga}}, \bibinfo
  {author} {\bibfnamefont {R.}~\bibnamefont {Quitzow-James}}, \bibinfo {author}
  {\bibfnamefont {F.~J.}\ \bibnamefont {Raab}}, \bibinfo {author}
  {\bibfnamefont {D.~S.}\ \bibnamefont {Rabeling}}, \bibinfo {author}
  {\bibfnamefont {H.}~\bibnamefont {Radkins}}, \bibinfo {author} {\bibfnamefont
  {P.}~\bibnamefont {Raffai}}, \bibinfo {author} {\bibfnamefont
  {S.}~\bibnamefont {Raja}}, \bibinfo {author} {\bibfnamefont {G.}~\bibnamefont
  {Rajalakshmi}}, \bibinfo {author} {\bibfnamefont {M.}~\bibnamefont
  {Rakhmanov}}, \bibinfo {author} {\bibfnamefont {K.}~\bibnamefont {Ramirez}},
  \bibinfo {author} {\bibfnamefont {V.}~\bibnamefont {Raymond}}, \bibinfo
  {author} {\bibfnamefont {C.~M.}\ \bibnamefont {Reed}}, \bibinfo {author}
  {\bibfnamefont {S.}~\bibnamefont {Reid}}, \bibinfo {author} {\bibfnamefont
  {D.~H.}\ \bibnamefont {Reitze}}, \bibinfo {author} {\bibfnamefont
  {O.}~\bibnamefont {Reula}}, \bibinfo {author} {\bibfnamefont
  {K.}~\bibnamefont {Riles}}, \bibinfo {author} {\bibfnamefont {N.~A.}\
  \bibnamefont {Robertson}}, \bibinfo {author} {\bibfnamefont {R.}~\bibnamefont
  {Robie}}, \bibinfo {author} {\bibfnamefont {J.~G.}\ \bibnamefont {Rollins}},
  \bibinfo {author} {\bibfnamefont {V.}~\bibnamefont {Roma}}, \bibinfo {author}
  {\bibfnamefont {J.~D.}\ \bibnamefont {Romano}}, \bibinfo {author}
  {\bibfnamefont {G.}~\bibnamefont {Romanov}}, \bibinfo {author} {\bibfnamefont
  {J.~H.}\ \bibnamefont {Romie}}, \bibinfo {author} {\bibfnamefont
  {S.}~\bibnamefont {Rowan}}, \bibinfo {author} {\bibfnamefont
  {A.}~\bibnamefont {RÃŒdiger}}, \bibinfo {author} {\bibfnamefont
  {K.}~\bibnamefont {Ryan}}, \bibinfo {author} {\bibfnamefont {S.}~\bibnamefont
  {Sachdev}}, \bibinfo {author} {\bibfnamefont {T.}~\bibnamefont {Sadecki}},
  \bibinfo {author} {\bibfnamefont {L.}~\bibnamefont {Sadeghian}}, \bibinfo
  {author} {\bibfnamefont {M.}~\bibnamefont {Saleem}}, \bibinfo {author}
  {\bibfnamefont {F.}~\bibnamefont {Salemi}}, \bibinfo {author} {\bibfnamefont
  {L.}~\bibnamefont {Sammut}}, \bibinfo {author} {\bibfnamefont
  {V.}~\bibnamefont {Sandberg}}, \bibinfo {author} {\bibfnamefont {J.~R.}\
  \bibnamefont {Sanders}}, \bibinfo {author} {\bibfnamefont {V.}~\bibnamefont
  {Sannibale}}, \bibinfo {author} {\bibfnamefont {I.}~\bibnamefont
  {Santiago-Prieto}}, \bibinfo {author} {\bibfnamefont {B.~S.}\ \bibnamefont
  {Sathyaprakash}}, \bibinfo {author} {\bibfnamefont {P.~R.}\ \bibnamefont
  {Saulson}}, \bibinfo {author} {\bibfnamefont {R.}~\bibnamefont {Savage}},
  \bibinfo {author} {\bibfnamefont {A.}~\bibnamefont {Sawadsky}}, \bibinfo
  {author} {\bibfnamefont {J.}~\bibnamefont {Scheuer}}, \bibinfo {author}
  {\bibfnamefont {R.}~\bibnamefont {Schilling}}, \bibinfo {author}
  {\bibfnamefont {P.}~\bibnamefont {Schmidt}}, \bibinfo {author} {\bibfnamefont
  {R.}~\bibnamefont {Schnabel}}, \bibinfo {author} {\bibfnamefont {R.~M.~S.}\
  \bibnamefont {Schofield}}, \bibinfo {author} {\bibfnamefont {E.}~\bibnamefont
  {Schreiber}}, \bibinfo {author} {\bibfnamefont {D.}~\bibnamefont {Schuette}},
  \bibinfo {author} {\bibfnamefont {B.~F.}\ \bibnamefont {Schutz}}, \bibinfo
  {author} {\bibfnamefont {J.}~\bibnamefont {Scott}}, \bibinfo {author}
  {\bibfnamefont {S.~M.}\ \bibnamefont {Scott}}, \bibinfo {author}
  {\bibfnamefont {D.}~\bibnamefont {Sellers}}, \bibinfo {author} {\bibfnamefont
  {A.~S.}\ \bibnamefont {Sengupta}}, \bibinfo {author} {\bibfnamefont
  {A.}~\bibnamefont {Sergeev}}, \bibinfo {author} {\bibfnamefont
  {G.}~\bibnamefont {Serna}}, \bibinfo {author} {\bibfnamefont
  {A.}~\bibnamefont {Sevigny}}, \bibinfo {author} {\bibfnamefont {D.~A.}\
  \bibnamefont {Shaddock}}, \bibinfo {author} {\bibfnamefont {M.~S.}\
  \bibnamefont {Shahriar}}, \bibinfo {author} {\bibfnamefont {M.}~\bibnamefont
  {Shaltev}}, \bibinfo {author} {\bibfnamefont {Z.}~\bibnamefont {Shao}},
  \bibinfo {author} {\bibfnamefont {B.}~\bibnamefont {Shapiro}}, \bibinfo
  {author} {\bibfnamefont {P.}~\bibnamefont {Shawhan}}, \bibinfo {author}
  {\bibfnamefont {D.~H.}\ \bibnamefont {Shoemaker}}, \bibinfo {author}
  {\bibfnamefont {T.~L.}\ \bibnamefont {Sidery}}, \bibinfo {author}
  {\bibfnamefont {X.}~\bibnamefont {Siemens}}, \bibinfo {author} {\bibfnamefont
  {D.}~\bibnamefont {Sigg}}, \bibinfo {author} {\bibfnamefont {A.~D.}\
  \bibnamefont {Silva}}, \bibinfo {author} {\bibfnamefont {D.}~\bibnamefont
  {Simakov}}, \bibinfo {author} {\bibfnamefont {A.}~\bibnamefont {Singer}},
  \bibinfo {author} {\bibfnamefont {L.}~\bibnamefont {Singer}}, \bibinfo
  {author} {\bibfnamefont {R.}~\bibnamefont {Singh}}, \bibinfo {author}
  {\bibfnamefont {A.~M.}\ \bibnamefont {Sintes}}, \bibinfo {author}
  {\bibfnamefont {B.~J.~J.}\ \bibnamefont {Slagmolen}}, \bibinfo {author}
  {\bibfnamefont {J.~R.}\ \bibnamefont {Smith}}, \bibinfo {author}
  {\bibfnamefont {M.~R.}\ \bibnamefont {Smith}}, \bibinfo {author}
  {\bibfnamefont {R.~J.~E.}\ \bibnamefont {Smith}}, \bibinfo {author}
  {\bibfnamefont {N.~D.}\ \bibnamefont {Smith-Lefebvre}}, \bibinfo {author}
  {\bibfnamefont {E.~J.}\ \bibnamefont {Son}}, \bibinfo {author} {\bibfnamefont
  {B.}~\bibnamefont {Sorazu}}, \bibinfo {author} {\bibfnamefont
  {T.}~\bibnamefont {Souradeep}}, \bibinfo {author} {\bibfnamefont
  {A.}~\bibnamefont {Staley}}, \bibinfo {author} {\bibfnamefont
  {J.}~\bibnamefont {Stebbins}}, \bibinfo {author} {\bibfnamefont
  {M.}~\bibnamefont {Steinke}}, \bibinfo {author} {\bibfnamefont
  {J.}~\bibnamefont {Steinlechner}}, \bibinfo {author} {\bibfnamefont
  {S.}~\bibnamefont {Steinlechner}}, \bibinfo {author} {\bibfnamefont
  {D.}~\bibnamefont {Steinmeyer}}, \bibinfo {author} {\bibfnamefont {B.~C.}\
  \bibnamefont {Stephens}}, \bibinfo {author} {\bibfnamefont {S.}~\bibnamefont
  {Steplewski}}, \bibinfo {author} {\bibfnamefont {S.}~\bibnamefont
  {Stevenson}}, \bibinfo {author} {\bibfnamefont {R.}~\bibnamefont {Stone}},
  \bibinfo {author} {\bibfnamefont {K.~A.}\ \bibnamefont {Strain}}, \bibinfo
  {author} {\bibfnamefont {S.}~\bibnamefont {Strigin}}, \bibinfo {author}
  {\bibfnamefont {R.}~\bibnamefont {Sturani}}, \bibinfo {author} {\bibfnamefont
  {A.~L.}\ \bibnamefont {Stuver}}, \bibinfo {author} {\bibfnamefont {T.~Z.}\
  \bibnamefont {Summerscales}}, \bibinfo {author} {\bibfnamefont {P.~J.}\
  \bibnamefont {Sutton}}, \bibinfo {author} {\bibfnamefont {M.}~\bibnamefont
  {Szczepanczyk}}, \bibinfo {author} {\bibfnamefont {G.}~\bibnamefont
  {Szeifert}}, \bibinfo {author} {\bibfnamefont {D.}~\bibnamefont {Talukder}},
  \bibinfo {author} {\bibfnamefont {D.~B.}\ \bibnamefont {Tanner}}, \bibinfo
  {author} {\bibfnamefont {M.}~\bibnamefont {T{\'{a}}pai}}, \bibinfo {author}
  {\bibfnamefont {S.~P.}\ \bibnamefont {Tarabrin}}, \bibinfo {author}
  {\bibfnamefont {A.}~\bibnamefont {Taracchini}}, \bibinfo {author}
  {\bibfnamefont {R.}~\bibnamefont {Taylor}}, \bibinfo {author} {\bibfnamefont
  {G.}~\bibnamefont {Tellez}}, \bibinfo {author} {\bibfnamefont
  {T.}~\bibnamefont {Theeg}}, \bibinfo {author} {\bibfnamefont {M.~P.}\
  \bibnamefont {Thirugnanasambandam}}, \bibinfo {author} {\bibfnamefont
  {M.}~\bibnamefont {Thomas}}, \bibinfo {author} {\bibfnamefont
  {P.}~\bibnamefont {Thomas}}, \bibinfo {author} {\bibfnamefont {K.~A.}\
  \bibnamefont {Thorne}}, \bibinfo {author} {\bibfnamefont {K.~S.}\
  \bibnamefont {Thorne}}, \bibinfo {author} {\bibfnamefont {E.}~\bibnamefont
  {Thrane}}, \bibinfo {author} {\bibfnamefont {V.}~\bibnamefont {Tiwari}},
  \bibinfo {author} {\bibfnamefont {C.}~\bibnamefont {Tomlinson}}, \bibinfo
  {author} {\bibfnamefont {C.~V.}\ \bibnamefont {Torres}}, \bibinfo {author}
  {\bibfnamefont {C.~I.}\ \bibnamefont {Torrie}}, \bibinfo {author}
  {\bibfnamefont {G.}~\bibnamefont {Traylor}}, \bibinfo {author} {\bibfnamefont
  {M.}~\bibnamefont {Tse}}, \bibinfo {author} {\bibfnamefont {D.}~\bibnamefont
  {Tshilumba}}, \bibinfo {author} {\bibfnamefont {D.}~\bibnamefont {Ugolini}},
  \bibinfo {author} {\bibfnamefont {C.~S.}\ \bibnamefont {Unnikrishnan}},
  \bibinfo {author} {\bibfnamefont {A.~L.}\ \bibnamefont {Urban}}, \bibinfo
  {author} {\bibfnamefont {S.~A.}\ \bibnamefont {Usman}}, \bibinfo {author}
  {\bibfnamefont {H.}~\bibnamefont {Vahlbruch}}, \bibinfo {author}
  {\bibfnamefont {G.}~\bibnamefont {Vajente}}, \bibinfo {author} {\bibfnamefont
  {G.}~\bibnamefont {Valdes}}, \bibinfo {author} {\bibfnamefont
  {M.}~\bibnamefont {Vallisneri}}, \bibinfo {author} {\bibfnamefont {A.~A.}\
  \bibnamefont {van Veggel}}, \bibinfo {author} {\bibfnamefont
  {S.}~\bibnamefont {Vass}}, \bibinfo {author} {\bibfnamefont {R.}~\bibnamefont
  {Vaulin}}, \bibinfo {author} {\bibfnamefont {A.}~\bibnamefont {Vecchio}},
  \bibinfo {author} {\bibfnamefont {J.}~\bibnamefont {Veitch}}, \bibinfo
  {author} {\bibfnamefont {P.~J.}\ \bibnamefont {Veitch}}, \bibinfo {author}
  {\bibfnamefont {K.}~\bibnamefont {Venkateswara}}, \bibinfo {author}
  {\bibfnamefont {R.}~\bibnamefont {Vincent-Finley}}, \bibinfo {author}
  {\bibfnamefont {S.}~\bibnamefont {Vitale}}, \bibinfo {author} {\bibfnamefont
  {T.}~\bibnamefont {Vo}}, \bibinfo {author} {\bibfnamefont {C.}~\bibnamefont
  {Vorvick}}, \bibinfo {author} {\bibfnamefont {W.~D.}\ \bibnamefont
  {Vousden}}, \bibinfo {author} {\bibfnamefont {S.~P.}\ \bibnamefont
  {Vyatchanin}}, \bibinfo {author} {\bibfnamefont {A.~R.}\ \bibnamefont
  {Wade}}, \bibinfo {author} {\bibfnamefont {L.}~\bibnamefont {Wade}}, \bibinfo
  {author} {\bibfnamefont {M.}~\bibnamefont {Wade}}, \bibinfo {author}
  {\bibfnamefont {M.}~\bibnamefont {Walker}}, \bibinfo {author} {\bibfnamefont
  {L.}~\bibnamefont {Wallace}}, \bibinfo {author} {\bibfnamefont
  {S.}~\bibnamefont {Walsh}}, \bibinfo {author} {\bibfnamefont
  {H.}~\bibnamefont {Wang}}, \bibinfo {author} {\bibfnamefont {M.}~\bibnamefont
  {Wang}}, \bibinfo {author} {\bibfnamefont {X.}~\bibnamefont {Wang}}, \bibinfo
  {author} {\bibfnamefont {R.~L.}\ \bibnamefont {Ward}}, \bibinfo {author}
  {\bibfnamefont {J.}~\bibnamefont {Warner}}, \bibinfo {author} {\bibfnamefont
  {M.}~\bibnamefont {Was}}, \bibinfo {author} {\bibfnamefont {B.}~\bibnamefont
  {Weaver}}, \bibinfo {author} {\bibfnamefont {M.}~\bibnamefont {Weinert}},
  \bibinfo {author} {\bibfnamefont {A.~J.}\ \bibnamefont {Weinstein}}, \bibinfo
  {author} {\bibfnamefont {R.}~\bibnamefont {Weiss}}, \bibinfo {author}
  {\bibfnamefont {T.}~\bibnamefont {Welborn}}, \bibinfo {author} {\bibfnamefont
  {L.}~\bibnamefont {Wen}}, \bibinfo {author} {\bibfnamefont {P.}~\bibnamefont
  {Wessels}}, \bibinfo {author} {\bibfnamefont {T.}~\bibnamefont {Westphal}},
  \bibinfo {author} {\bibfnamefont {K.}~\bibnamefont {Wette}}, \bibinfo
  {author} {\bibfnamefont {J.~T.}\ \bibnamefont {Whelan}}, \bibinfo {author}
  {\bibfnamefont {S.~E.}\ \bibnamefont {Whitcomb}}, \bibinfo {author}
  {\bibfnamefont {D.~J.}\ \bibnamefont {White}}, \bibinfo {author}
  {\bibfnamefont {B.~F.}\ \bibnamefont {Whiting}}, \bibinfo {author}
  {\bibfnamefont {C.}~\bibnamefont {Wilkinson}}, \bibinfo {author}
  {\bibfnamefont {L.}~\bibnamefont {Williams}}, \bibinfo {author}
  {\bibfnamefont {R.}~\bibnamefont {Williams}}, \bibinfo {author}
  {\bibfnamefont {A.~R.}\ \bibnamefont {Williamson}}, \bibinfo {author}
  {\bibfnamefont {J.~L.}\ \bibnamefont {Willis}}, \bibinfo {author}
  {\bibfnamefont {B.}~\bibnamefont {Willke}}, \bibinfo {author} {\bibfnamefont
  {M.}~\bibnamefont {Wimmer}}, \bibinfo {author} {\bibfnamefont
  {W.}~\bibnamefont {Winkler}}, \bibinfo {author} {\bibfnamefont {C.~C.}\
  \bibnamefont {Wipf}}, \bibinfo {author} {\bibfnamefont {H.}~\bibnamefont
  {Wittel}}, \bibinfo {author} {\bibfnamefont {G.}~\bibnamefont {Woan}},
  \bibinfo {author} {\bibfnamefont {J.}~\bibnamefont {Worden}}, \bibinfo
  {author} {\bibfnamefont {S.}~\bibnamefont {Xie}}, \bibinfo {author}
  {\bibfnamefont {J.}~\bibnamefont {Yablon}}, \bibinfo {author} {\bibfnamefont
  {I.}~\bibnamefont {Yakushin}}, \bibinfo {author} {\bibfnamefont
  {W.}~\bibnamefont {Yam}}, \bibinfo {author} {\bibfnamefont {H.}~\bibnamefont
  {Yamamoto}}, \bibinfo {author} {\bibfnamefont {C.~C.}\ \bibnamefont
  {Yancey}}, \bibinfo {author} {\bibfnamefont {Q.}~\bibnamefont {Yang}},
  \bibinfo {author} {\bibfnamefont {M.}~\bibnamefont {Zanolin}}, \bibinfo
  {author} {\bibfnamefont {F.}~\bibnamefont {Zhang}}, \bibinfo {author}
  {\bibfnamefont {L.}~\bibnamefont {Zhang}}, \bibinfo {author} {\bibfnamefont
  {M.}~\bibnamefont {Zhang}}, \bibinfo {author} {\bibfnamefont
  {Y.}~\bibnamefont {Zhang}}, \bibinfo {author} {\bibfnamefont
  {C.}~\bibnamefont {Zhao}}, \bibinfo {author} {\bibfnamefont {M.}~\bibnamefont
  {Zhou}}, \bibinfo {author} {\bibfnamefont {X.~J.}\ \bibnamefont {Zhu}},
  \bibinfo {author} {\bibfnamefont {M.~E.}\ \bibnamefont {Zucker}}, \bibinfo
  {author} {\bibfnamefont {S.}~\bibnamefont {Zuraw}}, \ and\ \bibinfo {author}
  {\bibfnamefont {J.}~\bibnamefont {Zweizig}},\ }\href {\doibase
  10.1088/0264-9381/32/7/074001} {\bibfield  {journal} {\bibinfo  {journal}
  {Classical and Quantum Gravity}\ }\textbf {\bibinfo {volume} {32}},\ \bibinfo
  {pages} {074001} (\bibinfo {year} {2015})}\BibitemShut {NoStop}%
\bibitem [{\citenamefont {{Acernese}}\ \emph {et~al.}(2015)\citenamefont
  {{Acernese}}, \citenamefont {{Agathos}}, \citenamefont {{Agatsuma}},
  \citenamefont {{Aisa}}, \citenamefont {{Allemandou}}, \citenamefont
  {{Allocca}}, \citenamefont {{Amarni}}, \citenamefont {{Astone}},
  \citenamefont {{Balestri}}, \citenamefont {{Ballardin}}, \citenamefont
  {{Barone}}, \citenamefont {{Baronick}}, \citenamefont {{Barsuglia}},
  \citenamefont {{Basti}}, \citenamefont {{Basti}}, \citenamefont {{Bauer}},
  \citenamefont {{Bavigadda}}, \citenamefont {{Bejger}}, \citenamefont
  {{Beker}}, \citenamefont {{Belczynski}}, \citenamefont {{Bersanetti}},
  \citenamefont {{Bertolini}}, \citenamefont {{Bitossi}}, \citenamefont
  {{Bizouard}}, \citenamefont {{Bloemen}}, \citenamefont {{Blom}},
  \citenamefont {{Boer}}, \citenamefont {{Bogaert}}, \citenamefont {{Bondi}},
  \citenamefont {{Bondu}}, \citenamefont {{Bonelli}}, \citenamefont
  {{Bonnand}}, \citenamefont {{Boschi}}, \citenamefont {{Bosi}}, \citenamefont
  {{Bouedo}}, \citenamefont {{Bradaschia}}, \citenamefont {{Branchesi}},
  \citenamefont {{Briant}}, \citenamefont {{Brillet}}, \citenamefont
  {{Brisson}}, \citenamefont {{Bulik}}, \citenamefont {{Bulten}}, \citenamefont
  {{Buskulic}}, \citenamefont {{Buy}}, \citenamefont {{Cagnoli}}, \citenamefont
  {{Calloni}}, \citenamefont {{Campeggi}}, \citenamefont {{Canuel}},
  \citenamefont {{Carbognani}}, \citenamefont {{Cavalier}}, \citenamefont
  {{Cavalieri}}, \citenamefont {{Cella}}, \citenamefont {{Cesarini}},
  \citenamefont {{Chassande-Mottin}}, \citenamefont {{Chincarini}},
  \citenamefont {{Chiummo}}, \citenamefont {{Chua}}, \citenamefont {{Cleva}},
  \citenamefont {{Coccia}}, \citenamefont {{Cohadon}}, \citenamefont {{Colla}},
  \citenamefont {{Colombini}}, \citenamefont {{Conte}}, \citenamefont
  {{Coulon}}, \citenamefont {{Cuoco}}, \citenamefont {{Dalmaz}}, \citenamefont
  {{D'Antonio}}, \citenamefont {{Dattilo}}, \citenamefont {{Davier}},
  \citenamefont {{Day}}, \citenamefont {{Debreczeni}}, \citenamefont
  {{Degallaix}}, \citenamefont {{Del{\'e}glise}}, \citenamefont {{Del Pozzo}},
  \citenamefont {{Dereli}}, \citenamefont {{De Rosa}}, \citenamefont {{Di
  Fiore}}, \citenamefont {{Di Lieto}}, \citenamefont {{Di Virgilio}},
  \citenamefont {{Doets}}, \citenamefont {{Dolique}}, \citenamefont {{Drago}},
  \citenamefont {{Ducrot}}, \citenamefont {{Endr{\H{o}}czi}}, \citenamefont
  {{Fafone}}, \citenamefont {{Farinon}}, \citenamefont {{Ferrante}},
  \citenamefont {{Ferrini}}, \citenamefont {{Fidecaro}}, \citenamefont
  {{Fiori}}, \citenamefont {{Flaminio}}, \citenamefont {{Fournier}},
  \citenamefont {{Franco}}, \citenamefont {{Frasca}}, \citenamefont
  {{Frasconi}}, \citenamefont {{Gammaitoni}}, \citenamefont {{Garufi}},
  \citenamefont {{Gaspard}}, \citenamefont {{Gatto}}, \citenamefont {{Gemme}},
  \citenamefont {{Gendre}}, \citenamefont {{Genin}}, \citenamefont {{Gennai}},
  \citenamefont {{Ghosh}}, \citenamefont {{Giacobone}}, \citenamefont
  {{Giazotto}}, \citenamefont {{Gouaty}}, \citenamefont {{Granata}},
  \citenamefont {{Greco}}, \citenamefont {{Groot}}, \citenamefont {{Guidi}},
  \citenamefont {{Harms}}, \citenamefont {{Heidmann}}, \citenamefont
  {{Heitmann}}, \citenamefont {{Hello}}, \citenamefont {{Hemming}},
  \citenamefont {{Hennes}}, \citenamefont {{Hofman}}, \citenamefont
  {{Jaranowski}}, \citenamefont {{Jonker}}, \citenamefont {{Kasprzack}},
  \citenamefont {{K{\'e}f{\'e}lian}}, \citenamefont {{Kowalska}}, \citenamefont
  {{Kraan}}, \citenamefont {{Kr{\'o}lak}}, \citenamefont {{Kutynia}},
  \citenamefont {{Lazzaro}}, \citenamefont {{Leonardi}}, \citenamefont
  {{Leroy}}, \citenamefont {{Letendre}}, \citenamefont {{Li}}, \citenamefont
  {{Lieunard}}, \citenamefont {{Lorenzini}}, \citenamefont {{Loriette}},
  \citenamefont {{Losurdo}}, \citenamefont {{Magazz{\`u}}}, \citenamefont
  {{Majorana}}, \citenamefont {{Maksimovic}}, \citenamefont {{Malvezzi}},
  \citenamefont {{Man}}, \citenamefont {{Mangano}}, \citenamefont
  {{Mantovani}}, \citenamefont {{Marchesoni}}, \citenamefont {{Marion}},
  \citenamefont {{Marque}}, \citenamefont {{Martelli}}, \citenamefont
  {{Martellini}}, \citenamefont {{Masserot}}, \citenamefont {{Meacher}},
  \citenamefont {{Meidam}}, \citenamefont {{Mezzani}}, \citenamefont
  {{Michel}}, \citenamefont {{Milano}}, \citenamefont {{Minenkov}},
  \citenamefont {{Moggi}}, \citenamefont {{Mohan}}, \citenamefont {{Montani}},
  \citenamefont {{Morgado}}, \citenamefont {{Mours}}, \citenamefont {{Mul}},
  \citenamefont {{Nagy}}, \citenamefont {{Nardecchia}}, \citenamefont
  {{Naticchioni}}, \citenamefont {{Nelemans}}, \citenamefont {{Neri}},
  \citenamefont {{Neri}}, \citenamefont {{Nocera}}, \citenamefont {{Pacaud}},
  \citenamefont {{Palomba}}, \citenamefont {{Paoletti}}, \citenamefont
  {{Paoli}}, \citenamefont {{Pasqualetti}}, \citenamefont {{Passaquieti}},
  \citenamefont {{Passuello}}, \citenamefont {{Perciballi}}, \citenamefont
  {{Petit}}, \citenamefont {{Pichot}}, \citenamefont {{Piergiovanni}},
  \citenamefont {{Pillant}}, \citenamefont {{Piluso}}, \citenamefont
  {{Pinard}}, \citenamefont {{Poggiani}}, \citenamefont {{Prijatelj}},
  \citenamefont {{Prodi}}, \citenamefont {{Punturo}}, \citenamefont {{Puppo}},
  \citenamefont {{Rabeling}}, \citenamefont {{R{\'a}cz}}, \citenamefont
  {{Rapagnani}}, \citenamefont {{Razzano}}, \citenamefont {{Re}}, \citenamefont
  {{Regimbau}}, \citenamefont {{Ricci}}, \citenamefont {{Robinet}},
  \citenamefont {{Rocchi}}, \citenamefont {{Rolland}}, \citenamefont
  {{Romano}}, \citenamefont {{Rosi{\'n}ska}}, \citenamefont {{Ruggi}},
  \citenamefont {{Saracco}}, \citenamefont {{Sassolas}}, \citenamefont
  {{Schimmel}}, \citenamefont {{Sentenac}}, \citenamefont {{Sequino}},
  \citenamefont {{Shah}}, \citenamefont {{Siellez}}, \citenamefont
  {{Straniero}}, \citenamefont {{Swinkels}}, \citenamefont {{Tacca}},
  \citenamefont {{Tonelli}}, \citenamefont {{Travasso}}, \citenamefont
  {{Turconi}}, \citenamefont {{Vajente}}, \citenamefont {{van Bakel}},
  \citenamefont {{van Beuzekom}}, \citenamefont {{van den Brand}},
  \citenamefont {{Van Den Broeck}}, \citenamefont {{van der Sluys}},
  \citenamefont {{van Heijningen}}, \citenamefont {{Vas{\'u}th}}, \citenamefont
  {{Vedovato}}, \citenamefont {{Veitch}}, \citenamefont {{Verkindt}},
  \citenamefont {{Vetrano}}, \citenamefont {{Vicer{\'e}}}, \citenamefont
  {{Vinet}}, \citenamefont {{Visser}}, \citenamefont {{Vocca}}, \citenamefont
  {{Ward}}, \citenamefont {{Was}}, \citenamefont {{Wei}}, \citenamefont
  {{Yvert}}, \citenamefont {{Zadro {\.z}ny}},\ and\ \citenamefont
  {{Zendri}}}]{AdvancedVirgo15}%
  \BibitemOpen
  \bibfield  {author} {\bibinfo {author} {\bibfnamefont {F.}~\bibnamefont
  {{Acernese}}}, \bibinfo {author} {\bibfnamefont {M.}~\bibnamefont
  {{Agathos}}}, \bibinfo {author} {\bibfnamefont {K.}~\bibnamefont
  {{Agatsuma}}}, \bibinfo {author} {\bibfnamefont {D.}~\bibnamefont {{Aisa}}},
  \bibinfo {author} {\bibfnamefont {N.}~\bibnamefont {{Allemandou}}}, \bibinfo
  {author} {\bibfnamefont {A.}~\bibnamefont {{Allocca}}}, \bibinfo {author}
  {\bibfnamefont {J.}~\bibnamefont {{Amarni}}}, \bibinfo {author}
  {\bibfnamefont {P.}~\bibnamefont {{Astone}}}, \bibinfo {author}
  {\bibfnamefont {G.}~\bibnamefont {{Balestri}}}, \bibinfo {author}
  {\bibfnamefont {G.}~\bibnamefont {{Ballardin}}}, \bibinfo {author}
  {\bibfnamefont {F.}~\bibnamefont {{Barone}}}, \bibinfo {author}
  {\bibfnamefont {J.~P.}\ \bibnamefont {{Baronick}}}, \bibinfo {author}
  {\bibfnamefont {M.}~\bibnamefont {{Barsuglia}}}, \bibinfo {author}
  {\bibfnamefont {A.}~\bibnamefont {{Basti}}}, \bibinfo {author} {\bibfnamefont
  {F.}~\bibnamefont {{Basti}}}, \bibinfo {author} {\bibfnamefont {T.~S.}\
  \bibnamefont {{Bauer}}}, \bibinfo {author} {\bibfnamefont {V.}~\bibnamefont
  {{Bavigadda}}}, \bibinfo {author} {\bibfnamefont {M.}~\bibnamefont
  {{Bejger}}}, \bibinfo {author} {\bibfnamefont {M.~G.}\ \bibnamefont
  {{Beker}}}, \bibinfo {author} {\bibfnamefont {C.}~\bibnamefont
  {{Belczynski}}}, \bibinfo {author} {\bibfnamefont {D.}~\bibnamefont
  {{Bersanetti}}}, \bibinfo {author} {\bibfnamefont {A.}~\bibnamefont
  {{Bertolini}}}, \bibinfo {author} {\bibfnamefont {M.}~\bibnamefont
  {{Bitossi}}}, \bibinfo {author} {\bibfnamefont {M.~A.}\ \bibnamefont
  {{Bizouard}}}, \bibinfo {author} {\bibfnamefont {S.}~\bibnamefont
  {{Bloemen}}}, \bibinfo {author} {\bibfnamefont {M.}~\bibnamefont {{Blom}}},
  \bibinfo {author} {\bibfnamefont {M.}~\bibnamefont {{Boer}}}, \bibinfo
  {author} {\bibfnamefont {G.}~\bibnamefont {{Bogaert}}}, \bibinfo {author}
  {\bibfnamefont {D.}~\bibnamefont {{Bondi}}}, \bibinfo {author} {\bibfnamefont
  {F.}~\bibnamefont {{Bondu}}}, \bibinfo {author} {\bibfnamefont
  {L.}~\bibnamefont {{Bonelli}}}, \bibinfo {author} {\bibfnamefont
  {R.}~\bibnamefont {{Bonnand}}}, \bibinfo {author} {\bibfnamefont
  {V.}~\bibnamefont {{Boschi}}}, \bibinfo {author} {\bibfnamefont
  {L.}~\bibnamefont {{Bosi}}}, \bibinfo {author} {\bibfnamefont
  {T.}~\bibnamefont {{Bouedo}}}, \bibinfo {author} {\bibfnamefont
  {C.}~\bibnamefont {{Bradaschia}}}, \bibinfo {author} {\bibfnamefont
  {M.}~\bibnamefont {{Branchesi}}}, \bibinfo {author} {\bibfnamefont
  {T.}~\bibnamefont {{Briant}}}, \bibinfo {author} {\bibfnamefont
  {A.}~\bibnamefont {{Brillet}}}, \bibinfo {author} {\bibfnamefont
  {V.}~\bibnamefont {{Brisson}}}, \bibinfo {author} {\bibfnamefont
  {T.}~\bibnamefont {{Bulik}}}, \bibinfo {author} {\bibfnamefont {H.~J.}\
  \bibnamefont {{Bulten}}}, \bibinfo {author} {\bibfnamefont {D.}~\bibnamefont
  {{Buskulic}}}, \bibinfo {author} {\bibfnamefont {C.}~\bibnamefont {{Buy}}},
  \bibinfo {author} {\bibfnamefont {G.}~\bibnamefont {{Cagnoli}}}, \bibinfo
  {author} {\bibfnamefont {E.}~\bibnamefont {{Calloni}}}, \bibinfo {author}
  {\bibfnamefont {C.}~\bibnamefont {{Campeggi}}}, \bibinfo {author}
  {\bibfnamefont {B.}~\bibnamefont {{Canuel}}}, \bibinfo {author}
  {\bibfnamefont {F.}~\bibnamefont {{Carbognani}}}, \bibinfo {author}
  {\bibfnamefont {F.}~\bibnamefont {{Cavalier}}}, \bibinfo {author}
  {\bibfnamefont {R.}~\bibnamefont {{Cavalieri}}}, \bibinfo {author}
  {\bibfnamefont {G.}~\bibnamefont {{Cella}}}, \bibinfo {author} {\bibfnamefont
  {E.}~\bibnamefont {{Cesarini}}}, \bibinfo {author} {\bibfnamefont
  {E.}~\bibnamefont {{Chassande-Mottin}}}, \bibinfo {author} {\bibfnamefont
  {A.}~\bibnamefont {{Chincarini}}}, \bibinfo {author} {\bibfnamefont
  {A.}~\bibnamefont {{Chiummo}}}, \bibinfo {author} {\bibfnamefont
  {S.}~\bibnamefont {{Chua}}}, \bibinfo {author} {\bibfnamefont
  {F.}~\bibnamefont {{Cleva}}}, \bibinfo {author} {\bibfnamefont
  {E.}~\bibnamefont {{Coccia}}}, \bibinfo {author} {\bibfnamefont {P.~F.}\
  \bibnamefont {{Cohadon}}}, \bibinfo {author} {\bibfnamefont {A.}~\bibnamefont
  {{Colla}}}, \bibinfo {author} {\bibfnamefont {M.}~\bibnamefont
  {{Colombini}}}, \bibinfo {author} {\bibfnamefont {A.}~\bibnamefont
  {{Conte}}}, \bibinfo {author} {\bibfnamefont {J.~P.}\ \bibnamefont
  {{Coulon}}}, \bibinfo {author} {\bibfnamefont {E.}~\bibnamefont {{Cuoco}}},
  \bibinfo {author} {\bibfnamefont {A.}~\bibnamefont {{Dalmaz}}}, \bibinfo
  {author} {\bibfnamefont {S.}~\bibnamefont {{D'Antonio}}}, \bibinfo {author}
  {\bibfnamefont {V.}~\bibnamefont {{Dattilo}}}, \bibinfo {author}
  {\bibfnamefont {M.}~\bibnamefont {{Davier}}}, \bibinfo {author}
  {\bibfnamefont {R.}~\bibnamefont {{Day}}}, \bibinfo {author} {\bibfnamefont
  {G.}~\bibnamefont {{Debreczeni}}}, \bibinfo {author} {\bibfnamefont
  {J.}~\bibnamefont {{Degallaix}}}, \bibinfo {author} {\bibfnamefont
  {S.}~\bibnamefont {{Del{\'e}glise}}}, \bibinfo {author} {\bibfnamefont
  {W.}~\bibnamefont {{Del Pozzo}}}, \bibinfo {author} {\bibfnamefont
  {H.}~\bibnamefont {{Dereli}}}, \bibinfo {author} {\bibfnamefont
  {R.}~\bibnamefont {{De Rosa}}}, \bibinfo {author} {\bibfnamefont
  {L.}~\bibnamefont {{Di Fiore}}}, \bibinfo {author} {\bibfnamefont
  {A.}~\bibnamefont {{Di Lieto}}}, \bibinfo {author} {\bibfnamefont
  {A.}~\bibnamefont {{Di Virgilio}}}, \bibinfo {author} {\bibfnamefont
  {M.}~\bibnamefont {{Doets}}}, \bibinfo {author} {\bibfnamefont
  {V.}~\bibnamefont {{Dolique}}}, \bibinfo {author} {\bibfnamefont
  {M.}~\bibnamefont {{Drago}}}, \bibinfo {author} {\bibfnamefont
  {M.}~\bibnamefont {{Ducrot}}}, \bibinfo {author} {\bibfnamefont
  {G.}~\bibnamefont {{Endr{\H{o}}czi}}}, \bibinfo {author} {\bibfnamefont
  {V.}~\bibnamefont {{Fafone}}}, \bibinfo {author} {\bibfnamefont
  {S.}~\bibnamefont {{Farinon}}}, \bibinfo {author} {\bibfnamefont
  {I.}~\bibnamefont {{Ferrante}}}, \bibinfo {author} {\bibfnamefont
  {F.}~\bibnamefont {{Ferrini}}}, \bibinfo {author} {\bibfnamefont
  {F.}~\bibnamefont {{Fidecaro}}}, \bibinfo {author} {\bibfnamefont
  {I.}~\bibnamefont {{Fiori}}}, \bibinfo {author} {\bibfnamefont
  {R.}~\bibnamefont {{Flaminio}}}, \bibinfo {author} {\bibfnamefont {J.~D.}\
  \bibnamefont {{Fournier}}}, \bibinfo {author} {\bibfnamefont
  {S.}~\bibnamefont {{Franco}}}, \bibinfo {author} {\bibfnamefont
  {S.}~\bibnamefont {{Frasca}}}, \bibinfo {author} {\bibfnamefont
  {F.}~\bibnamefont {{Frasconi}}}, \bibinfo {author} {\bibfnamefont
  {L.}~\bibnamefont {{Gammaitoni}}}, \bibinfo {author} {\bibfnamefont
  {F.}~\bibnamefont {{Garufi}}}, \bibinfo {author} {\bibfnamefont
  {M.}~\bibnamefont {{Gaspard}}}, \bibinfo {author} {\bibfnamefont
  {A.}~\bibnamefont {{Gatto}}}, \bibinfo {author} {\bibfnamefont
  {G.}~\bibnamefont {{Gemme}}}, \bibinfo {author} {\bibfnamefont
  {B.}~\bibnamefont {{Gendre}}}, \bibinfo {author} {\bibfnamefont
  {E.}~\bibnamefont {{Genin}}}, \bibinfo {author} {\bibfnamefont
  {A.}~\bibnamefont {{Gennai}}}, \bibinfo {author} {\bibfnamefont
  {S.}~\bibnamefont {{Ghosh}}}, \bibinfo {author} {\bibfnamefont
  {L.}~\bibnamefont {{Giacobone}}}, \bibinfo {author} {\bibfnamefont
  {A.}~\bibnamefont {{Giazotto}}}, \bibinfo {author} {\bibfnamefont
  {R.}~\bibnamefont {{Gouaty}}}, \bibinfo {author} {\bibfnamefont
  {M.}~\bibnamefont {{Granata}}}, \bibinfo {author} {\bibfnamefont
  {G.}~\bibnamefont {{Greco}}}, \bibinfo {author} {\bibfnamefont
  {P.}~\bibnamefont {{Groot}}}, \bibinfo {author} {\bibfnamefont {G.~M.}\
  \bibnamefont {{Guidi}}}, \bibinfo {author} {\bibfnamefont {J.}~\bibnamefont
  {{Harms}}}, \bibinfo {author} {\bibfnamefont {A.}~\bibnamefont {{Heidmann}}},
  \bibinfo {author} {\bibfnamefont {H.}~\bibnamefont {{Heitmann}}}, \bibinfo
  {author} {\bibfnamefont {P.}~\bibnamefont {{Hello}}}, \bibinfo {author}
  {\bibfnamefont {G.}~\bibnamefont {{Hemming}}}, \bibinfo {author}
  {\bibfnamefont {E.}~\bibnamefont {{Hennes}}}, \bibinfo {author}
  {\bibfnamefont {D.}~\bibnamefont {{Hofman}}}, \bibinfo {author}
  {\bibfnamefont {P.}~\bibnamefont {{Jaranowski}}}, \bibinfo {author}
  {\bibfnamefont {R.~J.~G.}\ \bibnamefont {{Jonker}}}, \bibinfo {author}
  {\bibfnamefont {M.}~\bibnamefont {{Kasprzack}}}, \bibinfo {author}
  {\bibfnamefont {F.}~\bibnamefont {{K{\'e}f{\'e}lian}}}, \bibinfo {author}
  {\bibfnamefont {I.}~\bibnamefont {{Kowalska}}}, \bibinfo {author}
  {\bibfnamefont {M.}~\bibnamefont {{Kraan}}}, \bibinfo {author} {\bibfnamefont
  {A.}~\bibnamefont {{Kr{\'o}lak}}}, \bibinfo {author} {\bibfnamefont
  {A.}~\bibnamefont {{Kutynia}}}, \bibinfo {author} {\bibfnamefont
  {C.}~\bibnamefont {{Lazzaro}}}, \bibinfo {author} {\bibfnamefont
  {M.}~\bibnamefont {{Leonardi}}}, \bibinfo {author} {\bibfnamefont
  {N.}~\bibnamefont {{Leroy}}}, \bibinfo {author} {\bibfnamefont
  {N.}~\bibnamefont {{Letendre}}}, \bibinfo {author} {\bibfnamefont {T.~G.~F.}\
  \bibnamefont {{Li}}}, \bibinfo {author} {\bibfnamefont {B.}~\bibnamefont
  {{Lieunard}}}, \bibinfo {author} {\bibfnamefont {M.}~\bibnamefont
  {{Lorenzini}}}, \bibinfo {author} {\bibfnamefont {V.}~\bibnamefont
  {{Loriette}}}, \bibinfo {author} {\bibfnamefont {G.}~\bibnamefont
  {{Losurdo}}}, \bibinfo {author} {\bibfnamefont {C.}~\bibnamefont
  {{Magazz{\`u}}}}, \bibinfo {author} {\bibfnamefont {E.}~\bibnamefont
  {{Majorana}}}, \bibinfo {author} {\bibfnamefont {I.}~\bibnamefont
  {{Maksimovic}}}, \bibinfo {author} {\bibfnamefont {V.}~\bibnamefont
  {{Malvezzi}}}, \bibinfo {author} {\bibfnamefont {N.}~\bibnamefont {{Man}}},
  \bibinfo {author} {\bibfnamefont {V.}~\bibnamefont {{Mangano}}}, \bibinfo
  {author} {\bibfnamefont {M.}~\bibnamefont {{Mantovani}}}, \bibinfo {author}
  {\bibfnamefont {F.}~\bibnamefont {{Marchesoni}}}, \bibinfo {author}
  {\bibfnamefont {F.}~\bibnamefont {{Marion}}}, \bibinfo {author}
  {\bibfnamefont {J.}~\bibnamefont {{Marque}}}, \bibinfo {author}
  {\bibfnamefont {F.}~\bibnamefont {{Martelli}}}, \bibinfo {author}
  {\bibfnamefont {L.}~\bibnamefont {{Martellini}}}, \bibinfo {author}
  {\bibfnamefont {A.}~\bibnamefont {{Masserot}}}, \bibinfo {author}
  {\bibfnamefont {D.}~\bibnamefont {{Meacher}}}, \bibinfo {author}
  {\bibfnamefont {J.}~\bibnamefont {{Meidam}}}, \bibinfo {author}
  {\bibfnamefont {F.}~\bibnamefont {{Mezzani}}}, \bibinfo {author}
  {\bibfnamefont {C.}~\bibnamefont {{Michel}}}, \bibinfo {author}
  {\bibfnamefont {L.}~\bibnamefont {{Milano}}}, \bibinfo {author}
  {\bibfnamefont {Y.}~\bibnamefont {{Minenkov}}}, \bibinfo {author}
  {\bibfnamefont {A.}~\bibnamefont {{Moggi}}}, \bibinfo {author} {\bibfnamefont
  {M.}~\bibnamefont {{Mohan}}}, \bibinfo {author} {\bibfnamefont
  {M.}~\bibnamefont {{Montani}}}, \bibinfo {author} {\bibfnamefont
  {N.}~\bibnamefont {{Morgado}}}, \bibinfo {author} {\bibfnamefont
  {B.}~\bibnamefont {{Mours}}}, \bibinfo {author} {\bibfnamefont
  {F.}~\bibnamefont {{Mul}}}, \bibinfo {author} {\bibfnamefont {M.~F.}\
  \bibnamefont {{Nagy}}}, \bibinfo {author} {\bibfnamefont {I.}~\bibnamefont
  {{Nardecchia}}}, \bibinfo {author} {\bibfnamefont {L.}~\bibnamefont
  {{Naticchioni}}}, \bibinfo {author} {\bibfnamefont {G.}~\bibnamefont
  {{Nelemans}}}, \bibinfo {author} {\bibfnamefont {I.}~\bibnamefont {{Neri}}},
  \bibinfo {author} {\bibfnamefont {M.}~\bibnamefont {{Neri}}}, \bibinfo
  {author} {\bibfnamefont {F.}~\bibnamefont {{Nocera}}}, \bibinfo {author}
  {\bibfnamefont {E.}~\bibnamefont {{Pacaud}}}, \bibinfo {author}
  {\bibfnamefont {C.}~\bibnamefont {{Palomba}}}, \bibinfo {author}
  {\bibfnamefont {F.}~\bibnamefont {{Paoletti}}}, \bibinfo {author}
  {\bibfnamefont {A.}~\bibnamefont {{Paoli}}}, \bibinfo {author} {\bibfnamefont
  {A.}~\bibnamefont {{Pasqualetti}}}, \bibinfo {author} {\bibfnamefont
  {R.}~\bibnamefont {{Passaquieti}}}, \bibinfo {author} {\bibfnamefont
  {D.}~\bibnamefont {{Passuello}}}, \bibinfo {author} {\bibfnamefont
  {M.}~\bibnamefont {{Perciballi}}}, \bibinfo {author} {\bibfnamefont
  {S.}~\bibnamefont {{Petit}}}, \bibinfo {author} {\bibfnamefont
  {M.}~\bibnamefont {{Pichot}}}, \bibinfo {author} {\bibfnamefont
  {F.}~\bibnamefont {{Piergiovanni}}}, \bibinfo {author} {\bibfnamefont
  {G.}~\bibnamefont {{Pillant}}}, \bibinfo {author} {\bibfnamefont
  {A.}~\bibnamefont {{Piluso}}}, \bibinfo {author} {\bibfnamefont
  {L.}~\bibnamefont {{Pinard}}}, \bibinfo {author} {\bibfnamefont
  {R.}~\bibnamefont {{Poggiani}}}, \bibinfo {author} {\bibfnamefont
  {M.}~\bibnamefont {{Prijatelj}}}, \bibinfo {author} {\bibfnamefont {G.~A.}\
  \bibnamefont {{Prodi}}}, \bibinfo {author} {\bibfnamefont {M.}~\bibnamefont
  {{Punturo}}}, \bibinfo {author} {\bibfnamefont {P.}~\bibnamefont {{Puppo}}},
  \bibinfo {author} {\bibfnamefont {D.~S.}\ \bibnamefont {{Rabeling}}},
  \bibinfo {author} {\bibfnamefont {I.}~\bibnamefont {{R{\'a}cz}}}, \bibinfo
  {author} {\bibfnamefont {P.}~\bibnamefont {{Rapagnani}}}, \bibinfo {author}
  {\bibfnamefont {M.}~\bibnamefont {{Razzano}}}, \bibinfo {author}
  {\bibfnamefont {V.}~\bibnamefont {{Re}}}, \bibinfo {author} {\bibfnamefont
  {T.}~\bibnamefont {{Regimbau}}}, \bibinfo {author} {\bibfnamefont
  {F.}~\bibnamefont {{Ricci}}}, \bibinfo {author} {\bibfnamefont
  {F.}~\bibnamefont {{Robinet}}}, \bibinfo {author} {\bibfnamefont
  {A.}~\bibnamefont {{Rocchi}}}, \bibinfo {author} {\bibfnamefont
  {L.}~\bibnamefont {{Rolland}}}, \bibinfo {author} {\bibfnamefont
  {R.}~\bibnamefont {{Romano}}}, \bibinfo {author} {\bibfnamefont
  {D.}~\bibnamefont {{Rosi{\'n}ska}}}, \bibinfo {author} {\bibfnamefont
  {P.}~\bibnamefont {{Ruggi}}}, \bibinfo {author} {\bibfnamefont
  {E.}~\bibnamefont {{Saracco}}}, \bibinfo {author} {\bibfnamefont
  {B.}~\bibnamefont {{Sassolas}}}, \bibinfo {author} {\bibfnamefont
  {F.}~\bibnamefont {{Schimmel}}}, \bibinfo {author} {\bibfnamefont
  {D.}~\bibnamefont {{Sentenac}}}, \bibinfo {author} {\bibfnamefont
  {V.}~\bibnamefont {{Sequino}}}, \bibinfo {author} {\bibfnamefont
  {S.}~\bibnamefont {{Shah}}}, \bibinfo {author} {\bibfnamefont
  {K.}~\bibnamefont {{Siellez}}}, \bibinfo {author} {\bibfnamefont
  {N.}~\bibnamefont {{Straniero}}}, \bibinfo {author} {\bibfnamefont
  {B.}~\bibnamefont {{Swinkels}}}, \bibinfo {author} {\bibfnamefont
  {M.}~\bibnamefont {{Tacca}}}, \bibinfo {author} {\bibfnamefont
  {M.}~\bibnamefont {{Tonelli}}}, \bibinfo {author} {\bibfnamefont
  {F.}~\bibnamefont {{Travasso}}}, \bibinfo {author} {\bibfnamefont
  {M.}~\bibnamefont {{Turconi}}}, \bibinfo {author} {\bibfnamefont
  {G.}~\bibnamefont {{Vajente}}}, \bibinfo {author} {\bibfnamefont
  {N.}~\bibnamefont {{van Bakel}}}, \bibinfo {author} {\bibfnamefont
  {M.}~\bibnamefont {{van Beuzekom}}}, \bibinfo {author} {\bibfnamefont
  {J.~F.~J.}\ \bibnamefont {{van den Brand}}}, \bibinfo {author} {\bibfnamefont
  {C.}~\bibnamefont {{Van Den Broeck}}}, \bibinfo {author} {\bibfnamefont
  {M.~V.}\ \bibnamefont {{van der Sluys}}}, \bibinfo {author} {\bibfnamefont
  {J.}~\bibnamefont {{van Heijningen}}}, \bibinfo {author} {\bibfnamefont
  {M.}~\bibnamefont {{Vas{\'u}th}}}, \bibinfo {author} {\bibfnamefont
  {G.}~\bibnamefont {{Vedovato}}}, \bibinfo {author} {\bibfnamefont
  {J.}~\bibnamefont {{Veitch}}}, \bibinfo {author} {\bibfnamefont
  {D.}~\bibnamefont {{Verkindt}}}, \bibinfo {author} {\bibfnamefont
  {F.}~\bibnamefont {{Vetrano}}}, \bibinfo {author} {\bibfnamefont
  {A.}~\bibnamefont {{Vicer{\'e}}}}, \bibinfo {author} {\bibfnamefont {J.~Y.}\
  \bibnamefont {{Vinet}}}, \bibinfo {author} {\bibfnamefont {G.}~\bibnamefont
  {{Visser}}}, \bibinfo {author} {\bibfnamefont {H.}~\bibnamefont {{Vocca}}},
  \bibinfo {author} {\bibfnamefont {R.}~\bibnamefont {{Ward}}}, \bibinfo
  {author} {\bibfnamefont {M.}~\bibnamefont {{Was}}}, \bibinfo {author}
  {\bibfnamefont {L.~W.}\ \bibnamefont {{Wei}}}, \bibinfo {author}
  {\bibfnamefont {M.}~\bibnamefont {{Yvert}}}, \bibinfo {author} {\bibfnamefont
  {A.}~\bibnamefont {{Zadro {\.z}ny}}}, \ and\ \bibinfo {author} {\bibfnamefont
  {J.~P.}\ \bibnamefont {{Zendri}}},\ }\href
  {http://stacks.iop.org/0264-9381/32/i=2/a=024001} {\bibfield  {journal}
  {\bibinfo  {journal} {Class. Quantum Grav.}\ }\textbf {\bibinfo {volume}
  {32}},\ \bibinfo {pages} {024001} (\bibinfo {year} {2015})}\BibitemShut
  {NoStop}%
\bibitem [{\citenamefont {Ludlow}\ \emph {et~al.}(2015)\citenamefont {Ludlow},
  \citenamefont {Boyd}, \citenamefont {Ye}, \citenamefont {Peik},\ and\
  \citenamefont {Schmidt}}]{Ludlow15}%
  \BibitemOpen
  \bibfield  {author} {\bibinfo {author} {\bibfnamefont {A.~D.}\ \bibnamefont
  {Ludlow}}, \bibinfo {author} {\bibfnamefont {M.~M.}\ \bibnamefont {Boyd}},
  \bibinfo {author} {\bibfnamefont {J.}~\bibnamefont {Ye}}, \bibinfo {author}
  {\bibfnamefont {E.}~\bibnamefont {Peik}}, \ and\ \bibinfo {author}
  {\bibfnamefont {P.~O.}\ \bibnamefont {Schmidt}},\ }\href {\doibase
  10.1103/RevModPhys.87.637} {\bibfield  {journal} {\bibinfo  {journal} {Rev.
  Mod. Phys.}\ }\textbf {\bibinfo {volume} {87}},\ \bibinfo {pages} {637}
  (\bibinfo {year} {2015})}\BibitemShut {NoStop}%
\bibitem [{\citenamefont {Grote}\ \emph {et~al.}(2013)\citenamefont {Grote},
  \citenamefont {Danzmann}, \citenamefont {Dooley}, \citenamefont {Schnabel},
  \citenamefont {Slutsky},\ and\ \citenamefont {Vahlbruch}}]{Grote2013}%
  \BibitemOpen
  \bibfield  {author} {\bibinfo {author} {\bibfnamefont {H.}~\bibnamefont
  {Grote}}, \bibinfo {author} {\bibfnamefont {K.}~\bibnamefont {Danzmann}},
  \bibinfo {author} {\bibfnamefont {K.~L.}\ \bibnamefont {Dooley}}, \bibinfo
  {author} {\bibfnamefont {R.}~\bibnamefont {Schnabel}}, \bibinfo {author}
  {\bibfnamefont {J.}~\bibnamefont {Slutsky}}, \ and\ \bibinfo {author}
  {\bibfnamefont {H.}~\bibnamefont {Vahlbruch}},\ }\href
  {https://doi.org/10.1103/physrevlett.110.181101} {\bibfield  {journal}
  {\bibinfo  {journal} {Phys. Rev. Lett.}\ }\textbf {\bibinfo {volume} {110}}
  (\bibinfo {year} {2013})}\BibitemShut {NoStop}%
\bibitem [{\citenamefont {Acernese}\ \emph {et~al.}(2019)\citenamefont
  {Acernese}, \citenamefont {Agathos}, \citenamefont {Aiello}, \citenamefont
  {Allocca}, \citenamefont {Amato}, \citenamefont {Ansoldi}, \citenamefont
  {Antier}, \citenamefont {Ar\`ene}, \citenamefont {Arnaud}, \citenamefont
  {Ascenzi}, \citenamefont {Astone}, \citenamefont {Aubin}, \citenamefont
  {Babak}, \citenamefont {Bacon}, \citenamefont {Badaracco}, \citenamefont
  {Bader}, \citenamefont {Baird}, \citenamefont {Baldaccini}, \citenamefont
  {Ballardin}, \citenamefont {Baltus}, \citenamefont {Barbieri}, \citenamefont
  {Barneo}, \citenamefont {Barone}, \citenamefont {Barsuglia}, \citenamefont
  {Barta}, \citenamefont {Basti}, \citenamefont {Bawaj}, \citenamefont
  {Bazzan}, \citenamefont {Bejger}, \citenamefont {Belahcene}, \citenamefont
  {Bernuzzi}, \citenamefont {Bersanetti}, \citenamefont {Bertolini},
  \citenamefont {Bischi}, \citenamefont {Bitossi}, \citenamefont {Bizouard},
  \citenamefont {Bobba}, \citenamefont {Boer}, \citenamefont {Bogaert},
  \citenamefont {Bondu}, \citenamefont {Bonnand}, \citenamefont {Boom},
  \citenamefont {Boschi}, \citenamefont {Bouffanais}, \citenamefont {Bozzi},
  \citenamefont {Bradaschia}, \citenamefont {Branchesi}, \citenamefont
  {Breschi}, \citenamefont {Briant}, \citenamefont {Brighenti}, \citenamefont
  {Brillet}, \citenamefont {Brooks}, \citenamefont {Bruno}, \citenamefont
  {Bulik}, \citenamefont {Bulten}, \citenamefont {Buskulic}, \citenamefont
  {Cagnoli}, \citenamefont {Calloni}, \citenamefont {Canepa}, \citenamefont
  {Carapella}, \citenamefont {Carbognani}, \citenamefont {Carullo},
  \citenamefont {Casanueva~Diaz}, \citenamefont {Casentini}, \citenamefont
  {Casta\~neda}, \citenamefont {Caudill}, \citenamefont {Cavalier},
  \citenamefont {Cavalieri}, \citenamefont {Cella}, \citenamefont
  {Cerd\'a-Dur\'an}, \citenamefont {Cesarini}, \citenamefont {Chaibi},
  \citenamefont {Chassande-Mottin}, \citenamefont {Chiadini}, \citenamefont
  {Chierici}, \citenamefont {Chincarini}, \citenamefont {Chiummo},
  \citenamefont {Christensen}, \citenamefont {Chua}, \citenamefont {Ciani},
  \citenamefont {Ciecielag}, \citenamefont {Cie\ifmmode~\acute{s}\else
  \'{s}\fi{}lar}, \citenamefont {Ciolfi}, \citenamefont {Cipriano},
  \citenamefont {Cirone}, \citenamefont {Clesse}, \citenamefont {Cleva},
  \citenamefont {Coccia}, \citenamefont {Cohadon}, \citenamefont {Cohen},
  \citenamefont {Colpi}, \citenamefont {Conti}, \citenamefont
  {Cordero-Carri\'on}, \citenamefont {Corezzi}, \citenamefont {Corre},
  \citenamefont {Cortese}, \citenamefont {Coulon}, \citenamefont {Croquette},
  \citenamefont {Cudell}, \citenamefont {Cuoco}, \citenamefont {Curylo},
  \citenamefont {D'Angelo}, \citenamefont {D'Antonio}, \citenamefont {Dattilo},
  \citenamefont {Davier}, \citenamefont {Degallaix}, \citenamefont
  {De~Laurentis}, \citenamefont {Del\'eglise}, \citenamefont {Del~Pozzo},
  \citenamefont {De~Pietri}, \citenamefont {De~Rosa}, \citenamefont {De~Rossi},
  \citenamefont {Dietrich}, \citenamefont {Di~Fiore}, \citenamefont
  {Di~Giorgio}, \citenamefont {Di~Giovanni}, \citenamefont {Di~Giovanni},
  \citenamefont {Di~Girolamo}, \citenamefont {Di~Lieto}, \citenamefont
  {Di~Pace}, \citenamefont {Di~Palma}, \citenamefont {Di~Renzo}, \citenamefont
  {Drago}, \citenamefont {Ducoin}, \citenamefont {Durante}, \citenamefont
  {D'Urso}, \citenamefont {Eisenmann}, \citenamefont {Errico}, \citenamefont
  {Estevez}, \citenamefont {Fafone}, \citenamefont {Farinon}, \citenamefont
  {Feng}, \citenamefont {Fenyvesi}, \citenamefont {Ferrante}, \citenamefont
  {Fidecaro}, \citenamefont {Fiori}, \citenamefont {Fiorucci}, \citenamefont
  {Fittipaldi}, \citenamefont {Fiumara}, \citenamefont {Flaminio},
  \citenamefont {Font}, \citenamefont {Fournier}, \citenamefont {Frasca},
  \citenamefont {Frasconi}, \citenamefont {Frey}, \citenamefont {Fronz\`e},
  \citenamefont {Garufi}, \citenamefont {Gemme}, \citenamefont {Genin},
  \citenamefont {Gennai}, \citenamefont {Ghosh}, \citenamefont {Giacomazzo},
  \citenamefont {Gosselin}, \citenamefont {Gouaty}, \citenamefont {Grado},
  \citenamefont {Granata}, \citenamefont {Greco}, \citenamefont {Grignani},
  \citenamefont {Grimaldi}, \citenamefont {Grimm}, \citenamefont {Gruning},
  \citenamefont {Guidi}, \citenamefont {Guix\'e}, \citenamefont {Guo},
  \citenamefont {Gupta}, \citenamefont {Halim}, \citenamefont {Harder},
  \citenamefont {Harms}, \citenamefont {Heidmann}, \citenamefont {Heitmann},
  \citenamefont {Hello}, \citenamefont {Hemming}, \citenamefont {Hennes},
  \citenamefont {Hinderer}, \citenamefont {Hofman}, \citenamefont {Huet},
  \citenamefont {Hui}, \citenamefont {Idzkowski}, \citenamefont {Iess},
  \citenamefont {Intini}, \citenamefont {Isac}, \citenamefont {Jacqmin},
  \citenamefont {Jaranowski}, \citenamefont {Jonker}, \citenamefont
  {Katsanevas}, \citenamefont {K\'ef\'elian}, \citenamefont {Khan},
  \citenamefont {Khetan}, \citenamefont {Koekoek}, \citenamefont {Koley},
  \citenamefont {Kr\'olak}, \citenamefont {Kutynia}, \citenamefont {Laghi},
  \citenamefont {Lamberts}, \citenamefont {La~Rosa}, \citenamefont
  {Lartaux-Vollard}, \citenamefont {Lazzaro}, \citenamefont {Leaci},
  \citenamefont {Leroy}, \citenamefont {Letendre}, \citenamefont {Linde},
  \citenamefont {Llorens-Monteagudo}, \citenamefont {Longo}, \citenamefont
  {Lorenzini}, \citenamefont {Loriette}, \citenamefont {Losurdo}, \citenamefont
  {Lumaca}, \citenamefont {Macquet}, \citenamefont {Majorana}, \citenamefont
  {Maksimovic}, \citenamefont {Man}, \citenamefont {Mangano}, \citenamefont
  {Mantovani}, \citenamefont {Mapelli}, \citenamefont {Marchesoni},
  \citenamefont {Marion}, \citenamefont {Marquina}, \citenamefont {Marsat},
  \citenamefont {Martelli}, \citenamefont {Martinez}, \citenamefont {Masserot},
  \citenamefont {Mastrogiovanni}, \citenamefont {Mejuto~Villa}, \citenamefont
  {Mereni}, \citenamefont {Merzougui}, \citenamefont {Metzdorff}, \citenamefont
  {Miani}, \citenamefont {Michel}, \citenamefont {Milano}, \citenamefont
  {Miller}, \citenamefont {Milotti}, \citenamefont {Minazzoli}, \citenamefont
  {Minenkov}, \citenamefont {Montani}, \citenamefont {Morawski}, \citenamefont
  {Mours}, \citenamefont {Muciaccia}, \citenamefont {Nagar}, \citenamefont
  {Nardecchia}, \citenamefont {Naticchioni}, \citenamefont {Neilson},
  \citenamefont {Nelemans}, \citenamefont {Nguyen}, \citenamefont {Nichols},
  \citenamefont {Nissanke}, \citenamefont {Nocera}, \citenamefont {Oganesyan},
  \citenamefont {Olivetto}, \citenamefont {Pagano}, \citenamefont {Pagliaroli},
  \citenamefont {Palomba}, \citenamefont {Pang}, \citenamefont {Pannarale},
  \citenamefont {Paoletti}, \citenamefont {Paoli}, \citenamefont {Pascucci},
  \citenamefont {Pasqualetti}, \citenamefont {Passaquieti}, \citenamefont
  {Passuello}, \citenamefont {Patricelli}, \citenamefont {Perego},
  \citenamefont {Pegoraro}, \citenamefont {P\'erigois}, \citenamefont
  {Perreca}, \citenamefont {Perri\`es}, \citenamefont {Phukon}, \citenamefont
  {Piccinni}, \citenamefont {Pichot}, \citenamefont {Piendibene}, \citenamefont
  {Piergiovanni}, \citenamefont {Pierro}, \citenamefont {Pillant},
  \citenamefont {Pinard}, \citenamefont {Pinto}, \citenamefont {Piotrzkowski},
  \citenamefont {Plastino}, \citenamefont {Poggiani}, \citenamefont
  {Popolizio}, \citenamefont {Porter}, \citenamefont {Prevedelli},
  \citenamefont {Principe}, \citenamefont {Prodi}, \citenamefont {Punturo},
  \citenamefont {Puppo}, \citenamefont {Raaijmakers}, \citenamefont
  {Radulesco}, \citenamefont {Rapagnani}, \citenamefont {Razzano},
  \citenamefont {Regimbau}, \citenamefont {Rei}, \citenamefont {Rettegno},
  \citenamefont {Ricci}, \citenamefont {Riemenschneider}, \citenamefont
  {Robinet}, \citenamefont {Rocchi}, \citenamefont {Rolland}, \citenamefont
  {Romanelli}, \citenamefont {Romano}, \citenamefont
  {Rosi\ifmmode~\acute{n}\else \'{n}\fi{}ska}, \citenamefont {Ruggi},
  \citenamefont {Salafia}, \citenamefont {Salconi}, \citenamefont {Samajdar},
  \citenamefont {Sanchis-Gual}, \citenamefont {Santos}, \citenamefont
  {Sassolas}, \citenamefont {Sauter}, \citenamefont {Sayah}, \citenamefont
  {Sentenac}, \citenamefont {Sequino}, \citenamefont {Sharma}, \citenamefont
  {Sieniawska}, \citenamefont {Singh}, \citenamefont {Singhal}, \citenamefont
  {Sipala}, \citenamefont {Sordini}, \citenamefont {Sorrentino}, \citenamefont
  {Spera}, \citenamefont {Stachie}, \citenamefont {Steer}, \citenamefont
  {Stratta}, \citenamefont {Sur}, \citenamefont {Swinkels}, \citenamefont
  {Tacca}, \citenamefont {Tanasijczuk}, \citenamefont {Tapia San~Martin},
  \citenamefont {Tiwari}, \citenamefont {Tonelli}, \citenamefont
  {Torres-Forn\'e}, \citenamefont {Tosta~e Melo}, \citenamefont {Travasso},
  \citenamefont {Tringali}, \citenamefont {Trovato}, \citenamefont {Tsang},
  \citenamefont {Turconi}, \citenamefont {Valentini}, \citenamefont {van
  Bakel}, \citenamefont {van Beuzekom}, \citenamefont {van~den Brand},
  \citenamefont {Van Den~Broeck}, \citenamefont {van~der Schaaf}, \citenamefont
  {Vardaro}, \citenamefont {Vas\'uth}, \citenamefont {Vedovato}, \citenamefont
  {Verkindt}, \citenamefont {Vetrano}, \citenamefont {Vicer\'e}, \citenamefont
  {Vinet}, \citenamefont {Vocca}, \citenamefont {Walet}, \citenamefont {Was},
  \citenamefont {Zadro\ifmmode~\dot{z}\else \.{z}\fi{}ny}, \citenamefont
  {Zelenova}, \citenamefont {Zendri}, \citenamefont {Vahlbruch}, \citenamefont
  {Mehmet}, \citenamefont {L\"uck},\ and\ \citenamefont
  {Danzmann}}]{Acernese2019}%
  \BibitemOpen
  \bibfield  {author} {\bibinfo {author} {\bibfnamefont {F.}~\bibnamefont
  {Acernese}}, \bibinfo {author} {\bibfnamefont {M.}~\bibnamefont {Agathos}},
  \bibinfo {author} {\bibfnamefont {L.}~\bibnamefont {Aiello}}, \bibinfo
  {author} {\bibfnamefont {A.}~\bibnamefont {Allocca}}, \bibinfo {author}
  {\bibfnamefont {A.}~\bibnamefont {Amato}}, \bibinfo {author} {\bibfnamefont
  {S.}~\bibnamefont {Ansoldi}}, \bibinfo {author} {\bibfnamefont
  {S.}~\bibnamefont {Antier}}, \bibinfo {author} {\bibfnamefont
  {M.}~\bibnamefont {Ar\`ene}}, \bibinfo {author} {\bibfnamefont
  {N.}~\bibnamefont {Arnaud}}, \bibinfo {author} {\bibfnamefont
  {S.}~\bibnamefont {Ascenzi}}, \bibinfo {author} {\bibfnamefont
  {P.}~\bibnamefont {Astone}}, \bibinfo {author} {\bibfnamefont
  {F.}~\bibnamefont {Aubin}}, \bibinfo {author} {\bibfnamefont
  {S.}~\bibnamefont {Babak}}, \bibinfo {author} {\bibfnamefont
  {P.}~\bibnamefont {Bacon}}, \bibinfo {author} {\bibfnamefont
  {F.}~\bibnamefont {Badaracco}}, \bibinfo {author} {\bibfnamefont {M.~K.~M.}\
  \bibnamefont {Bader}}, \bibinfo {author} {\bibfnamefont {J.}~\bibnamefont
  {Baird}}, \bibinfo {author} {\bibfnamefont {F.}~\bibnamefont {Baldaccini}},
  \bibinfo {author} {\bibfnamefont {G.}~\bibnamefont {Ballardin}}, \bibinfo
  {author} {\bibfnamefont {G.}~\bibnamefont {Baltus}}, \bibinfo {author}
  {\bibfnamefont {C.}~\bibnamefont {Barbieri}}, \bibinfo {author}
  {\bibfnamefont {P.}~\bibnamefont {Barneo}}, \bibinfo {author} {\bibfnamefont
  {F.}~\bibnamefont {Barone}}, \bibinfo {author} {\bibfnamefont
  {M.}~\bibnamefont {Barsuglia}}, \bibinfo {author} {\bibfnamefont
  {D.}~\bibnamefont {Barta}}, \bibinfo {author} {\bibfnamefont
  {A.}~\bibnamefont {Basti}}, \bibinfo {author} {\bibfnamefont
  {M.}~\bibnamefont {Bawaj}}, \bibinfo {author} {\bibfnamefont
  {M.}~\bibnamefont {Bazzan}}, \bibinfo {author} {\bibfnamefont
  {M.}~\bibnamefont {Bejger}}, \bibinfo {author} {\bibfnamefont
  {I.}~\bibnamefont {Belahcene}}, \bibinfo {author} {\bibfnamefont
  {S.}~\bibnamefont {Bernuzzi}}, \bibinfo {author} {\bibfnamefont
  {D.}~\bibnamefont {Bersanetti}}, \bibinfo {author} {\bibfnamefont
  {A.}~\bibnamefont {Bertolini}}, \bibinfo {author} {\bibfnamefont
  {M.}~\bibnamefont {Bischi}}, \bibinfo {author} {\bibfnamefont
  {M.}~\bibnamefont {Bitossi}}, \bibinfo {author} {\bibfnamefont {M.~A.}\
  \bibnamefont {Bizouard}}, \bibinfo {author} {\bibfnamefont {F.}~\bibnamefont
  {Bobba}}, \bibinfo {author} {\bibfnamefont {M.}~\bibnamefont {Boer}},
  \bibinfo {author} {\bibfnamefont {G.}~\bibnamefont {Bogaert}}, \bibinfo
  {author} {\bibfnamefont {F.}~\bibnamefont {Bondu}}, \bibinfo {author}
  {\bibfnamefont {R.}~\bibnamefont {Bonnand}}, \bibinfo {author} {\bibfnamefont
  {B.~A.}\ \bibnamefont {Boom}}, \bibinfo {author} {\bibfnamefont
  {V.}~\bibnamefont {Boschi}}, \bibinfo {author} {\bibfnamefont
  {Y.}~\bibnamefont {Bouffanais}}, \bibinfo {author} {\bibfnamefont
  {A.}~\bibnamefont {Bozzi}}, \bibinfo {author} {\bibfnamefont
  {C.}~\bibnamefont {Bradaschia}}, \bibinfo {author} {\bibfnamefont
  {M.}~\bibnamefont {Branchesi}}, \bibinfo {author} {\bibfnamefont
  {M.}~\bibnamefont {Breschi}}, \bibinfo {author} {\bibfnamefont
  {T.}~\bibnamefont {Briant}}, \bibinfo {author} {\bibfnamefont
  {F.}~\bibnamefont {Brighenti}}, \bibinfo {author} {\bibfnamefont
  {A.}~\bibnamefont {Brillet}}, \bibinfo {author} {\bibfnamefont
  {J.}~\bibnamefont {Brooks}}, \bibinfo {author} {\bibfnamefont
  {G.}~\bibnamefont {Bruno}}, \bibinfo {author} {\bibfnamefont
  {T.}~\bibnamefont {Bulik}}, \bibinfo {author} {\bibfnamefont {H.~J.}\
  \bibnamefont {Bulten}}, \bibinfo {author} {\bibfnamefont {D.}~\bibnamefont
  {Buskulic}}, \bibinfo {author} {\bibfnamefont {G.}~\bibnamefont {Cagnoli}},
  \bibinfo {author} {\bibfnamefont {E.}~\bibnamefont {Calloni}}, \bibinfo
  {author} {\bibfnamefont {M.}~\bibnamefont {Canepa}}, \bibinfo {author}
  {\bibfnamefont {G.}~\bibnamefont {Carapella}}, \bibinfo {author}
  {\bibfnamefont {F.}~\bibnamefont {Carbognani}}, \bibinfo {author}
  {\bibfnamefont {G.}~\bibnamefont {Carullo}}, \bibinfo {author} {\bibfnamefont
  {J.}~\bibnamefont {Casanueva~Diaz}}, \bibinfo {author} {\bibfnamefont
  {C.}~\bibnamefont {Casentini}}, \bibinfo {author} {\bibfnamefont
  {J.}~\bibnamefont {Casta\~neda}}, \bibinfo {author} {\bibfnamefont
  {S.}~\bibnamefont {Caudill}}, \bibinfo {author} {\bibfnamefont
  {F.}~\bibnamefont {Cavalier}}, \bibinfo {author} {\bibfnamefont
  {R.}~\bibnamefont {Cavalieri}}, \bibinfo {author} {\bibfnamefont
  {G.}~\bibnamefont {Cella}}, \bibinfo {author} {\bibfnamefont
  {P.}~\bibnamefont {Cerd\'a-Dur\'an}}, \bibinfo {author} {\bibfnamefont
  {E.}~\bibnamefont {Cesarini}}, \bibinfo {author} {\bibfnamefont
  {O.}~\bibnamefont {Chaibi}}, \bibinfo {author} {\bibfnamefont
  {E.}~\bibnamefont {Chassande-Mottin}}, \bibinfo {author} {\bibfnamefont
  {F.}~\bibnamefont {Chiadini}}, \bibinfo {author} {\bibfnamefont
  {R.}~\bibnamefont {Chierici}}, \bibinfo {author} {\bibfnamefont
  {A.}~\bibnamefont {Chincarini}}, \bibinfo {author} {\bibfnamefont
  {A.}~\bibnamefont {Chiummo}}, \bibinfo {author} {\bibfnamefont
  {N.}~\bibnamefont {Christensen}}, \bibinfo {author} {\bibfnamefont
  {S.}~\bibnamefont {Chua}}, \bibinfo {author} {\bibfnamefont {G.}~\bibnamefont
  {Ciani}}, \bibinfo {author} {\bibfnamefont {P.}~\bibnamefont {Ciecielag}},
  \bibinfo {author} {\bibfnamefont {M.}~\bibnamefont
  {Cie\ifmmode~\acute{s}\else \'{s}\fi{}lar}}, \bibinfo {author} {\bibfnamefont
  {R.}~\bibnamefont {Ciolfi}}, \bibinfo {author} {\bibfnamefont
  {F.}~\bibnamefont {Cipriano}}, \bibinfo {author} {\bibfnamefont
  {A.}~\bibnamefont {Cirone}}, \bibinfo {author} {\bibfnamefont
  {S.}~\bibnamefont {Clesse}}, \bibinfo {author} {\bibfnamefont
  {F.}~\bibnamefont {Cleva}}, \bibinfo {author} {\bibfnamefont
  {E.}~\bibnamefont {Coccia}}, \bibinfo {author} {\bibfnamefont {P.-F.}\
  \bibnamefont {Cohadon}}, \bibinfo {author} {\bibfnamefont {D.}~\bibnamefont
  {Cohen}}, \bibinfo {author} {\bibfnamefont {M.}~\bibnamefont {Colpi}},
  \bibinfo {author} {\bibfnamefont {L.}~\bibnamefont {Conti}}, \bibinfo
  {author} {\bibfnamefont {I.}~\bibnamefont {Cordero-Carri\'on}}, \bibinfo
  {author} {\bibfnamefont {S.}~\bibnamefont {Corezzi}}, \bibinfo {author}
  {\bibfnamefont {D.}~\bibnamefont {Corre}}, \bibinfo {author} {\bibfnamefont
  {S.}~\bibnamefont {Cortese}}, \bibinfo {author} {\bibfnamefont {J.-P.}\
  \bibnamefont {Coulon}}, \bibinfo {author} {\bibfnamefont {M.}~\bibnamefont
  {Croquette}}, \bibinfo {author} {\bibfnamefont {J.-R.}\ \bibnamefont
  {Cudell}}, \bibinfo {author} {\bibfnamefont {E.}~\bibnamefont {Cuoco}},
  \bibinfo {author} {\bibfnamefont {M.}~\bibnamefont {Curylo}}, \bibinfo
  {author} {\bibfnamefont {B.}~\bibnamefont {D'Angelo}}, \bibinfo {author}
  {\bibfnamefont {S.}~\bibnamefont {D'Antonio}}, \bibinfo {author}
  {\bibfnamefont {V.}~\bibnamefont {Dattilo}}, \bibinfo {author} {\bibfnamefont
  {M.}~\bibnamefont {Davier}}, \bibinfo {author} {\bibfnamefont
  {J.}~\bibnamefont {Degallaix}}, \bibinfo {author} {\bibfnamefont
  {M.}~\bibnamefont {De~Laurentis}}, \bibinfo {author} {\bibfnamefont
  {S.}~\bibnamefont {Del\'eglise}}, \bibinfo {author} {\bibfnamefont
  {W.}~\bibnamefont {Del~Pozzo}}, \bibinfo {author} {\bibfnamefont
  {R.}~\bibnamefont {De~Pietri}}, \bibinfo {author} {\bibfnamefont
  {R.}~\bibnamefont {De~Rosa}}, \bibinfo {author} {\bibfnamefont
  {C.}~\bibnamefont {De~Rossi}}, \bibinfo {author} {\bibfnamefont
  {T.}~\bibnamefont {Dietrich}}, \bibinfo {author} {\bibfnamefont
  {L.}~\bibnamefont {Di~Fiore}}, \bibinfo {author} {\bibfnamefont
  {C.}~\bibnamefont {Di~Giorgio}}, \bibinfo {author} {\bibfnamefont
  {F.}~\bibnamefont {Di~Giovanni}}, \bibinfo {author} {\bibfnamefont
  {M.}~\bibnamefont {Di~Giovanni}}, \bibinfo {author} {\bibfnamefont
  {T.}~\bibnamefont {Di~Girolamo}}, \bibinfo {author} {\bibfnamefont
  {A.}~\bibnamefont {Di~Lieto}}, \bibinfo {author} {\bibfnamefont
  {S.}~\bibnamefont {Di~Pace}}, \bibinfo {author} {\bibfnamefont
  {I.}~\bibnamefont {Di~Palma}}, \bibinfo {author} {\bibfnamefont
  {F.}~\bibnamefont {Di~Renzo}}, \bibinfo {author} {\bibfnamefont
  {M.}~\bibnamefont {Drago}}, \bibinfo {author} {\bibfnamefont {J.-G.}\
  \bibnamefont {Ducoin}}, \bibinfo {author} {\bibfnamefont {O.}~\bibnamefont
  {Durante}}, \bibinfo {author} {\bibfnamefont {D.}~\bibnamefont {D'Urso}},
  \bibinfo {author} {\bibfnamefont {M.}~\bibnamefont {Eisenmann}}, \bibinfo
  {author} {\bibfnamefont {L.}~\bibnamefont {Errico}}, \bibinfo {author}
  {\bibfnamefont {D.}~\bibnamefont {Estevez}}, \bibinfo {author} {\bibfnamefont
  {V.}~\bibnamefont {Fafone}}, \bibinfo {author} {\bibfnamefont
  {S.}~\bibnamefont {Farinon}}, \bibinfo {author} {\bibfnamefont
  {F.}~\bibnamefont {Feng}}, \bibinfo {author} {\bibfnamefont {E.}~\bibnamefont
  {Fenyvesi}}, \bibinfo {author} {\bibfnamefont {I.}~\bibnamefont {Ferrante}},
  \bibinfo {author} {\bibfnamefont {F.}~\bibnamefont {Fidecaro}}, \bibinfo
  {author} {\bibfnamefont {I.}~\bibnamefont {Fiori}}, \bibinfo {author}
  {\bibfnamefont {D.}~\bibnamefont {Fiorucci}}, \bibinfo {author}
  {\bibfnamefont {R.}~\bibnamefont {Fittipaldi}}, \bibinfo {author}
  {\bibfnamefont {V.}~\bibnamefont {Fiumara}}, \bibinfo {author} {\bibfnamefont
  {R.}~\bibnamefont {Flaminio}}, \bibinfo {author} {\bibfnamefont {J.~A.}\
  \bibnamefont {Font}}, \bibinfo {author} {\bibfnamefont {J.-D.}\ \bibnamefont
  {Fournier}}, \bibinfo {author} {\bibfnamefont {S.}~\bibnamefont {Frasca}},
  \bibinfo {author} {\bibfnamefont {F.}~\bibnamefont {Frasconi}}, \bibinfo
  {author} {\bibfnamefont {V.}~\bibnamefont {Frey}}, \bibinfo {author}
  {\bibfnamefont {G.}~\bibnamefont {Fronz\`e}}, \bibinfo {author}
  {\bibfnamefont {F.}~\bibnamefont {Garufi}}, \bibinfo {author} {\bibfnamefont
  {G.}~\bibnamefont {Gemme}}, \bibinfo {author} {\bibfnamefont
  {E.}~\bibnamefont {Genin}}, \bibinfo {author} {\bibfnamefont
  {A.}~\bibnamefont {Gennai}}, \bibinfo {author} {\bibfnamefont
  {A.}~\bibnamefont {Ghosh}}, \bibinfo {author} {\bibfnamefont
  {B.}~\bibnamefont {Giacomazzo}}, \bibinfo {author} {\bibfnamefont
  {M.}~\bibnamefont {Gosselin}}, \bibinfo {author} {\bibfnamefont
  {R.}~\bibnamefont {Gouaty}}, \bibinfo {author} {\bibfnamefont
  {A.}~\bibnamefont {Grado}}, \bibinfo {author} {\bibfnamefont
  {M.}~\bibnamefont {Granata}}, \bibinfo {author} {\bibfnamefont
  {G.}~\bibnamefont {Greco}}, \bibinfo {author} {\bibfnamefont
  {G.}~\bibnamefont {Grignani}}, \bibinfo {author} {\bibfnamefont
  {A.}~\bibnamefont {Grimaldi}}, \bibinfo {author} {\bibfnamefont {S.~J.}\
  \bibnamefont {Grimm}}, \bibinfo {author} {\bibfnamefont {P.}~\bibnamefont
  {Gruning}}, \bibinfo {author} {\bibfnamefont {G.~M.}\ \bibnamefont {Guidi}},
  \bibinfo {author} {\bibfnamefont {G.}~\bibnamefont {Guix\'e}}, \bibinfo
  {author} {\bibfnamefont {Y.}~\bibnamefont {Guo}}, \bibinfo {author}
  {\bibfnamefont {P.}~\bibnamefont {Gupta}}, \bibinfo {author} {\bibfnamefont
  {O.}~\bibnamefont {Halim}}, \bibinfo {author} {\bibfnamefont
  {T.}~\bibnamefont {Harder}}, \bibinfo {author} {\bibfnamefont
  {J.}~\bibnamefont {Harms}}, \bibinfo {author} {\bibfnamefont
  {A.}~\bibnamefont {Heidmann}}, \bibinfo {author} {\bibfnamefont
  {H.}~\bibnamefont {Heitmann}}, \bibinfo {author} {\bibfnamefont
  {P.}~\bibnamefont {Hello}}, \bibinfo {author} {\bibfnamefont
  {G.}~\bibnamefont {Hemming}}, \bibinfo {author} {\bibfnamefont
  {E.}~\bibnamefont {Hennes}}, \bibinfo {author} {\bibfnamefont
  {T.}~\bibnamefont {Hinderer}}, \bibinfo {author} {\bibfnamefont
  {D.}~\bibnamefont {Hofman}}, \bibinfo {author} {\bibfnamefont
  {D.}~\bibnamefont {Huet}}, \bibinfo {author} {\bibfnamefont {V.}~\bibnamefont
  {Hui}}, \bibinfo {author} {\bibfnamefont {B.}~\bibnamefont {Idzkowski}},
  \bibinfo {author} {\bibfnamefont {A.}~\bibnamefont {Iess}}, \bibinfo {author}
  {\bibfnamefont {G.}~\bibnamefont {Intini}}, \bibinfo {author} {\bibfnamefont
  {J.-M.}\ \bibnamefont {Isac}}, \bibinfo {author} {\bibfnamefont
  {T.}~\bibnamefont {Jacqmin}}, \bibinfo {author} {\bibfnamefont
  {P.}~\bibnamefont {Jaranowski}}, \bibinfo {author} {\bibfnamefont {R.~J.~G.}\
  \bibnamefont {Jonker}}, \bibinfo {author} {\bibfnamefont {S.}~\bibnamefont
  {Katsanevas}}, \bibinfo {author} {\bibfnamefont {F.}~\bibnamefont
  {K\'ef\'elian}}, \bibinfo {author} {\bibfnamefont {I.}~\bibnamefont {Khan}},
  \bibinfo {author} {\bibfnamefont {N.}~\bibnamefont {Khetan}}, \bibinfo
  {author} {\bibfnamefont {G.}~\bibnamefont {Koekoek}}, \bibinfo {author}
  {\bibfnamefont {S.}~\bibnamefont {Koley}}, \bibinfo {author} {\bibfnamefont
  {A.}~\bibnamefont {Kr\'olak}}, \bibinfo {author} {\bibfnamefont
  {A.}~\bibnamefont {Kutynia}}, \bibinfo {author} {\bibfnamefont
  {D.}~\bibnamefont {Laghi}}, \bibinfo {author} {\bibfnamefont
  {A.}~\bibnamefont {Lamberts}}, \bibinfo {author} {\bibfnamefont
  {I.}~\bibnamefont {La~Rosa}}, \bibinfo {author} {\bibfnamefont
  {A.}~\bibnamefont {Lartaux-Vollard}}, \bibinfo {author} {\bibfnamefont
  {C.}~\bibnamefont {Lazzaro}}, \bibinfo {author} {\bibfnamefont
  {P.}~\bibnamefont {Leaci}}, \bibinfo {author} {\bibfnamefont
  {N.}~\bibnamefont {Leroy}}, \bibinfo {author} {\bibfnamefont
  {N.}~\bibnamefont {Letendre}}, \bibinfo {author} {\bibfnamefont
  {F.}~\bibnamefont {Linde}}, \bibinfo {author} {\bibfnamefont
  {M.}~\bibnamefont {Llorens-Monteagudo}}, \bibinfo {author} {\bibfnamefont
  {A.}~\bibnamefont {Longo}}, \bibinfo {author} {\bibfnamefont
  {M.}~\bibnamefont {Lorenzini}}, \bibinfo {author} {\bibfnamefont
  {V.}~\bibnamefont {Loriette}}, \bibinfo {author} {\bibfnamefont
  {G.}~\bibnamefont {Losurdo}}, \bibinfo {author} {\bibfnamefont
  {D.}~\bibnamefont {Lumaca}}, \bibinfo {author} {\bibfnamefont
  {A.}~\bibnamefont {Macquet}}, \bibinfo {author} {\bibfnamefont
  {E.}~\bibnamefont {Majorana}}, \bibinfo {author} {\bibfnamefont
  {I.}~\bibnamefont {Maksimovic}}, \bibinfo {author} {\bibfnamefont
  {N.}~\bibnamefont {Man}}, \bibinfo {author} {\bibfnamefont {V.}~\bibnamefont
  {Mangano}}, \bibinfo {author} {\bibfnamefont {M.}~\bibnamefont {Mantovani}},
  \bibinfo {author} {\bibfnamefont {M.}~\bibnamefont {Mapelli}}, \bibinfo
  {author} {\bibfnamefont {F.}~\bibnamefont {Marchesoni}}, \bibinfo {author}
  {\bibfnamefont {F.}~\bibnamefont {Marion}}, \bibinfo {author} {\bibfnamefont
  {A.}~\bibnamefont {Marquina}}, \bibinfo {author} {\bibfnamefont
  {S.}~\bibnamefont {Marsat}}, \bibinfo {author} {\bibfnamefont
  {F.}~\bibnamefont {Martelli}}, \bibinfo {author} {\bibfnamefont
  {V.}~\bibnamefont {Martinez}}, \bibinfo {author} {\bibfnamefont
  {A.}~\bibnamefont {Masserot}}, \bibinfo {author} {\bibfnamefont
  {S.}~\bibnamefont {Mastrogiovanni}}, \bibinfo {author} {\bibfnamefont
  {E.}~\bibnamefont {Mejuto~Villa}}, \bibinfo {author} {\bibfnamefont
  {L.}~\bibnamefont {Mereni}}, \bibinfo {author} {\bibfnamefont
  {M.}~\bibnamefont {Merzougui}}, \bibinfo {author} {\bibfnamefont
  {R.}~\bibnamefont {Metzdorff}}, \bibinfo {author} {\bibfnamefont
  {A.}~\bibnamefont {Miani}}, \bibinfo {author} {\bibfnamefont
  {C.}~\bibnamefont {Michel}}, \bibinfo {author} {\bibfnamefont
  {L.}~\bibnamefont {Milano}}, \bibinfo {author} {\bibfnamefont
  {A.}~\bibnamefont {Miller}}, \bibinfo {author} {\bibfnamefont
  {E.}~\bibnamefont {Milotti}}, \bibinfo {author} {\bibfnamefont
  {O.}~\bibnamefont {Minazzoli}}, \bibinfo {author} {\bibfnamefont
  {Y.}~\bibnamefont {Minenkov}}, \bibinfo {author} {\bibfnamefont
  {M.}~\bibnamefont {Montani}}, \bibinfo {author} {\bibfnamefont
  {F.}~\bibnamefont {Morawski}}, \bibinfo {author} {\bibfnamefont
  {B.}~\bibnamefont {Mours}}, \bibinfo {author} {\bibfnamefont
  {F.}~\bibnamefont {Muciaccia}}, \bibinfo {author} {\bibfnamefont
  {A.}~\bibnamefont {Nagar}}, \bibinfo {author} {\bibfnamefont
  {I.}~\bibnamefont {Nardecchia}}, \bibinfo {author} {\bibfnamefont
  {L.}~\bibnamefont {Naticchioni}}, \bibinfo {author} {\bibfnamefont
  {J.}~\bibnamefont {Neilson}}, \bibinfo {author} {\bibfnamefont
  {G.}~\bibnamefont {Nelemans}}, \bibinfo {author} {\bibfnamefont
  {C.}~\bibnamefont {Nguyen}}, \bibinfo {author} {\bibfnamefont
  {D.}~\bibnamefont {Nichols}}, \bibinfo {author} {\bibfnamefont
  {S.}~\bibnamefont {Nissanke}}, \bibinfo {author} {\bibfnamefont
  {F.}~\bibnamefont {Nocera}}, \bibinfo {author} {\bibfnamefont
  {G.}~\bibnamefont {Oganesyan}}, \bibinfo {author} {\bibfnamefont
  {C.}~\bibnamefont {Olivetto}}, \bibinfo {author} {\bibfnamefont
  {G.}~\bibnamefont {Pagano}}, \bibinfo {author} {\bibfnamefont
  {G.}~\bibnamefont {Pagliaroli}}, \bibinfo {author} {\bibfnamefont
  {C.}~\bibnamefont {Palomba}}, \bibinfo {author} {\bibfnamefont {P.~T.~H.}\
  \bibnamefont {Pang}}, \bibinfo {author} {\bibfnamefont {F.}~\bibnamefont
  {Pannarale}}, \bibinfo {author} {\bibfnamefont {F.}~\bibnamefont {Paoletti}},
  \bibinfo {author} {\bibfnamefont {A.}~\bibnamefont {Paoli}}, \bibinfo
  {author} {\bibfnamefont {D.}~\bibnamefont {Pascucci}}, \bibinfo {author}
  {\bibfnamefont {A.}~\bibnamefont {Pasqualetti}}, \bibinfo {author}
  {\bibfnamefont {R.}~\bibnamefont {Passaquieti}}, \bibinfo {author}
  {\bibfnamefont {D.}~\bibnamefont {Passuello}}, \bibinfo {author}
  {\bibfnamefont {B.}~\bibnamefont {Patricelli}}, \bibinfo {author}
  {\bibfnamefont {A.}~\bibnamefont {Perego}}, \bibinfo {author} {\bibfnamefont
  {M.}~\bibnamefont {Pegoraro}}, \bibinfo {author} {\bibfnamefont
  {C.}~\bibnamefont {P\'erigois}}, \bibinfo {author} {\bibfnamefont
  {A.}~\bibnamefont {Perreca}}, \bibinfo {author} {\bibfnamefont
  {S.}~\bibnamefont {Perri\`es}}, \bibinfo {author} {\bibfnamefont {K.~S.}\
  \bibnamefont {Phukon}}, \bibinfo {author} {\bibfnamefont {O.~J.}\
  \bibnamefont {Piccinni}}, \bibinfo {author} {\bibfnamefont {M.}~\bibnamefont
  {Pichot}}, \bibinfo {author} {\bibfnamefont {M.}~\bibnamefont {Piendibene}},
  \bibinfo {author} {\bibfnamefont {F.}~\bibnamefont {Piergiovanni}}, \bibinfo
  {author} {\bibfnamefont {V.}~\bibnamefont {Pierro}}, \bibinfo {author}
  {\bibfnamefont {G.}~\bibnamefont {Pillant}}, \bibinfo {author} {\bibfnamefont
  {L.}~\bibnamefont {Pinard}}, \bibinfo {author} {\bibfnamefont {I.~M.}\
  \bibnamefont {Pinto}}, \bibinfo {author} {\bibfnamefont {K.}~\bibnamefont
  {Piotrzkowski}}, \bibinfo {author} {\bibfnamefont {W.}~\bibnamefont
  {Plastino}}, \bibinfo {author} {\bibfnamefont {R.}~\bibnamefont {Poggiani}},
  \bibinfo {author} {\bibfnamefont {P.}~\bibnamefont {Popolizio}}, \bibinfo
  {author} {\bibfnamefont {E.~K.}\ \bibnamefont {Porter}}, \bibinfo {author}
  {\bibfnamefont {M.}~\bibnamefont {Prevedelli}}, \bibinfo {author}
  {\bibfnamefont {M.}~\bibnamefont {Principe}}, \bibinfo {author}
  {\bibfnamefont {G.~A.}\ \bibnamefont {Prodi}}, \bibinfo {author}
  {\bibfnamefont {M.}~\bibnamefont {Punturo}}, \bibinfo {author} {\bibfnamefont
  {P.}~\bibnamefont {Puppo}}, \bibinfo {author} {\bibfnamefont
  {G.}~\bibnamefont {Raaijmakers}}, \bibinfo {author} {\bibfnamefont
  {N.}~\bibnamefont {Radulesco}}, \bibinfo {author} {\bibfnamefont
  {P.}~\bibnamefont {Rapagnani}}, \bibinfo {author} {\bibfnamefont
  {M.}~\bibnamefont {Razzano}}, \bibinfo {author} {\bibfnamefont
  {T.}~\bibnamefont {Regimbau}}, \bibinfo {author} {\bibfnamefont
  {L.}~\bibnamefont {Rei}}, \bibinfo {author} {\bibfnamefont {P.}~\bibnamefont
  {Rettegno}}, \bibinfo {author} {\bibfnamefont {F.}~\bibnamefont {Ricci}},
  \bibinfo {author} {\bibfnamefont {G.}~\bibnamefont {Riemenschneider}},
  \bibinfo {author} {\bibfnamefont {F.}~\bibnamefont {Robinet}}, \bibinfo
  {author} {\bibfnamefont {A.}~\bibnamefont {Rocchi}}, \bibinfo {author}
  {\bibfnamefont {L.}~\bibnamefont {Rolland}}, \bibinfo {author} {\bibfnamefont
  {M.}~\bibnamefont {Romanelli}}, \bibinfo {author} {\bibfnamefont
  {R.}~\bibnamefont {Romano}}, \bibinfo {author} {\bibfnamefont
  {D.}~\bibnamefont {Rosi\ifmmode~\acute{n}\else \'{n}\fi{}ska}}, \bibinfo
  {author} {\bibfnamefont {P.}~\bibnamefont {Ruggi}}, \bibinfo {author}
  {\bibfnamefont {O.~S.}\ \bibnamefont {Salafia}}, \bibinfo {author}
  {\bibfnamefont {L.}~\bibnamefont {Salconi}}, \bibinfo {author} {\bibfnamefont
  {A.}~\bibnamefont {Samajdar}}, \bibinfo {author} {\bibfnamefont
  {N.}~\bibnamefont {Sanchis-Gual}}, \bibinfo {author} {\bibfnamefont
  {E.}~\bibnamefont {Santos}}, \bibinfo {author} {\bibfnamefont
  {B.}~\bibnamefont {Sassolas}}, \bibinfo {author} {\bibfnamefont
  {O.}~\bibnamefont {Sauter}}, \bibinfo {author} {\bibfnamefont
  {S.}~\bibnamefont {Sayah}}, \bibinfo {author} {\bibfnamefont
  {D.}~\bibnamefont {Sentenac}}, \bibinfo {author} {\bibfnamefont
  {V.}~\bibnamefont {Sequino}}, \bibinfo {author} {\bibfnamefont
  {A.}~\bibnamefont {Sharma}}, \bibinfo {author} {\bibfnamefont
  {M.}~\bibnamefont {Sieniawska}}, \bibinfo {author} {\bibfnamefont
  {N.}~\bibnamefont {Singh}}, \bibinfo {author} {\bibfnamefont
  {A.}~\bibnamefont {Singhal}}, \bibinfo {author} {\bibfnamefont
  {V.}~\bibnamefont {Sipala}}, \bibinfo {author} {\bibfnamefont
  {V.}~\bibnamefont {Sordini}}, \bibinfo {author} {\bibfnamefont
  {F.}~\bibnamefont {Sorrentino}}, \bibinfo {author} {\bibfnamefont
  {M.}~\bibnamefont {Spera}}, \bibinfo {author} {\bibfnamefont
  {C.}~\bibnamefont {Stachie}}, \bibinfo {author} {\bibfnamefont {D.~A.}\
  \bibnamefont {Steer}}, \bibinfo {author} {\bibfnamefont {G.}~\bibnamefont
  {Stratta}}, \bibinfo {author} {\bibfnamefont {A.}~\bibnamefont {Sur}},
  \bibinfo {author} {\bibfnamefont {B.~L.}\ \bibnamefont {Swinkels}}, \bibinfo
  {author} {\bibfnamefont {M.}~\bibnamefont {Tacca}}, \bibinfo {author}
  {\bibfnamefont {A.~J.}\ \bibnamefont {Tanasijczuk}}, \bibinfo {author}
  {\bibfnamefont {E.~N.}\ \bibnamefont {Tapia San~Martin}}, \bibinfo {author}
  {\bibfnamefont {S.}~\bibnamefont {Tiwari}}, \bibinfo {author} {\bibfnamefont
  {M.}~\bibnamefont {Tonelli}}, \bibinfo {author} {\bibfnamefont
  {A.}~\bibnamefont {Torres-Forn\'e}}, \bibinfo {author} {\bibfnamefont
  {I.}~\bibnamefont {Tosta~e Melo}}, \bibinfo {author} {\bibfnamefont
  {F.}~\bibnamefont {Travasso}}, \bibinfo {author} {\bibfnamefont {M.~C.}\
  \bibnamefont {Tringali}}, \bibinfo {author} {\bibfnamefont {A.}~\bibnamefont
  {Trovato}}, \bibinfo {author} {\bibfnamefont {K.~W.}\ \bibnamefont {Tsang}},
  \bibinfo {author} {\bibfnamefont {M.}~\bibnamefont {Turconi}}, \bibinfo
  {author} {\bibfnamefont {M.}~\bibnamefont {Valentini}}, \bibinfo {author}
  {\bibfnamefont {N.}~\bibnamefont {van Bakel}}, \bibinfo {author}
  {\bibfnamefont {M.}~\bibnamefont {van Beuzekom}}, \bibinfo {author}
  {\bibfnamefont {J.~F.~J.}\ \bibnamefont {van~den Brand}}, \bibinfo {author}
  {\bibfnamefont {C.}~\bibnamefont {Van Den~Broeck}}, \bibinfo {author}
  {\bibfnamefont {L.}~\bibnamefont {van~der Schaaf}}, \bibinfo {author}
  {\bibfnamefont {M.}~\bibnamefont {Vardaro}}, \bibinfo {author} {\bibfnamefont
  {M.}~\bibnamefont {Vas\'uth}}, \bibinfo {author} {\bibfnamefont
  {G.}~\bibnamefont {Vedovato}}, \bibinfo {author} {\bibfnamefont
  {D.}~\bibnamefont {Verkindt}}, \bibinfo {author} {\bibfnamefont
  {F.}~\bibnamefont {Vetrano}}, \bibinfo {author} {\bibfnamefont
  {A.}~\bibnamefont {Vicer\'e}}, \bibinfo {author} {\bibfnamefont {J.-Y.}\
  \bibnamefont {Vinet}}, \bibinfo {author} {\bibfnamefont {H.}~\bibnamefont
  {Vocca}}, \bibinfo {author} {\bibfnamefont {R.}~\bibnamefont {Walet}},
  \bibinfo {author} {\bibfnamefont {M.}~\bibnamefont {Was}}, \bibinfo {author}
  {\bibfnamefont {A.}~\bibnamefont {Zadro\ifmmode~\dot{z}\else \.{z}\fi{}ny}},
  \bibinfo {author} {\bibfnamefont {T.}~\bibnamefont {Zelenova}}, \bibinfo
  {author} {\bibfnamefont {J.-P.}\ \bibnamefont {Zendri}}, \bibinfo {author}
  {\bibfnamefont {H.}~\bibnamefont {Vahlbruch}}, \bibinfo {author}
  {\bibfnamefont {M.}~\bibnamefont {Mehmet}}, \bibinfo {author} {\bibfnamefont
  {H.}~\bibnamefont {L\"uck}}, \ and\ \bibinfo {author} {\bibfnamefont
  {K.}~\bibnamefont {Danzmann}} (\bibinfo {collaboration} {Virgo
  Collaboration}),\ }\href {\doibase 10.1103/PhysRevLett.123.231108} {\bibfield
   {journal} {\bibinfo  {journal} {Phys. Rev. Lett.}\ }\textbf {\bibinfo
  {volume} {123}},\ \bibinfo {pages} {231108} (\bibinfo {year}
  {2019})}\BibitemShut {NoStop}%
\bibitem [{\citenamefont {Tse}\ \emph {et~al.}(2019)\citenamefont {Tse},
  \citenamefont {Yu}, \citenamefont {Kijbunchoo}, \citenamefont
  {Fernandez-Galiana}, \citenamefont {Dupej}, \citenamefont {Barsotti},
  \citenamefont {Blair}, \citenamefont {Brown}, \citenamefont {Dwyer},
  \citenamefont {Effler}, \citenamefont {Evans}, \citenamefont {Fritschel},
  \citenamefont {Frolov}, \citenamefont {Green}, \citenamefont {Mansell},
  \citenamefont {Matichard}, \citenamefont {Mavalvala}, \citenamefont
  {McClelland}, \citenamefont {McCuller}, \citenamefont {McRae}, \citenamefont
  {Miller}, \citenamefont {Mullavey}, \citenamefont {Oelker}, \citenamefont
  {Phinney}, \citenamefont {Sigg}, \citenamefont {Slagmolen}, \citenamefont
  {Vo}, \citenamefont {Ward}, \citenamefont {Whittle}, \citenamefont {Abbott},
  \citenamefont {Adams}, \citenamefont {Adhikari}, \citenamefont {Ananyeva},
  \citenamefont {Appert}, \citenamefont {Arai}, \citenamefont {Areeda},
  \citenamefont {Asali}, \citenamefont {Aston}, \citenamefont {Austin},
  \citenamefont {Baer}, \citenamefont {Ball}, \citenamefont {Ballmer},
  \citenamefont {Banagiri}, \citenamefont {Barker}, \citenamefont {Bartlett},
  \citenamefont {Berger}, \citenamefont {Betzwieser}, \citenamefont
  {Bhattacharjee}, \citenamefont {Billingsley}, \citenamefont {Biscans},
  \citenamefont {Blair}, \citenamefont {Bode}, \citenamefont {Booker},
  \citenamefont {Bork}, \citenamefont {Bramley}, \citenamefont {Brooks},
  \citenamefont {Buikema}, \citenamefont {Cahillane}, \citenamefont {Cannon},
  \citenamefont {Chen}, \citenamefont {Ciobanu}, \citenamefont {Clara},
  \citenamefont {Cooper}, \citenamefont {Corley}, \citenamefont {Countryman},
  \citenamefont {Covas}, \citenamefont {Coyne}, \citenamefont {Datrier},
  \citenamefont {Davis}, \citenamefont {Di~Fronzo}, \citenamefont {Driggers},
  \citenamefont {Etzel}, \citenamefont {Evans}, \citenamefont {Feicht},
  \citenamefont {Fulda}, \citenamefont {Fyffe}, \citenamefont {Giaime},
  \citenamefont {Giardina}, \citenamefont {Godwin}, \citenamefont {Goetz},
  \citenamefont {Gras}, \citenamefont {Gray}, \citenamefont {Gray},
  \citenamefont {Gupta}, \citenamefont {Gustafson}, \citenamefont {Gustafson},
  \citenamefont {Hanks}, \citenamefont {Hanson}, \citenamefont {Hardwick},
  \citenamefont {Hasskew}, \citenamefont {Heintze}, \citenamefont
  {Helmling-Cornell}, \citenamefont {Holland}, \citenamefont {Jones},
  \citenamefont {Kandhasamy}, \citenamefont {Karki}, \citenamefont {Kasprzack},
  \citenamefont {Kawabe}, \citenamefont {King}, \citenamefont {Kissel},
  \citenamefont {Kumar}, \citenamefont {Landry}, \citenamefont {Lane},
  \citenamefont {Lantz}, \citenamefont {Laxen}, \citenamefont {Lecoeuche},
  \citenamefont {Leviton}, \citenamefont {Liu}, \citenamefont {Lormand},
  \citenamefont {Lundgren}, \citenamefont {Macas}, \citenamefont {MacInnis},
  \citenamefont {Macleod}, \citenamefont {M\'arka}, \citenamefont {M\'arka},
  \citenamefont {Martynov}, \citenamefont {Mason}, \citenamefont {Massinger},
  \citenamefont {McCarthy}, \citenamefont {McCormick}, \citenamefont {McIver},
  \citenamefont {Mendell}, \citenamefont {Merfeld}, \citenamefont {Merilh},
  \citenamefont {Meylahn}, \citenamefont {Mistry}, \citenamefont {Mittleman},
  \citenamefont {Moreno}, \citenamefont {Mow-Lowry}, \citenamefont {Mozzon},
  \citenamefont {Nelson}, \citenamefont {Nguyen}, \citenamefont {Nuttall},
  \citenamefont {Oberling}, \citenamefont {Oram}, \citenamefont {O'Reilly},
  \citenamefont {Osthelder}, \citenamefont {Ottaway}, \citenamefont {Overmier},
  \citenamefont {Palamos}, \citenamefont {Parker}, \citenamefont {Payne},
  \citenamefont {Pele}, \citenamefont {Perez}, \citenamefont {Pirello},
  \citenamefont {Radkins}, \citenamefont {Ramirez}, \citenamefont {Richardson},
  \citenamefont {Riles}, \citenamefont {Robertson}, \citenamefont {Rollins},
  \citenamefont {Romel}, \citenamefont {Romie}, \citenamefont {Ross},
  \citenamefont {Ryan}, \citenamefont {Sadecki}, \citenamefont {Sanchez},
  \citenamefont {Sanchez}, \citenamefont {Saravanan}, \citenamefont {Savage},
  \citenamefont {Schaetzl}, \citenamefont {Schnabel}, \citenamefont
  {Schofield}, \citenamefont {Schwartz}, \citenamefont {Sellers}, \citenamefont
  {Shaffer}, \citenamefont {Smith}, \citenamefont {Soni}, \citenamefont
  {Sorazu}, \citenamefont {Spencer}, \citenamefont {Strain}, \citenamefont
  {Sun}, \citenamefont {Szczepa\ifmmode~\acute{n}\else \'{n}\fi{}czyk},
  \citenamefont {Thomas}, \citenamefont {Thomas}, \citenamefont {Thorne},
  \citenamefont {Toland}, \citenamefont {Torrie}, \citenamefont {Traylor},
  \citenamefont {Urban}, \citenamefont {Vajente}, \citenamefont {Valdes},
  \citenamefont {Vander-Hyde}, \citenamefont {Veitch}, \citenamefont
  {Venkateswara}, \citenamefont {Venugopalan}, \citenamefont {Viets},
  \citenamefont {Vorvick}, \citenamefont {Wade}, \citenamefont {Warner},
  \citenamefont {Weaver}, \citenamefont {Weiss}, \citenamefont {Willke},
  \citenamefont {Wipf}, \citenamefont {Xiao}, \citenamefont {Yamamoto},
  \citenamefont {Yap}, \citenamefont {Yu}, \citenamefont {Zhang}, \citenamefont
  {Zucker},\ and\ \citenamefont {Zweizig}}]{Tse2019}%
  \BibitemOpen
  \bibfield  {author} {\bibinfo {author} {\bibfnamefont {M.}~\bibnamefont
  {Tse}}, \bibinfo {author} {\bibfnamefont {H.}~\bibnamefont {Yu}}, \bibinfo
  {author} {\bibfnamefont {N.}~\bibnamefont {Kijbunchoo}}, \bibinfo {author}
  {\bibfnamefont {A.}~\bibnamefont {Fernandez-Galiana}}, \bibinfo {author}
  {\bibfnamefont {P.}~\bibnamefont {Dupej}}, \bibinfo {author} {\bibfnamefont
  {L.}~\bibnamefont {Barsotti}}, \bibinfo {author} {\bibfnamefont {C.~D.}\
  \bibnamefont {Blair}}, \bibinfo {author} {\bibfnamefont {D.~D.}\ \bibnamefont
  {Brown}}, \bibinfo {author} {\bibfnamefont {S.~E.}\ \bibnamefont {Dwyer}},
  \bibinfo {author} {\bibfnamefont {A.}~\bibnamefont {Effler}}, \bibinfo
  {author} {\bibfnamefont {M.}~\bibnamefont {Evans}}, \bibinfo {author}
  {\bibfnamefont {P.}~\bibnamefont {Fritschel}}, \bibinfo {author}
  {\bibfnamefont {V.~V.}\ \bibnamefont {Frolov}}, \bibinfo {author}
  {\bibfnamefont {A.~C.}\ \bibnamefont {Green}}, \bibinfo {author}
  {\bibfnamefont {G.~L.}\ \bibnamefont {Mansell}}, \bibinfo {author}
  {\bibfnamefont {F.}~\bibnamefont {Matichard}}, \bibinfo {author}
  {\bibfnamefont {N.}~\bibnamefont {Mavalvala}}, \bibinfo {author}
  {\bibfnamefont {D.~E.}\ \bibnamefont {McClelland}}, \bibinfo {author}
  {\bibfnamefont {L.}~\bibnamefont {McCuller}}, \bibinfo {author}
  {\bibfnamefont {T.}~\bibnamefont {McRae}}, \bibinfo {author} {\bibfnamefont
  {J.}~\bibnamefont {Miller}}, \bibinfo {author} {\bibfnamefont
  {A.}~\bibnamefont {Mullavey}}, \bibinfo {author} {\bibfnamefont
  {E.}~\bibnamefont {Oelker}}, \bibinfo {author} {\bibfnamefont {I.~Y.}\
  \bibnamefont {Phinney}}, \bibinfo {author} {\bibfnamefont {D.}~\bibnamefont
  {Sigg}}, \bibinfo {author} {\bibfnamefont {B.~J.~J.}\ \bibnamefont
  {Slagmolen}}, \bibinfo {author} {\bibfnamefont {T.}~\bibnamefont {Vo}},
  \bibinfo {author} {\bibfnamefont {R.~L.}\ \bibnamefont {Ward}}, \bibinfo
  {author} {\bibfnamefont {C.}~\bibnamefont {Whittle}}, \bibinfo {author}
  {\bibfnamefont {R.}~\bibnamefont {Abbott}}, \bibinfo {author} {\bibfnamefont
  {C.}~\bibnamefont {Adams}}, \bibinfo {author} {\bibfnamefont {R.~X.}\
  \bibnamefont {Adhikari}}, \bibinfo {author} {\bibfnamefont {A.}~\bibnamefont
  {Ananyeva}}, \bibinfo {author} {\bibfnamefont {S.}~\bibnamefont {Appert}},
  \bibinfo {author} {\bibfnamefont {K.}~\bibnamefont {Arai}}, \bibinfo {author}
  {\bibfnamefont {J.~S.}\ \bibnamefont {Areeda}}, \bibinfo {author}
  {\bibfnamefont {Y.}~\bibnamefont {Asali}}, \bibinfo {author} {\bibfnamefont
  {S.~M.}\ \bibnamefont {Aston}}, \bibinfo {author} {\bibfnamefont
  {C.}~\bibnamefont {Austin}}, \bibinfo {author} {\bibfnamefont {A.~M.}\
  \bibnamefont {Baer}}, \bibinfo {author} {\bibfnamefont {M.}~\bibnamefont
  {Ball}}, \bibinfo {author} {\bibfnamefont {S.~W.}\ \bibnamefont {Ballmer}},
  \bibinfo {author} {\bibfnamefont {S.}~\bibnamefont {Banagiri}}, \bibinfo
  {author} {\bibfnamefont {D.}~\bibnamefont {Barker}}, \bibinfo {author}
  {\bibfnamefont {J.}~\bibnamefont {Bartlett}}, \bibinfo {author}
  {\bibfnamefont {B.~K.}\ \bibnamefont {Berger}}, \bibinfo {author}
  {\bibfnamefont {J.}~\bibnamefont {Betzwieser}}, \bibinfo {author}
  {\bibfnamefont {D.}~\bibnamefont {Bhattacharjee}}, \bibinfo {author}
  {\bibfnamefont {G.}~\bibnamefont {Billingsley}}, \bibinfo {author}
  {\bibfnamefont {S.}~\bibnamefont {Biscans}}, \bibinfo {author} {\bibfnamefont
  {R.~M.}\ \bibnamefont {Blair}}, \bibinfo {author} {\bibfnamefont
  {N.}~\bibnamefont {Bode}}, \bibinfo {author} {\bibfnamefont {P.}~\bibnamefont
  {Booker}}, \bibinfo {author} {\bibfnamefont {R.}~\bibnamefont {Bork}},
  \bibinfo {author} {\bibfnamefont {A.}~\bibnamefont {Bramley}}, \bibinfo
  {author} {\bibfnamefont {A.~F.}\ \bibnamefont {Brooks}}, \bibinfo {author}
  {\bibfnamefont {A.}~\bibnamefont {Buikema}}, \bibinfo {author} {\bibfnamefont
  {C.}~\bibnamefont {Cahillane}}, \bibinfo {author} {\bibfnamefont {K.~C.}\
  \bibnamefont {Cannon}}, \bibinfo {author} {\bibfnamefont {X.}~\bibnamefont
  {Chen}}, \bibinfo {author} {\bibfnamefont {A.~A.}\ \bibnamefont {Ciobanu}},
  \bibinfo {author} {\bibfnamefont {F.}~\bibnamefont {Clara}}, \bibinfo
  {author} {\bibfnamefont {S.~J.}\ \bibnamefont {Cooper}}, \bibinfo {author}
  {\bibfnamefont {K.~R.}\ \bibnamefont {Corley}}, \bibinfo {author}
  {\bibfnamefont {S.~T.}\ \bibnamefont {Countryman}}, \bibinfo {author}
  {\bibfnamefont {P.~B.}\ \bibnamefont {Covas}}, \bibinfo {author}
  {\bibfnamefont {D.~C.}\ \bibnamefont {Coyne}}, \bibinfo {author}
  {\bibfnamefont {L.~E.~H.}\ \bibnamefont {Datrier}}, \bibinfo {author}
  {\bibfnamefont {D.}~\bibnamefont {Davis}}, \bibinfo {author} {\bibfnamefont
  {C.}~\bibnamefont {Di~Fronzo}}, \bibinfo {author} {\bibfnamefont {J.~C.}\
  \bibnamefont {Driggers}}, \bibinfo {author} {\bibfnamefont {T.}~\bibnamefont
  {Etzel}}, \bibinfo {author} {\bibfnamefont {T.~M.}\ \bibnamefont {Evans}},
  \bibinfo {author} {\bibfnamefont {J.}~\bibnamefont {Feicht}}, \bibinfo
  {author} {\bibfnamefont {P.}~\bibnamefont {Fulda}}, \bibinfo {author}
  {\bibfnamefont {M.}~\bibnamefont {Fyffe}}, \bibinfo {author} {\bibfnamefont
  {J.~A.}\ \bibnamefont {Giaime}}, \bibinfo {author} {\bibfnamefont {K.~D.}\
  \bibnamefont {Giardina}}, \bibinfo {author} {\bibfnamefont {P.}~\bibnamefont
  {Godwin}}, \bibinfo {author} {\bibfnamefont {E.}~\bibnamefont {Goetz}},
  \bibinfo {author} {\bibfnamefont {S.}~\bibnamefont {Gras}}, \bibinfo {author}
  {\bibfnamefont {C.}~\bibnamefont {Gray}}, \bibinfo {author} {\bibfnamefont
  {R.}~\bibnamefont {Gray}}, \bibinfo {author} {\bibfnamefont {A.}~\bibnamefont
  {Gupta}}, \bibinfo {author} {\bibfnamefont {E.~K.}\ \bibnamefont
  {Gustafson}}, \bibinfo {author} {\bibfnamefont {R.}~\bibnamefont
  {Gustafson}}, \bibinfo {author} {\bibfnamefont {J.}~\bibnamefont {Hanks}},
  \bibinfo {author} {\bibfnamefont {J.}~\bibnamefont {Hanson}}, \bibinfo
  {author} {\bibfnamefont {T.}~\bibnamefont {Hardwick}}, \bibinfo {author}
  {\bibfnamefont {R.~K.}\ \bibnamefont {Hasskew}}, \bibinfo {author}
  {\bibfnamefont {M.~C.}\ \bibnamefont {Heintze}}, \bibinfo {author}
  {\bibfnamefont {A.~F.}\ \bibnamefont {Helmling-Cornell}}, \bibinfo {author}
  {\bibfnamefont {N.~A.}\ \bibnamefont {Holland}}, \bibinfo {author}
  {\bibfnamefont {J.~D.}\ \bibnamefont {Jones}}, \bibinfo {author}
  {\bibfnamefont {S.}~\bibnamefont {Kandhasamy}}, \bibinfo {author}
  {\bibfnamefont {S.}~\bibnamefont {Karki}}, \bibinfo {author} {\bibfnamefont
  {M.}~\bibnamefont {Kasprzack}}, \bibinfo {author} {\bibfnamefont
  {K.}~\bibnamefont {Kawabe}}, \bibinfo {author} {\bibfnamefont {P.~J.}\
  \bibnamefont {King}}, \bibinfo {author} {\bibfnamefont {J.~S.}\ \bibnamefont
  {Kissel}}, \bibinfo {author} {\bibfnamefont {R.}~\bibnamefont {Kumar}},
  \bibinfo {author} {\bibfnamefont {M.}~\bibnamefont {Landry}}, \bibinfo
  {author} {\bibfnamefont {B.~B.}\ \bibnamefont {Lane}}, \bibinfo {author}
  {\bibfnamefont {B.}~\bibnamefont {Lantz}}, \bibinfo {author} {\bibfnamefont
  {M.}~\bibnamefont {Laxen}}, \bibinfo {author} {\bibfnamefont {Y.~K.}\
  \bibnamefont {Lecoeuche}}, \bibinfo {author} {\bibfnamefont {J.}~\bibnamefont
  {Leviton}}, \bibinfo {author} {\bibfnamefont {J.}~\bibnamefont {Liu}},
  \bibinfo {author} {\bibfnamefont {M.}~\bibnamefont {Lormand}}, \bibinfo
  {author} {\bibfnamefont {A.~P.}\ \bibnamefont {Lundgren}}, \bibinfo {author}
  {\bibfnamefont {R.}~\bibnamefont {Macas}}, \bibinfo {author} {\bibfnamefont
  {M.}~\bibnamefont {MacInnis}}, \bibinfo {author} {\bibfnamefont {D.~M.}\
  \bibnamefont {Macleod}}, \bibinfo {author} {\bibfnamefont {S.}~\bibnamefont
  {M\'arka}}, \bibinfo {author} {\bibfnamefont {Z.}~\bibnamefont {M\'arka}},
  \bibinfo {author} {\bibfnamefont {D.~V.}\ \bibnamefont {Martynov}}, \bibinfo
  {author} {\bibfnamefont {K.}~\bibnamefont {Mason}}, \bibinfo {author}
  {\bibfnamefont {T.~J.}\ \bibnamefont {Massinger}}, \bibinfo {author}
  {\bibfnamefont {R.}~\bibnamefont {McCarthy}}, \bibinfo {author}
  {\bibfnamefont {S.}~\bibnamefont {McCormick}}, \bibinfo {author}
  {\bibfnamefont {J.}~\bibnamefont {McIver}}, \bibinfo {author} {\bibfnamefont
  {G.}~\bibnamefont {Mendell}}, \bibinfo {author} {\bibfnamefont
  {K.}~\bibnamefont {Merfeld}}, \bibinfo {author} {\bibfnamefont {E.~L.}\
  \bibnamefont {Merilh}}, \bibinfo {author} {\bibfnamefont {F.}~\bibnamefont
  {Meylahn}}, \bibinfo {author} {\bibfnamefont {T.}~\bibnamefont {Mistry}},
  \bibinfo {author} {\bibfnamefont {R.}~\bibnamefont {Mittleman}}, \bibinfo
  {author} {\bibfnamefont {G.}~\bibnamefont {Moreno}}, \bibinfo {author}
  {\bibfnamefont {C.~M.}\ \bibnamefont {Mow-Lowry}}, \bibinfo {author}
  {\bibfnamefont {S.}~\bibnamefont {Mozzon}}, \bibinfo {author} {\bibfnamefont
  {T.~J.~N.}\ \bibnamefont {Nelson}}, \bibinfo {author} {\bibfnamefont
  {P.}~\bibnamefont {Nguyen}}, \bibinfo {author} {\bibfnamefont {L.~K.}\
  \bibnamefont {Nuttall}}, \bibinfo {author} {\bibfnamefont {J.}~\bibnamefont
  {Oberling}}, \bibinfo {author} {\bibfnamefont {R.~J.}\ \bibnamefont {Oram}},
  \bibinfo {author} {\bibfnamefont {B.}~\bibnamefont {O'Reilly}}, \bibinfo
  {author} {\bibfnamefont {C.}~\bibnamefont {Osthelder}}, \bibinfo {author}
  {\bibfnamefont {D.~J.}\ \bibnamefont {Ottaway}}, \bibinfo {author}
  {\bibfnamefont {H.}~\bibnamefont {Overmier}}, \bibinfo {author}
  {\bibfnamefont {J.~R.}\ \bibnamefont {Palamos}}, \bibinfo {author}
  {\bibfnamefont {W.}~\bibnamefont {Parker}}, \bibinfo {author} {\bibfnamefont
  {E.}~\bibnamefont {Payne}}, \bibinfo {author} {\bibfnamefont
  {A.}~\bibnamefont {Pele}}, \bibinfo {author} {\bibfnamefont {C.~J.}\
  \bibnamefont {Perez}}, \bibinfo {author} {\bibfnamefont {M.}~\bibnamefont
  {Pirello}}, \bibinfo {author} {\bibfnamefont {H.}~\bibnamefont {Radkins}},
  \bibinfo {author} {\bibfnamefont {K.~E.}\ \bibnamefont {Ramirez}}, \bibinfo
  {author} {\bibfnamefont {J.~W.}\ \bibnamefont {Richardson}}, \bibinfo
  {author} {\bibfnamefont {K.}~\bibnamefont {Riles}}, \bibinfo {author}
  {\bibfnamefont {N.~A.}\ \bibnamefont {Robertson}}, \bibinfo {author}
  {\bibfnamefont {J.~G.}\ \bibnamefont {Rollins}}, \bibinfo {author}
  {\bibfnamefont {C.~L.}\ \bibnamefont {Romel}}, \bibinfo {author}
  {\bibfnamefont {J.~H.}\ \bibnamefont {Romie}}, \bibinfo {author}
  {\bibfnamefont {M.~P.}\ \bibnamefont {Ross}}, \bibinfo {author}
  {\bibfnamefont {K.}~\bibnamefont {Ryan}}, \bibinfo {author} {\bibfnamefont
  {T.}~\bibnamefont {Sadecki}}, \bibinfo {author} {\bibfnamefont {E.~J.}\
  \bibnamefont {Sanchez}}, \bibinfo {author} {\bibfnamefont {L.~E.}\
  \bibnamefont {Sanchez}}, \bibinfo {author} {\bibfnamefont {T.~R.}\
  \bibnamefont {Saravanan}}, \bibinfo {author} {\bibfnamefont {R.~L.}\
  \bibnamefont {Savage}}, \bibinfo {author} {\bibfnamefont {D.}~\bibnamefont
  {Schaetzl}}, \bibinfo {author} {\bibfnamefont {R.}~\bibnamefont {Schnabel}},
  \bibinfo {author} {\bibfnamefont {R.~M.~S.}\ \bibnamefont {Schofield}},
  \bibinfo {author} {\bibfnamefont {E.}~\bibnamefont {Schwartz}}, \bibinfo
  {author} {\bibfnamefont {D.}~\bibnamefont {Sellers}}, \bibinfo {author}
  {\bibfnamefont {T.~J.}\ \bibnamefont {Shaffer}}, \bibinfo {author}
  {\bibfnamefont {J.~R.}\ \bibnamefont {Smith}}, \bibinfo {author}
  {\bibfnamefont {S.}~\bibnamefont {Soni}}, \bibinfo {author} {\bibfnamefont
  {B.}~\bibnamefont {Sorazu}}, \bibinfo {author} {\bibfnamefont {A.~P.}\
  \bibnamefont {Spencer}}, \bibinfo {author} {\bibfnamefont {K.~A.}\
  \bibnamefont {Strain}}, \bibinfo {author} {\bibfnamefont {L.}~\bibnamefont
  {Sun}}, \bibinfo {author} {\bibfnamefont {M.~J.}\ \bibnamefont
  {Szczepa\ifmmode~\acute{n}\else \'{n}\fi{}czyk}}, \bibinfo {author}
  {\bibfnamefont {M.}~\bibnamefont {Thomas}}, \bibinfo {author} {\bibfnamefont
  {P.}~\bibnamefont {Thomas}}, \bibinfo {author} {\bibfnamefont {K.~A.}\
  \bibnamefont {Thorne}}, \bibinfo {author} {\bibfnamefont {K.}~\bibnamefont
  {Toland}}, \bibinfo {author} {\bibfnamefont {C.~I.}\ \bibnamefont {Torrie}},
  \bibinfo {author} {\bibfnamefont {G.}~\bibnamefont {Traylor}}, \bibinfo
  {author} {\bibfnamefont {A.~L.}\ \bibnamefont {Urban}}, \bibinfo {author}
  {\bibfnamefont {G.}~\bibnamefont {Vajente}}, \bibinfo {author} {\bibfnamefont
  {G.}~\bibnamefont {Valdes}}, \bibinfo {author} {\bibfnamefont {D.~C.}\
  \bibnamefont {Vander-Hyde}}, \bibinfo {author} {\bibfnamefont {P.~J.}\
  \bibnamefont {Veitch}}, \bibinfo {author} {\bibfnamefont {K.}~\bibnamefont
  {Venkateswara}}, \bibinfo {author} {\bibfnamefont {G.}~\bibnamefont
  {Venugopalan}}, \bibinfo {author} {\bibfnamefont {A.~D.}\ \bibnamefont
  {Viets}}, \bibinfo {author} {\bibfnamefont {C.}~\bibnamefont {Vorvick}},
  \bibinfo {author} {\bibfnamefont {M.}~\bibnamefont {Wade}}, \bibinfo {author}
  {\bibfnamefont {J.}~\bibnamefont {Warner}}, \bibinfo {author} {\bibfnamefont
  {B.}~\bibnamefont {Weaver}}, \bibinfo {author} {\bibfnamefont
  {R.}~\bibnamefont {Weiss}}, \bibinfo {author} {\bibfnamefont
  {B.}~\bibnamefont {Willke}}, \bibinfo {author} {\bibfnamefont {C.~C.}\
  \bibnamefont {Wipf}}, \bibinfo {author} {\bibfnamefont {L.}~\bibnamefont
  {Xiao}}, \bibinfo {author} {\bibfnamefont {H.}~\bibnamefont {Yamamoto}},
  \bibinfo {author} {\bibfnamefont {M.~J.}\ \bibnamefont {Yap}}, \bibinfo
  {author} {\bibfnamefont {H.}~\bibnamefont {Yu}}, \bibinfo {author}
  {\bibfnamefont {L.}~\bibnamefont {Zhang}}, \bibinfo {author} {\bibfnamefont
  {M.~E.}\ \bibnamefont {Zucker}}, \ and\ \bibinfo {author} {\bibfnamefont
  {J.}~\bibnamefont {Zweizig}},\ }\href {\doibase
  10.1103/PhysRevLett.123.231107} {\bibfield  {journal} {\bibinfo  {journal}
  {Phys. Rev. Lett.}\ }\textbf {\bibinfo {volume} {123}},\ \bibinfo {pages}
  {231107} (\bibinfo {year} {2019})}\BibitemShut {NoStop}%
\bibitem [{\citenamefont {{Mours}}\ \emph {et~al.}(2006)\citenamefont
  {{Mours}}, \citenamefont {{Tournefier}},\ and\ \citenamefont
  {{Vinet}}}]{Mours06}%
  \BibitemOpen
  \bibfield  {author} {\bibinfo {author} {\bibfnamefont {B.}~\bibnamefont
  {{Mours}}}, \bibinfo {author} {\bibfnamefont {E.}~\bibnamefont
  {{Tournefier}}}, \ and\ \bibinfo {author} {\bibfnamefont {J.-Y.}\
  \bibnamefont {{Vinet}}},\ }\href {\doibase 10.1088/0264-9381/23/20/001}
  {\bibfield  {journal} {\bibinfo  {journal} {Classical and Quantum Gravity}\
  }\textbf {\bibinfo {volume} {23}},\ \bibinfo {pages} {5777} (\bibinfo {year}
  {2006})}\BibitemShut {NoStop}%
\bibitem [{\citenamefont {Chelkowski}\ \emph {et~al.}(2009)\citenamefont
  {Chelkowski}, \citenamefont {Hild},\ and\ \citenamefont
  {Freise}}]{Chelkowski09}%
  \BibitemOpen
  \bibfield  {author} {\bibinfo {author} {\bibfnamefont {S.}~\bibnamefont
  {Chelkowski}}, \bibinfo {author} {\bibfnamefont {S.}~\bibnamefont {Hild}}, \
  and\ \bibinfo {author} {\bibfnamefont {A.}~\bibnamefont {Freise}},\ }\href
  {\doibase 10.1103/PhysRevD.79.122002} {\bibfield  {journal} {\bibinfo
  {journal} {Physical Review D (Particles, Fields, Gravitation, and
  Cosmology)}\ }\textbf {\bibinfo {volume} {79}},\ \bibinfo {eid} {122002}
  (\bibinfo {year} {2009})}\BibitemShut {NoStop}%
\bibitem [{\citenamefont {{Ast}}\ \emph {et~al.}(2019)\citenamefont {{Ast}},
  \citenamefont {{Di Pace}}, \citenamefont {{Millo}}, \citenamefont {{Pichot}},
  \citenamefont {{Turconi}},\ and\ \citenamefont {{Chaibi}}}]{Ast2019}%
  \BibitemOpen
  \bibfield  {author} {\bibinfo {author} {\bibfnamefont {S.}~\bibnamefont
  {{Ast}}}, \bibinfo {author} {\bibfnamefont {S.}~\bibnamefont {{Di Pace}}},
  \bibinfo {author} {\bibfnamefont {J.}~\bibnamefont {{Millo}}}, \bibinfo
  {author} {\bibfnamefont {M.}~\bibnamefont {{Pichot}}}, \bibinfo {author}
  {\bibfnamefont {M.}~\bibnamefont {{Turconi}}}, \ and\ \bibinfo {author}
  {\bibfnamefont {W.}~\bibnamefont {{Chaibi}}},\ }\href@noop {} {\bibfield
  {journal} {\bibinfo  {journal} {arXiv e-prints}\ ,\ \bibinfo {eid}
  {arXiv:1902.01671}} (\bibinfo {year} {2019})},\ \Eprint
  {http://arxiv.org/abs/1902.01671} {arXiv:1902.01671 [gr-qc]} \BibitemShut
  {NoStop}%
\bibitem [{\citenamefont {{Biscans}}\ \emph {et~al.}(2019)\citenamefont
  {{Biscans}}, \citenamefont {{Gras}}, \citenamefont {{Blair}}, \citenamefont
  {{Driggers}}, \citenamefont {{Evans}}, \citenamefont {{Fritschel}},
  \citenamefont {{Hardwick}},\ and\ \citenamefont {{Mansell}}}]{Biscans19}%
  \BibitemOpen
  \bibfield  {author} {\bibinfo {author} {\bibfnamefont {S.}~\bibnamefont
  {{Biscans}}}, \bibinfo {author} {\bibfnamefont {S.}~\bibnamefont {{Gras}}},
  \bibinfo {author} {\bibfnamefont {C.~D.}\ \bibnamefont {{Blair}}}, \bibinfo
  {author} {\bibfnamefont {J.}~\bibnamefont {{Driggers}}}, \bibinfo {author}
  {\bibfnamefont {M.}~\bibnamefont {{Evans}}}, \bibinfo {author} {\bibfnamefont
  {P.}~\bibnamefont {{Fritschel}}}, \bibinfo {author} {\bibfnamefont
  {T.}~\bibnamefont {{Hardwick}}}, \ and\ \bibinfo {author} {\bibfnamefont
  {G.}~\bibnamefont {{Mansell}}},\ }\href@noop {} {\bibfield  {journal}
  {\bibinfo  {journal} {arXiv e-prints}\ ,\ \bibinfo {eid} {arXiv:1909.07805}}
  (\bibinfo {year} {2019})},\ \Eprint {http://arxiv.org/abs/1909.07805}
  {arXiv:1909.07805 [physics.app-ph]} \BibitemShut {NoStop}%
\bibitem [{\citenamefont {Sorazu}\ \emph {et~al.}(2013)\citenamefont {Sorazu},
  \citenamefont {Fulda}, \citenamefont {Barr}, \citenamefont {Bell},
  \citenamefont {Bond}, \citenamefont {Carbone}, \citenamefont {Freise},
  \citenamefont {Hild}, \citenamefont {Huttner}, \citenamefont {Macarthur},\
  and\ \citenamefont {Strain}}]{Sorazu13}%
  \BibitemOpen
  \bibfield  {author} {\bibinfo {author} {\bibfnamefont {B.}~\bibnamefont
  {Sorazu}}, \bibinfo {author} {\bibfnamefont {P.~J.}\ \bibnamefont {Fulda}},
  \bibinfo {author} {\bibfnamefont {B.~W.}\ \bibnamefont {Barr}}, \bibinfo
  {author} {\bibfnamefont {A.~S.}\ \bibnamefont {Bell}}, \bibinfo {author}
  {\bibfnamefont {C.}~\bibnamefont {Bond}}, \bibinfo {author} {\bibfnamefont
  {L.}~\bibnamefont {Carbone}}, \bibinfo {author} {\bibfnamefont
  {A.}~\bibnamefont {Freise}}, \bibinfo {author} {\bibfnamefont
  {S.}~\bibnamefont {Hild}}, \bibinfo {author} {\bibfnamefont {S.~H.}\
  \bibnamefont {Huttner}}, \bibinfo {author} {\bibfnamefont {J.}~\bibnamefont
  {Macarthur}}, \ and\ \bibinfo {author} {\bibfnamefont {K.~A.}\ \bibnamefont
  {Strain}},\ }\href {http://stacks.iop.org/0264-9381/30/i=3/a=035004}
  {\bibfield  {journal} {\bibinfo  {journal} {Classical and Quantum Gravity}\
  }\textbf {\bibinfo {volume} {30}},\ \bibinfo {pages} {035004} (\bibinfo
  {year} {2013})}\BibitemShut {NoStop}%
\bibitem [{\citenamefont {Allocca}\ \emph {et~al.}(2015)\citenamefont
  {Allocca}, \citenamefont {Gatto}, \citenamefont {Tacca}, \citenamefont {Day},
  \citenamefont {Barsuglia}, \citenamefont {Pillant}, \citenamefont {Buy},\
  and\ \citenamefont {Vajente}}]{Allocca15}%
  \BibitemOpen
  \bibfield  {author} {\bibinfo {author} {\bibfnamefont {A.}~\bibnamefont
  {Allocca}}, \bibinfo {author} {\bibfnamefont {A.}~\bibnamefont {Gatto}},
  \bibinfo {author} {\bibfnamefont {M.}~\bibnamefont {Tacca}}, \bibinfo
  {author} {\bibfnamefont {R.~A.}\ \bibnamefont {Day}}, \bibinfo {author}
  {\bibfnamefont {M.}~\bibnamefont {Barsuglia}}, \bibinfo {author}
  {\bibfnamefont {G.}~\bibnamefont {Pillant}}, \bibinfo {author} {\bibfnamefont
  {C.}~\bibnamefont {Buy}}, \ and\ \bibinfo {author} {\bibfnamefont
  {G.}~\bibnamefont {Vajente}},\ }\href {\doibase 10.1103/PhysRevD.92.102002}
  {\bibfield  {journal} {\bibinfo  {journal} {Phys. Rev. D}\ }\textbf {\bibinfo
  {volume} {92}},\ \bibinfo {pages} {102002} (\bibinfo {year}
  {2015})}\BibitemShut {NoStop}%
\bibitem [{\citenamefont {Anderson}(1984)}]{anderson84}%
  \BibitemOpen
  \bibfield  {author} {\bibinfo {author} {\bibfnamefont {D.~Z.}\ \bibnamefont
  {Anderson}},\ }\href {\doibase 10.1364/AO.23.002944} {\bibfield  {journal}
  {\bibinfo  {journal} {Appl. Opt.}\ }\textbf {\bibinfo {volume} {23}},\
  \bibinfo {pages} {2944} (\bibinfo {year} {1984})}\BibitemShut {NoStop}%
\bibitem [{\citenamefont {Babusci}\ \emph {et~al.}(1997)\citenamefont
  {Babusci}, \citenamefont {Fang}, \citenamefont {Giordano}, \citenamefont
  {Matone}, \citenamefont {Matone},\ and\ \citenamefont
  {Sannibale}}]{Babusci97}%
  \BibitemOpen
  \bibfield  {author} {\bibinfo {author} {\bibfnamefont {D.}~\bibnamefont
  {Babusci}}, \bibinfo {author} {\bibfnamefont {H.}~\bibnamefont {Fang}},
  \bibinfo {author} {\bibfnamefont {G.}~\bibnamefont {Giordano}}, \bibinfo
  {author} {\bibfnamefont {G.}~\bibnamefont {Matone}}, \bibinfo {author}
  {\bibfnamefont {L.}~\bibnamefont {Matone}}, \ and\ \bibinfo {author}
  {\bibfnamefont {V.}~\bibnamefont {Sannibale}},\ }\href {\doibase
  https://doi.org/10.1016/S0375-9601(96)00907-3} {\bibfield  {journal}
  {\bibinfo  {journal} {Physics Letters A}\ }\textbf {\bibinfo {volume}
  {226}},\ \bibinfo {pages} {31 } (\bibinfo {year} {1997})}\BibitemShut
  {NoStop}%
\bibitem [{\citenamefont {Slagmolen}\ \emph {et~al.}(2005)\citenamefont
  {Slagmolen}, \citenamefont {Barton}, \citenamefont {Mow-Lowry}, \citenamefont
  {Vine}, \citenamefont {Rabeling}, \citenamefont {Chow}, \citenamefont
  {Romann}, \citenamefont {Zhao}, \citenamefont {Gray},\ and\ \citenamefont
  {McClelland}}]{Slagmolen2005}%
  \BibitemOpen
  \bibfield  {author} {\bibinfo {author} {\bibfnamefont {B.~J.~J.}\
  \bibnamefont {Slagmolen}}, \bibinfo {author} {\bibfnamefont {M.}~\bibnamefont
  {Barton}}, \bibinfo {author} {\bibfnamefont {C.}~\bibnamefont {Mow-Lowry}},
  \bibinfo {author} {\bibfnamefont {G.~d.}\ \bibnamefont {Vine}}, \bibinfo
  {author} {\bibfnamefont {D.~S.}\ \bibnamefont {Rabeling}}, \bibinfo {author}
  {\bibfnamefont {J.~H.}\ \bibnamefont {Chow}}, \bibinfo {author}
  {\bibfnamefont {A.}~\bibnamefont {Romann}}, \bibinfo {author} {\bibfnamefont
  {C.}~\bibnamefont {Zhao}}, \bibinfo {author} {\bibfnamefont {M.~B.}\
  \bibnamefont {Gray}}, \ and\ \bibinfo {author} {\bibfnamefont {D.~E.}\
  \bibnamefont {McClelland}},\ }\href {\doibase 10.1007/s10714-005-0144-6}
  {\bibfield  {journal} {\bibinfo  {journal} {General Relativity and
  Gravitation}\ }\textbf {\bibinfo {volume} {37}},\ \bibinfo {pages} {1601}
  (\bibinfo {year} {2005})}\BibitemShut {NoStop}%
\bibitem [{\citenamefont {Brooks}\ \emph {et~al.}(2015)\citenamefont {Brooks},
  \citenamefont {Adhikari}, \citenamefont {Ballmer}, \citenamefont {Barsotti},
  \citenamefont {Fulda}, \citenamefont {Perreca},\ and\ \citenamefont
  {Ottaway}}]{Brooks15}%
  \BibitemOpen
  \bibfield  {author} {\bibinfo {author} {\bibfnamefont {A.~F.}\ \bibnamefont
  {Brooks}}, \bibinfo {author} {\bibfnamefont {R.~X.}\ \bibnamefont
  {Adhikari}}, \bibinfo {author} {\bibfnamefont {S.}~\bibnamefont {Ballmer}},
  \bibinfo {author} {\bibfnamefont {L.}~\bibnamefont {Barsotti}}, \bibinfo
  {author} {\bibfnamefont {P.}~\bibnamefont {Fulda}}, \bibinfo {author}
  {\bibfnamefont {A.}~\bibnamefont {Perreca}}, \ and\ \bibinfo {author}
  {\bibfnamefont {D.}~\bibnamefont {Ottaway}},\ }\href
  {https://dcc.ligo.org/LIGO-T1500188/public} {\emph {\bibinfo {title} {Active
  wavefront control in and beyond Advanced LIGO (LIGO-T1500188)}}},\ \bibinfo
  {type} {techreport}\ (\bibinfo  {institution} {LIGO Scientific
  Collaboration},\ \bibinfo {year} {2015})\BibitemShut {NoStop}%
\bibitem [{\citenamefont {Mueller}\ \emph {et~al.}(2000)\citenamefont
  {Mueller}, \citenamefont {ze~Shu}, \citenamefont {Adhikari}, \citenamefont
  {Tanner}, \citenamefont {Reitze}, \citenamefont {Sigg}, \citenamefont
  {Mavalvala},\ and\ \citenamefont {Camp}}]{mueller2000}%
  \BibitemOpen
  \bibfield  {author} {\bibinfo {author} {\bibfnamefont {G.}~\bibnamefont
  {Mueller}}, \bibinfo {author} {\bibfnamefont {Q.}~\bibnamefont {ze~Shu}},
  \bibinfo {author} {\bibfnamefont {R.}~\bibnamefont {Adhikari}}, \bibinfo
  {author} {\bibfnamefont {D.~B.}\ \bibnamefont {Tanner}}, \bibinfo {author}
  {\bibfnamefont {D.}~\bibnamefont {Reitze}}, \bibinfo {author} {\bibfnamefont
  {D.}~\bibnamefont {Sigg}}, \bibinfo {author} {\bibfnamefont {N.}~\bibnamefont
  {Mavalvala}}, \ and\ \bibinfo {author} {\bibfnamefont {J.}~\bibnamefont
  {Camp}},\ }\href {\doibase 10.1364/OL.25.000266} {\bibfield  {journal}
  {\bibinfo  {journal} {Opt. Lett.}\ }\textbf {\bibinfo {volume} {25}},\
  \bibinfo {pages} {266} (\bibinfo {year} {2000})}\BibitemShut {NoStop}%
\bibitem [{\citenamefont {{Maga{\~n}a-Sandoval}}\ \emph
  {et~al.}(2019)\citenamefont {{Maga{\~n}a-Sandoval}}, \citenamefont {{Vo}},
  \citenamefont {{Vand er-Hyde}}, \citenamefont {{Sanders}},\ and\
  \citenamefont {{Ballmer}}}]{fabian_qpd}%
  \BibitemOpen
  \bibfield  {author} {\bibinfo {author} {\bibfnamefont {F.}~\bibnamefont
  {{Maga{\~n}a-Sandoval}}}, \bibinfo {author} {\bibfnamefont {T.}~\bibnamefont
  {{Vo}}}, \bibinfo {author} {\bibfnamefont {D.}~\bibnamefont {{Vand
  er-Hyde}}}, \bibinfo {author} {\bibfnamefont {J.~R.}\ \bibnamefont
  {{Sanders}}}, \ and\ \bibinfo {author} {\bibfnamefont {S.~W.}\ \bibnamefont
  {{Ballmer}}},\ }\href@noop {} {\bibfield  {journal} {\bibinfo  {journal}
  {arXiv e-prints}\ ,\ \bibinfo {eid} {arXiv:1909.08084}} (\bibinfo {year}
  {2019})},\ \Eprint {http://arxiv.org/abs/1909.08084} {arXiv:1909.08084
  [physics.optics]} \BibitemShut {NoStop}%
\bibitem [{\citenamefont {Brooks}\ \emph {et~al.}(2007)\citenamefont {Brooks},
  \citenamefont {Kelly}, \citenamefont {Veitch},\ and\ \citenamefont
  {Munch}}]{Brooks07}%
  \BibitemOpen
  \bibfield  {author} {\bibinfo {author} {\bibfnamefont {A.~F.}\ \bibnamefont
  {Brooks}}, \bibinfo {author} {\bibfnamefont {T.-L.}\ \bibnamefont {Kelly}},
  \bibinfo {author} {\bibfnamefont {P.~J.}\ \bibnamefont {Veitch}}, \ and\
  \bibinfo {author} {\bibfnamefont {J.}~\bibnamefont {Munch}},\ }\href
  {\doibase 10.1364/OE.15.010370} {\bibfield  {journal} {\bibinfo  {journal}
  {Opt. Express}\ }\textbf {\bibinfo {volume} {15}},\ \bibinfo {pages} {10370}
  (\bibinfo {year} {2007})}\BibitemShut {NoStop}%
\bibitem [{\citenamefont {Miller}\ and\ \citenamefont
  {Evans}(2014)}]{miller14}%
  \BibitemOpen
  \bibfield  {author} {\bibinfo {author} {\bibfnamefont {J.}~\bibnamefont
  {Miller}}\ and\ \bibinfo {author} {\bibfnamefont {M.}~\bibnamefont {Evans}},\
  }\href {\doibase 10.1364/OL.39.002495} {\bibfield  {journal} {\bibinfo
  {journal} {Optics Letters}\ }\textbf {\bibinfo {volume} {39}},\ \bibinfo
  {pages} {2495} (\bibinfo {year} {2014})}\BibitemShut {NoStop}%
\bibitem [{\citenamefont {Agatsuma}\ \emph {et~al.}(2019)\citenamefont
  {Agatsuma}, \citenamefont {van~der Schaaf}, \citenamefont {van Beuzekom},
  \citenamefont {Rabeling},\ and\ \citenamefont {van~den Brand}}]{Agatsuma19}%
  \BibitemOpen
  \bibfield  {author} {\bibinfo {author} {\bibfnamefont {K.}~\bibnamefont
  {Agatsuma}}, \bibinfo {author} {\bibfnamefont {L.}~\bibnamefont {van~der
  Schaaf}}, \bibinfo {author} {\bibfnamefont {M.}~\bibnamefont {van Beuzekom}},
  \bibinfo {author} {\bibfnamefont {D.}~\bibnamefont {Rabeling}}, \ and\
  \bibinfo {author} {\bibfnamefont {J.}~\bibnamefont {van~den Brand}},\ }\href
  {\doibase 10.1364/OE.27.018533} {\bibfield  {journal} {\bibinfo  {journal}
  {Opt. Express}\ }\textbf {\bibinfo {volume} {27}},\ \bibinfo {pages} {18533}
  (\bibinfo {year} {2019})}\BibitemShut {NoStop}%
\bibitem [{\citenamefont {{Tuong Cao}}\ \emph {et~al.}(2019)\citenamefont
  {{Tuong Cao}}, \citenamefont {{Brown}}, \citenamefont {{Veitch}},\ and\
  \citenamefont {{Ottaway}}}]{cao2019}%
  \BibitemOpen
  \bibfield  {author} {\bibinfo {author} {\bibfnamefont {H.}~\bibnamefont
  {{Tuong Cao}}}, \bibinfo {author} {\bibfnamefont {D.~D.}\ \bibnamefont
  {{Brown}}}, \bibinfo {author} {\bibfnamefont {P.}~\bibnamefont {{Veitch}}}, \
  and\ \bibinfo {author} {\bibfnamefont {D.~J.}\ \bibnamefont {{Ottaway}}},\
  }\href@noop {} {\bibfield  {journal} {\bibinfo  {journal} {arXiv e-prints}\
  ,\ \bibinfo {eid} {arXiv:1907.05224}} (\bibinfo {year} {2019})},\ \Eprint
  {http://arxiv.org/abs/1907.05224} {arXiv:1907.05224 [astro-ph.IM]}
  \BibitemShut {NoStop}%
\bibitem [{\citenamefont {Ralph}\ \emph {et~al.}(2017)\citenamefont {Ralph},
  \citenamefont {Altin}, \citenamefont {Rabeling}, \citenamefont {McClelland},\
  and\ \citenamefont {Shaddock}}]{Ralph17}%
  \BibitemOpen
  \bibfield  {author} {\bibinfo {author} {\bibfnamefont {D.~T.}\ \bibnamefont
  {Ralph}}, \bibinfo {author} {\bibfnamefont {P.~A.}\ \bibnamefont {Altin}},
  \bibinfo {author} {\bibfnamefont {D.~S.}\ \bibnamefont {Rabeling}}, \bibinfo
  {author} {\bibfnamefont {D.~E.}\ \bibnamefont {McClelland}}, \ and\ \bibinfo
  {author} {\bibfnamefont {D.~A.}\ \bibnamefont {Shaddock}},\ }\href {\doibase
  10.1364/AO.56.002353} {\bibfield  {journal} {\bibinfo  {journal} {Appl.
  Opt.}\ }\textbf {\bibinfo {volume} {56}},\ \bibinfo {pages} {2353} (\bibinfo
  {year} {2017})}\BibitemShut {NoStop}%
\bibitem [{\citenamefont {Kwee}\ \emph {et~al.}(2007)\citenamefont {Kwee},
  \citenamefont {Seifert}, \citenamefont {Willke},\ and\ \citenamefont
  {Danzmann}}]{KSWD07}%
  \BibitemOpen
  \bibfield  {author} {\bibinfo {author} {\bibfnamefont {P.}~\bibnamefont
  {Kwee}}, \bibinfo {author} {\bibfnamefont {F.}~\bibnamefont {Seifert}},
  \bibinfo {author} {\bibfnamefont {B.}~\bibnamefont {Willke}}, \ and\ \bibinfo
  {author} {\bibfnamefont {K.}~\bibnamefont {Danzmann}},\ }\href@noop {}
  {\bibfield  {journal} {\bibinfo  {journal} {Review of Scientific
  Instruments}\ }\textbf {\bibinfo {volume} {78}},\ \bibinfo {pages} {073103}
  (\bibinfo {year} {2007})}\BibitemShut {NoStop}%
\bibitem [{\citenamefont {Takeno}\ \emph {et~al.}(2011)\citenamefont {Takeno},
  \citenamefont {Ohmae}, \citenamefont {Mio},\ and\ \citenamefont
  {Shirai}}]{Takeno11}%
  \BibitemOpen
  \bibfield  {author} {\bibinfo {author} {\bibfnamefont {K.}~\bibnamefont
  {Takeno}}, \bibinfo {author} {\bibfnamefont {N.}~\bibnamefont {Ohmae}},
  \bibinfo {author} {\bibfnamefont {N.}~\bibnamefont {Mio}}, \ and\ \bibinfo
  {author} {\bibfnamefont {T.}~\bibnamefont {Shirai}},\ }\href {\doibase
  https://doi.org/10.1016/j.optcom.2011.03.012} {\bibfield  {journal} {\bibinfo
   {journal} {Optics Communications}\ }\textbf {\bibinfo {volume} {284}},\
  \bibinfo {pages} {3197 } (\bibinfo {year} {2011})}\BibitemShut {NoStop}%
\bibitem [{\citenamefont {Bogan}\ \emph {et~al.}(2015)\citenamefont {Bogan},
  \citenamefont {Kwee}, \citenamefont {Hild}, \citenamefont {Huttner},\ and\
  \citenamefont {Willke}}]{Bogan15}%
  \BibitemOpen
  \bibfield  {author} {\bibinfo {author} {\bibfnamefont {C.}~\bibnamefont
  {Bogan}}, \bibinfo {author} {\bibfnamefont {P.}~\bibnamefont {Kwee}},
  \bibinfo {author} {\bibfnamefont {S.}~\bibnamefont {Hild}}, \bibinfo {author}
  {\bibfnamefont {S.~H.}\ \bibnamefont {Huttner}}, \ and\ \bibinfo {author}
  {\bibfnamefont {B.}~\bibnamefont {Willke}},\ }\href {\doibase
  10.1364/OE.23.015380} {\bibfield  {journal} {\bibinfo  {journal} {Opt.
  Express}\ }\textbf {\bibinfo {volume} {23}},\ \bibinfo {pages} {15380}
  (\bibinfo {year} {2015})}\BibitemShut {NoStop}%
\bibitem [{\citenamefont {Golub}\ \emph {et~al.}(1982)\citenamefont {Golub},
  \citenamefont {Prokhorov}, \citenamefont {Sisakyan},\ and\ \citenamefont
  {So{\u{\i}}fer}}]{Golub82}%
  \BibitemOpen
  \bibfield  {author} {\bibinfo {author} {\bibfnamefont {M.~A.}\ \bibnamefont
  {Golub}}, \bibinfo {author} {\bibfnamefont {A.~M.}\ \bibnamefont
  {Prokhorov}}, \bibinfo {author} {\bibfnamefont {I.~N.}\ \bibnamefont
  {Sisakyan}}, \ and\ \bibinfo {author} {\bibfnamefont {V.~A.}\ \bibnamefont
  {So{\u{\i}}fer}},\ }\href {\doibase 10.1070/qe1982v012n09abeh005998}
  {\bibfield  {journal} {\bibinfo  {journal} {Soviet Journal of Quantum
  Electronics}\ }\textbf {\bibinfo {volume} {12}},\ \bibinfo {pages} {1208}
  (\bibinfo {year} {1982})}\BibitemShut {NoStop}%
\bibitem [{\citenamefont {Kaiser}\ \emph {et~al.}(2009)\citenamefont {Kaiser},
  \citenamefont {Flamm}, \citenamefont {Schr\"{o}ter},\ and\ \citenamefont
  {Duparr\'{e}}}]{Kaiser09}%
  \BibitemOpen
  \bibfield  {author} {\bibinfo {author} {\bibfnamefont {T.}~\bibnamefont
  {Kaiser}}, \bibinfo {author} {\bibfnamefont {D.}~\bibnamefont {Flamm}},
  \bibinfo {author} {\bibfnamefont {S.}~\bibnamefont {Schr\"{o}ter}}, \ and\
  \bibinfo {author} {\bibfnamefont {M.}~\bibnamefont {Duparr\'{e}}},\ }\href
  {\doibase 10.1364/OE.17.009347} {\bibfield  {journal} {\bibinfo  {journal}
  {Optics Express}\ }\textbf {\bibinfo {volume} {17}},\ \bibinfo {pages} {9347}
  (\bibinfo {year} {2009})}\BibitemShut {NoStop}%
\bibitem [{\citenamefont {Flamm}\ \emph {et~al.}(2012)\citenamefont {Flamm},
  \citenamefont {Naidoo}, \citenamefont {Schulze}, \citenamefont {Forbes},\
  and\ \citenamefont {Duparr\'{e}}}]{Flamm12}%
  \BibitemOpen
  \bibfield  {author} {\bibinfo {author} {\bibfnamefont {D.}~\bibnamefont
  {Flamm}}, \bibinfo {author} {\bibfnamefont {D.}~\bibnamefont {Naidoo}},
  \bibinfo {author} {\bibfnamefont {C.}~\bibnamefont {Schulze}}, \bibinfo
  {author} {\bibfnamefont {A.}~\bibnamefont {Forbes}}, \ and\ \bibinfo {author}
  {\bibfnamefont {M.}~\bibnamefont {Duparr\'{e}}},\ }\href {\doibase
  10.1364/OL.37.002478} {\bibfield  {journal} {\bibinfo  {journal} {Optics
  Letters}\ }\textbf {\bibinfo {volume} {37}},\ \bibinfo {pages} {2478}
  (\bibinfo {year} {2012})}\BibitemShut {NoStop}%
\bibitem [{\citenamefont {Dudley}\ \emph {et~al.}(2014)\citenamefont {Dudley},
  \citenamefont {Milione}, \citenamefont {Alfano},\ and\ \citenamefont
  {Forbes}}]{dudley14}%
  \BibitemOpen
  \bibfield  {author} {\bibinfo {author} {\bibfnamefont {A.}~\bibnamefont
  {Dudley}}, \bibinfo {author} {\bibfnamefont {G.}~\bibnamefont {Milione}},
  \bibinfo {author} {\bibfnamefont {R.~R.}\ \bibnamefont {Alfano}}, \ and\
  \bibinfo {author} {\bibfnamefont {A.}~\bibnamefont {Forbes}},\ }\href
  {\doibase 10.1364/OE.22.014031} {\bibfield  {journal} {\bibinfo  {journal}
  {Opt. Express}\ }\textbf {\bibinfo {volume} {22}},\ \bibinfo {pages} {14031}
  (\bibinfo {year} {2014})}\BibitemShut {NoStop}%
\bibitem [{\citenamefont {Wang}\ \emph {et~al.}(2018)\citenamefont {Wang},
  \citenamefont {Zhu}, \citenamefont {Wang}, \citenamefont {Ai}, \citenamefont
  {Chen},\ and\ \citenamefont {Wang}}]{Wang18}%
  \BibitemOpen
  \bibfield  {author} {\bibinfo {author} {\bibfnamefont {A.}~\bibnamefont
  {Wang}}, \bibinfo {author} {\bibfnamefont {L.}~\bibnamefont {Zhu}}, \bibinfo
  {author} {\bibfnamefont {L.}~\bibnamefont {Wang}}, \bibinfo {author}
  {\bibfnamefont {J.}~\bibnamefont {Ai}}, \bibinfo {author} {\bibfnamefont
  {S.}~\bibnamefont {Chen}}, \ and\ \bibinfo {author} {\bibfnamefont
  {J.}~\bibnamefont {Wang}},\ }\href {\doibase 10.1364/OE.26.010038} {\bibfield
   {journal} {\bibinfo  {journal} {Opt. Express}\ }\textbf {\bibinfo {volume}
  {26}},\ \bibinfo {pages} {10038} (\bibinfo {year} {2018})}\BibitemShut
  {NoStop}%
\bibitem [{\citenamefont {Richardson}\ \emph {et~al.}(2013)\citenamefont
  {Richardson}, \citenamefont {Fini},\ and\ \citenamefont
  {Nelson}}]{Richardson13}%
  \BibitemOpen
  \bibfield  {author} {\bibinfo {author} {\bibfnamefont {D.~J.}\ \bibnamefont
  {Richardson}}, \bibinfo {author} {\bibfnamefont {J.~M.}\ \bibnamefont
  {Fini}}, \ and\ \bibinfo {author} {\bibfnamefont {L.~E.}\ \bibnamefont
  {Nelson}},\ }\href {\doibase 10.1038/nphoton.2013.94} {\bibfield  {journal}
  {\bibinfo  {journal} {Nature Photon}\ }\textbf {\bibinfo {volume} {7}}
  (\bibinfo {year} {2013}),\ 10.1038/nphoton.2013.94}\BibitemShut {NoStop}%
\bibitem [{\citenamefont {Forbes}\ \emph {et~al.}(2016)\citenamefont {Forbes},
  \citenamefont {Dudley},\ and\ \citenamefont {McLaren}}]{Forbes16}%
  \BibitemOpen
  \bibfield  {author} {\bibinfo {author} {\bibfnamefont {A.}~\bibnamefont
  {Forbes}}, \bibinfo {author} {\bibfnamefont {A.}~\bibnamefont {Dudley}}, \
  and\ \bibinfo {author} {\bibfnamefont {M.}~\bibnamefont {McLaren}},\ }\href
  {\doibase 10.1364/AOP.8.000200} {\bibfield  {journal} {\bibinfo  {journal}
  {Adv. Opt. Photon.}\ }\textbf {\bibinfo {volume} {8}},\ \bibinfo {pages}
  {200} (\bibinfo {year} {2016})}\BibitemShut {NoStop}%
\bibitem [{\citenamefont {Bond}\ \emph {et~al.}(2017)\citenamefont {Bond},
  \citenamefont {Brown}, \citenamefont {Freise},\ and\ \citenamefont
  {Strain}}]{Bond2017}%
  \BibitemOpen
  \bibfield  {author} {\bibinfo {author} {\bibfnamefont {C.}~\bibnamefont
  {Bond}}, \bibinfo {author} {\bibfnamefont {D.}~\bibnamefont {Brown}},
  \bibinfo {author} {\bibfnamefont {A.}~\bibnamefont {Freise}}, \ and\ \bibinfo
  {author} {\bibfnamefont {K.~A.}\ \bibnamefont {Strain}},\ }\href {\doibase
  10.1007/s41114-016-0002-8} {\bibfield  {journal} {\bibinfo  {journal} {Living
  Reviews in Relativity}\ }\textbf {\bibinfo {volume} {19}},\ \bibinfo {pages}
  {3} (\bibinfo {year} {2017})}\BibitemShut {NoStop}%
\bibitem [{\citenamefont {{Uehara}}\ \emph {et~al.}(1997)\citenamefont
  {{Uehara}}, \citenamefont {{Gustafson}}, \citenamefont {{Fejer}},\ and\
  \citenamefont {{Byer}}}]{UeharaSPIE}%
  \BibitemOpen
  \bibfield  {author} {\bibinfo {author} {\bibfnamefont {N.}~\bibnamefont
  {{Uehara}}}, \bibinfo {author} {\bibfnamefont {E.~K.}\ \bibnamefont
  {{Gustafson}}}, \bibinfo {author} {\bibfnamefont {M.~M.}\ \bibnamefont
  {{Fejer}}}, \ and\ \bibinfo {author} {\bibfnamefont {R.~L.}\ \bibnamefont
  {{Byer}}},\ }in\ \href@noop {} {\emph {\bibinfo {booktitle} {Society of
  Photo-Optical Instrumentation Engineers (SPIE) Conference Series}}},\
  \bibinfo {series} {Presented at the Society of Photo-Optical Instrumentation
  Engineers (SPIE) Conference}, Vol.\ \bibinfo {volume} {2989},\ \bibinfo
  {editor} {edited by\ \bibinfo {editor} {\bibfnamefont {U.~O.}\ \bibnamefont
  {{Farrukh}}}\ and\ \bibinfo {editor} {\bibfnamefont {S.}~\bibnamefont
  {{Basu}}}}\ (\bibinfo {year} {1997})\ pp.\ \bibinfo {pages}
  {57--68}\BibitemShut {NoStop}%
\bibitem [{\citenamefont {Willke}\ \emph {et~al.}(1998)\citenamefont {Willke},
  \citenamefont {Uehara}, \citenamefont {Gustafson}, \citenamefont {Byer},
  \citenamefont {King}, \citenamefont {Seel},\ and\ \citenamefont
  {R.~L.~Savage}}]{WUGBKSS98}%
  \BibitemOpen
  \bibfield  {author} {\bibinfo {author} {\bibfnamefont {B.}~\bibnamefont
  {Willke}}, \bibinfo {author} {\bibfnamefont {N.}~\bibnamefont {Uehara}},
  \bibinfo {author} {\bibfnamefont {E.~K.}\ \bibnamefont {Gustafson}}, \bibinfo
  {author} {\bibfnamefont {R.~L.}\ \bibnamefont {Byer}}, \bibinfo {author}
  {\bibfnamefont {P.~J.}\ \bibnamefont {King}}, \bibinfo {author}
  {\bibfnamefont {S.~U.}\ \bibnamefont {Seel}}, \ and\ \bibinfo {author}
  {\bibfnamefont {J.}~\bibnamefont {R.~L.~Savage}},\ }\href
  {http://ol.osa.org/abstract.cfm?URI=ol-23-21-1704} {\bibfield  {journal}
  {\bibinfo  {journal} {Opt. Lett.}\ }\textbf {\bibinfo {volume} {23}},\
  \bibinfo {pages} {1704} (\bibinfo {year} {1998})}\BibitemShut {NoStop}%
\bibitem [{\citenamefont {Bolduc}\ \emph {et~al.}(2013)\citenamefont {Bolduc},
  \citenamefont {Bent}, \citenamefont {Santamato}, \citenamefont {Karimi},\
  and\ \citenamefont {Boyd}}]{Bolduc13}%
  \BibitemOpen
  \bibfield  {author} {\bibinfo {author} {\bibfnamefont {E.}~\bibnamefont
  {Bolduc}}, \bibinfo {author} {\bibfnamefont {N.}~\bibnamefont {Bent}},
  \bibinfo {author} {\bibfnamefont {E.}~\bibnamefont {Santamato}}, \bibinfo
  {author} {\bibfnamefont {E.}~\bibnamefont {Karimi}}, \ and\ \bibinfo {author}
  {\bibfnamefont {R.~W.}\ \bibnamefont {Boyd}},\ }\href {\doibase
  10.1364/OL.38.003546} {\bibfield  {journal} {\bibinfo  {journal} {Opt.
  Lett.}\ }\textbf {\bibinfo {volume} {38}},\ \bibinfo {pages} {3546} (\bibinfo
  {year} {2013})}\BibitemShut {NoStop}%
\bibitem [{\citenamefont {Brown}\ and\ \citenamefont
  {Freise}(2017)}]{Brown2017}%
  \BibitemOpen
  \bibfield  {author} {\bibinfo {author} {\bibfnamefont {D.}~\bibnamefont
  {Brown}}\ and\ \bibinfo {author} {\bibfnamefont {A.}~\bibnamefont {Freise}},\
  }\href {\doibase 10.5281/zenodo.821389} {\enquote {\bibinfo {title}
  {Pykat},}\ } (\bibinfo {year} {2017}),\ \bibinfo {note}
  {{\url{http://www.gwoptics.org/pykat}}}\BibitemShut {NoStop}%
\bibitem [{\citenamefont {{Bayer-Helms}}(1984)}]{bayer-helms}%
  \BibitemOpen
  \bibfield  {author} {\bibinfo {author} {\bibfnamefont {F.}~\bibnamefont
  {{Bayer-Helms}}},\ }\href@noop {} {\bibfield  {journal} {\bibinfo  {journal}
  {Appl. Opt.}\ }\textbf {\bibinfo {volume} {23}},\ \bibinfo {pages} {1369}
  (\bibinfo {year} {1984})}\BibitemShut {NoStop}%
\bibitem [{\citenamefont {{Virtanen}}\ \emph {et~al.}(2019)\citenamefont
  {{Virtanen}}, \citenamefont {{Gommers}}, \citenamefont {{Oliphant}},
  \citenamefont {{Haberland}}, \citenamefont {{Reddy}}, \citenamefont
  {{Cournapeau}}, \citenamefont {{Burovski}}, \citenamefont {{Peterson}},
  \citenamefont {{Weckesser}}, \citenamefont {{Bright}}, \citenamefont {{van
  der Walt}}, \citenamefont {{Brett}}, \citenamefont {{Wilson}}, \citenamefont
  {{Jarrod Millman}}, \citenamefont {{Mayorov}}, \citenamefont {{Nelson}},
  \citenamefont {{Jones}}, \citenamefont {{Kern}}, \citenamefont {{Larson}},
  \citenamefont {{Carey}}, \citenamefont {{Polat}}, \citenamefont {{Feng}},
  \citenamefont {{Moore}}, \citenamefont {{Vand erPlas}}, \citenamefont
  {{Laxalde}}, \citenamefont {{Perktold}}, \citenamefont {{Cimrman}},
  \citenamefont {{Henriksen}}, \citenamefont {{Quintero}}, \citenamefont
  {{Harris}}, \citenamefont {{Archibald}}, \citenamefont {{Ribeiro}},
  \citenamefont {{Pedregosa}}, \citenamefont {{van Mulbregt}},\ and\
  \citenamefont {{Contributors}}}]{scipy}%
  \BibitemOpen
  \bibfield  {author} {\bibinfo {author} {\bibfnamefont {P.}~\bibnamefont
  {{Virtanen}}}, \bibinfo {author} {\bibfnamefont {R.}~\bibnamefont
  {{Gommers}}}, \bibinfo {author} {\bibfnamefont {T.~E.}\ \bibnamefont
  {{Oliphant}}}, \bibinfo {author} {\bibfnamefont {M.}~\bibnamefont
  {{Haberland}}}, \bibinfo {author} {\bibfnamefont {T.}~\bibnamefont
  {{Reddy}}}, \bibinfo {author} {\bibfnamefont {D.}~\bibnamefont
  {{Cournapeau}}}, \bibinfo {author} {\bibfnamefont {E.}~\bibnamefont
  {{Burovski}}}, \bibinfo {author} {\bibfnamefont {P.}~\bibnamefont
  {{Peterson}}}, \bibinfo {author} {\bibfnamefont {W.}~\bibnamefont
  {{Weckesser}}}, \bibinfo {author} {\bibfnamefont {J.}~\bibnamefont
  {{Bright}}}, \bibinfo {author} {\bibfnamefont {S.~J.}\ \bibnamefont {{van der
  Walt}}}, \bibinfo {author} {\bibfnamefont {M.}~\bibnamefont {{Brett}}},
  \bibinfo {author} {\bibfnamefont {J.}~\bibnamefont {{Wilson}}}, \bibinfo
  {author} {\bibfnamefont {K.}~\bibnamefont {{Jarrod Millman}}}, \bibinfo
  {author} {\bibfnamefont {N.}~\bibnamefont {{Mayorov}}}, \bibinfo {author}
  {\bibfnamefont {A.~R.~J.}\ \bibnamefont {{Nelson}}}, \bibinfo {author}
  {\bibfnamefont {E.}~\bibnamefont {{Jones}}}, \bibinfo {author} {\bibfnamefont
  {R.}~\bibnamefont {{Kern}}}, \bibinfo {author} {\bibfnamefont
  {E.}~\bibnamefont {{Larson}}}, \bibinfo {author} {\bibfnamefont
  {C.}~\bibnamefont {{Carey}}}, \bibinfo {author} {\bibfnamefont
  {{\.I}.}~\bibnamefont {{Polat}}}, \bibinfo {author} {\bibfnamefont
  {Y.}~\bibnamefont {{Feng}}}, \bibinfo {author} {\bibfnamefont {E.~W.}\
  \bibnamefont {{Moore}}}, \bibinfo {author} {\bibfnamefont {J.}~\bibnamefont
  {{Vand erPlas}}}, \bibinfo {author} {\bibfnamefont {D.}~\bibnamefont
  {{Laxalde}}}, \bibinfo {author} {\bibfnamefont {J.}~\bibnamefont
  {{Perktold}}}, \bibinfo {author} {\bibfnamefont {R.}~\bibnamefont
  {{Cimrman}}}, \bibinfo {author} {\bibfnamefont {I.}~\bibnamefont
  {{Henriksen}}}, \bibinfo {author} {\bibfnamefont {E.~A.}\ \bibnamefont
  {{Quintero}}}, \bibinfo {author} {\bibfnamefont {C.~R.}\ \bibnamefont
  {{Harris}}}, \bibinfo {author} {\bibfnamefont {A.~M.}\ \bibnamefont
  {{Archibald}}}, \bibinfo {author} {\bibfnamefont {A.~H.}\ \bibnamefont
  {{Ribeiro}}}, \bibinfo {author} {\bibfnamefont {F.}~\bibnamefont
  {{Pedregosa}}}, \bibinfo {author} {\bibfnamefont {P.}~\bibnamefont {{van
  Mulbregt}}}, \ and\ \bibinfo {author} {\bibfnamefont {S.~.~.}\ \bibnamefont
  {{Contributors}}},\ }\href@noop {} {\bibfield  {journal} {\bibinfo  {journal}
  {arXiv e-prints}\ ,\ \bibinfo {eid} {arXiv:1907.10121}} (\bibinfo {year}
  {2019})},\ \Eprint {http://arxiv.org/abs/1907.10121} {arXiv:1907.10121
  [cs.MS]} \BibitemShut {NoStop}%
\bibitem [{\citenamefont {Mor{\'e}}(1978)}]{more1978levenberg}%
  \BibitemOpen
  \bibfield  {author} {\bibinfo {author} {\bibfnamefont {J.~J.}\ \bibnamefont
  {Mor{\'e}}},\ }in\ \href@noop {} {\emph {\bibinfo {booktitle} {Numerical
  analysis}}}\ (\bibinfo  {publisher} {Springer},\ \bibinfo {year} {1978})\
  pp.\ \bibinfo {pages} {105--116}\BibitemShut {NoStop}%
\bibitem [{\citenamefont {Jollivet}\ \emph {et~al.}(2014)\citenamefont
  {Jollivet}, \citenamefont {Flamm}, \citenamefont {Duparr\'{e}},\ and\
  \citenamefont {Sch\"{u}lzgen}}]{Jollivet14}%
  \BibitemOpen
  \bibfield  {author} {\bibinfo {author} {\bibfnamefont {C.}~\bibnamefont
  {Jollivet}}, \bibinfo {author} {\bibfnamefont {D.}~\bibnamefont {Flamm}},
  \bibinfo {author} {\bibfnamefont {M.}~\bibnamefont {Duparr\'{e}}}, \ and\
  \bibinfo {author} {\bibfnamefont {A.}~\bibnamefont {Sch\"{u}lzgen}},\ }\href
  {http://jlt.osa.org/abstract.cfm?URI=jlt-32-6-1068} {\bibfield  {journal}
  {\bibinfo  {journal} {J. Lightwave Technol.}\ }\textbf {\bibinfo {volume}
  {32}},\ \bibinfo {pages} {1068} (\bibinfo {year} {2014})}\BibitemShut
  {NoStop}%
\end{thebibliography}%

\end{document}